\documentclass{jfm}

\usepackage{graphicx}
\usepackage{newtxtext}
\usepackage{newtxmath}
\usepackage{natbib}
\usepackage{hyperref}
\usepackage{caption}
\usepackage{amssymb}
\hypersetup{
    colorlinks = true,
    urlcolor   = blue,
    citecolor  = black,
}

\newcommand{\RomanNumeralCaps}[1]

\title{On the algebraic stretching dynamics of variable-density mixing in shock-bubble interaction}

\author{Xu Han\aff{1},
  Bin Yu\aff{1}\corresp{\email{kianyu@sjtu.edu.cn}}
 \and Hong Liu\aff{1}\corresp{\email{hongliu@sjtu.edu.cn}}}

\affiliation{\aff{1}J.C. Wu Center for Aerodynamics, School of Aeronautics and Astronautics, Shanghai Jiao Tong University, Shanghai 200240, PR China}

\begin{document}
\maketitle

\makeatletter
\newcommand{\rmnum}[1]{\romannumeral #1}
\newcommand{\Rmnum}[1]{\expandafter\@slowromancap\romannumeral #1@}
\makeatother

\begin{abstract}

The mixing mechanism within a single-vortex has been a theoretical focus for decades, while it remains unclear especially under variable-density (VD) scenario. This study investigates canonical single-vortex VD mixing in shock-bubble interactions (SBI) through high-resolution numerical simulations. Special attention is paid to examine the stretching dynamics and its impact on VD mixing within a single-vortex, and this problem is investigated by quantitatively characterizing the scalar dissipation rate (SDR), namely mixing rate, and its time integral, referred to as mixedness. To study VD mixing, we first examine single-vortex passive-scalar (PS) mixing with the absence of density difference. Mixing originates from diffusion and is further enhanced by stretching dynamics. Under axisymmetry and zero diffusion assumptions, the single-vortex stretching rate illustrates an algebraic growth of the length of scalar strips over time. By incorporating diffusion process through the solution of the advection-diffusion equation along these stretched scalar strips, a PS mixing model for SDR is proposed based on the single-vortex algebraic stretching characteristic. Within this framework, density-gradient effects from two perspectives of stretching dynamics and diffusion process are discovered to challenge the extension of PS mixing model to VD mixing. First, the secondary baroclinic effect increases the VD stretching rate by the additional secondary baroclinic principal strain, while the algebraic stretching characteristic is still retained. Second, the density source effect, originating from the intrinsic nature of the density difference in the multi-component transport equation, suppresses the diffusion process. By accounting for both the secondary baroclinic effect on stretching and the density source effect on diffusion, a VD mixing model for SBI is further modified. This model establishes a quantitative relationship between the stretching dynamics and the evolution of the mixing rate and mixedness for single-vortex VD mixing over a broad range of Mach numbers. Furthermore, the essential role of stretching dynamics on the mixing rate is demonstrated by the derived dependence of the time-averaged mixing rate $\overline{\left\langle\chi\right\rangle}$ on the P\'{e}clet number $Pe$, which scales as $\overline{\left\langle \chi \right\rangle} \sim Pe^{\frac{2}{3}}$.

\end{abstract}

\begin{keywords}

\end{keywords}

\section{Introduction}
\label{Introduction}

Variable-density (VD) mixing, which occurs in fluids with varying densities, is a fundamental phenomenon observed in various contexts, such as the mixing of fluids with different densities in gravitational fields, exemplified by Rayleigh-Taylor instability \citep{rayleigh1882investigation,taylor1950instability}, the formation of supernova remnants \citep{kifonidis2006non,muller2020hydrodynamics} and the irreversible transport of salt or pollutants in density-stratified oceanic and atmospheric circulations \citep{fernando1991turbulent,wunsch2004vertical,caulfield2021layering}. A notable form of VD mixing occurs at the interface of fluids with different densities when subjected to shock waves, identified as the Ritchmyer-Meshkov instability (RMI) \citep{richtmyer1960taylor,meshkov1969instability,brouillette2002richtmyer,zhou2017rayleigh,zhou2017rayleigh2}. One typical RMI scenario involves the interaction between a shock wave and a circular gas bubble with a density contrast relative to the ambient gas. Unlike classical single-mode RMI, the high curvature of the density interface introduces strong nonlinear effects, complicating the extension of RM linear theory to such shock-bubble interactions (SBI) \citep{ranjan2011shock}. \textcolor{black}{As SBI serves as the intrinsic mechanism for supersonic mixing enhancement strategies \citep{marble1985growth,marble1990shock, yang1993applications, yang1994model,yu2020two}, and has significant applications in areas such as inertial confinement fusion (ICF) \citep{lindl1992progress} and supernova explosions \citep{klein1994hydrodynamic}, understanding SBI has become a central focus in VD mixing research.}

\textcolor{black}{With regard to the evolution of flow structures in SBI, \cite{haas1987interaction} conducted early experimental studies in a shock tube, revealing that vortex structures in SBI tend to evolve into symmetric or axisymmetric forms at the later stages. \cite{jacobs1992shock} utilized Planar Laser-Induced Fluorescence (PLIF) to visualize the interaction between a shock wave and a hydrogen gas column, focusing on the resulting flow structures and mixing dynamics. \cite{ranjan2007experimental} systematically investigated SBI at a higher Mach number $(Ma=2.88)$, identifying the formation of secondary vortex rings, which are absent in low Mach number scenarios. \cite{zhai2014interaction} conducted a carefully designed experiment to study shock–interface interactions and successfully captured three distinct types of Mach stem reflections, along with a detailed analysis of the corresponding shock wave system evolution. \cite{ding2017interaction,ding2018interaction} examined the morphology of a three-dimensional gas cylinder following interaction with a planar shock. In addition, \cite{liang2018interaction} investigated the interaction between a strong converging shock wave and an $\rm SF_6$ gas bubble inside a converging shock tube. From a numerical standpoint, \cite{picone1988vorticity} performed detailed two-dimensional simulations of SBI, demonstrating flow structures that closely resembled those observed in the experiments by \cite{haas1987interaction}. With the advancement of numerical techniques such as adaptive mesh refinement and high-order accurate schemes, numerical investigations of SBI have significantly improved in terms of resolution and physical fidelity. \cite{Zabusky1998shock} simulated SBI at $Ma=5$ and identified the generation of secondary baroclinic vortices. Furthermore, \cite{niederhaus2008computational} showed that for shock Mach numbers exceeding $5$ and Atwood numbers greater than $0.5$, the post-shock flow field within SBI transitions into turbulence, rather than remaining governed by vortex structures.}

\textcolor{black}{In the context of circulation models for SBI, early studies on vorticity distribution led to the development of the first circulation model by \cite{hawley1989vortex}. Building on this foundational research, subsequent studies introduced four types of circulation models: the ``PB model'' \citep{picone1988vorticity}, the ``Yang model'' \citep{yang1994model}, the ``SZ model'' \citep{sandoval1995dynamics}, and the ``1D model'' \citep{niederhaus2008computational}. These circulation models were later extended to more complex bubble shapes, including elliptical \citep{zhang2019numerical}, triangular \citep{zeng2018numerical}, and square bubbles \citep{igra2018numerical}. Furthermore, these circulation models do not account for viscous effect \citep{wang2018scaling}. By incorporating viscosity into the vorticity transport equation, more accurate vortex models can be developed for complex configurations \citep{liu2020contribution}.}

\textcolor{black}{From the perspective of mixing, the mixing evolution in SBI can be conceptually outlined as follows:} following shock interaction, the formation of a large-scale primary vortex benefits from the baroclinic vorticity deposition originating from misalignment of the pressure gradient $\nabla p$ of the shock and density gradient $\nabla \rho$ from the light/heavy density bubble. Driven by this main vortex, the initially concentrated bubble is tangentially stretched along the interface with the ambient air, leading to an increase in material line length. Simultaneously, as the interface undergoes stretching, the scalar gradient deposited along it is amplified, thereby intensifying molecular diffusion until the bubble is completely mixed with the surrounding air \citep{ranjan2011shock}. This simplified mixing process reveals mixing are governed by the coupled actions of stretching and diffusion processes, which is consistent with the one proposed by Villermaux et al. \citep{meunier2003vortices,villermaux2019mixing}.  According to this simplified mixing conception, the material line stretching rate and bubble area serve as the critical parameters characterizing mixing, and have been widely used in SBI mixing research. Early experiments conducted by \citet{jacobs1992shock} utilized the bubble area to depict the extent of mixing and revealed a constant mixing rate by investigating the temporal derivative of the bubble area ratio. \citet{yang1993applications} employed the material line stretching rate to characterize the mixing rate, and discovered an exponential stretching of the material line in the early stage of SBI mixing. Subsequent experiments on SBI of $\rm SF_6$ cylinders using the PLIF technique observed a strong dependence of the exponential stretching rate on the configuration and orientation of the cylinders \citep{kumar2005stretching}, while numerical simulations of mixing in three-dimensional shock spherical interactions employed the bubble volume fraction to describe the evolution of mixedness \citep{niederhaus2008computational}.

Moreover, mixing is characterized by the molecular diffusion of the mass fraction $Y$. Therefore, in addition to the material line stretching rate, mixing indicators can also be directly defined in terms of the mass fraction. In a study of single-vortex mixing, \citet{cetegen1993experiments} introduced the mixedness parameter,
\begin{equation}
    f=4Y(1-Y),
    \label{mixedness definition}
\end{equation}
to quantify the extent of mixing, which reaches its maximum when the bubble and surrounding air are in a 1:1 state. \textcolor{black}{The evolution of mixedness $f$ in passive-scalar (PS) mixing, where the density difference is absent, is governed by the following equation:
\begin{equation}
    \left(\frac{\partial }{\partial t} + u_j\frac{\partial}{\partial x_j}-\mathcal{D}\frac{\partial^2}{\partial x_j^2}\right)f = 8\mathcal{D}\frac{\partial Y}{\partial x_j}\frac{\partial Y}{\partial x_j}.
    \label{mixedness equation in PS mixing}
\end{equation}
The presence of $\chi = \nabla Y\cdot\nabla Y$ on the right-hand side term of this equation $8\mathcal{D}\frac{\partial Y}{\partial x_j}\frac{\partial Y}{\partial x_j}$ leads to the definition of the scalar dissipation rate (SDR) \citep{buch1996experimental}:
\begin{equation}
    \chi = \nabla Y \cdot \nabla Y,
    \label{SDR definition}
\end{equation}
which represents the mixing rate, and $\nabla Y$ denotes the scalar gradient.} Subsequent PLIF experiments investigated SDR in shock interactions with $\rm SF_6$ cylinders, revealing that SDR is primarily concentrated in the bridge structure \citep{tomkins2008experimental}. \citet{li2019gaussian} developed a Gaussian model to predict the late-time evolution of SBI based on the mixedness definition. Recognizing the differences between variable-density (VD) and passive-scalar (PS) mixing, \citet{yu2020scaling} introduced the density-accelerated mixing rate. Through an investigation of the scaling behavior of this mixing rate with respect to the Reynolds number $Re={\varGamma_{t}}/{\nu}$ and \textcolor{black}{Schmidt number $Sc={\nu}/{\mathcal{D}}$}, where $\varGamma_{t}$ is the total circulation of the primary vortex, $\nu$ is the kinematic viscosity, and $\mathcal{D}$ is the Fickian diffusivity, it was discovered that at a constant P\'{e}clet number $Pe={\varGamma_{t}}/{\mathcal{D}}$, the mixing rate experiences minimal variation with increasing Reynolds number once $Re$ reaches a sufficiently large value. This scaling behavior was further confirmed in homogeneous isotropic turbulence mixing \citep{buaria2021turbulence}. Therefore, extensive research in the field of SBI has provided valuable insights into characterizing stretching behavior and diffusion states from two key perspectives of the mixing process. Despite the widely acknowledged opinion that the stretching process plays the critical role in mixing, the precise connection between the stretching dynamics and VD mixing in SBI has not been established rigorously.

It is necessary to define the stretching rate within a reasonable framework supported by the governing equation for SDR. This equation, derived by \citet{corrsin1953remarks}, \citet{batchelor1959small}, and \citet{buch1996experimental}, has been widely used to study PS mixing, which involves the mixing of fluids with uniform density. The governing equation is expressed as:
\begin{equation}
    \left(\frac{\partial}{\partial t} + u_{j}\frac{\partial}{\partial x_{j}} - \mathcal{D}\frac{\partial}{\partial x_{j}^2}\right)\chi = -2\frac{\partial Y}{\partial x_{i}}S_{ij}\frac{\partial Y}{\partial x_{j}} - 2\mathcal{D}\frac{\partial}{\partial x_{i}}\left(\frac{\partial Y}{\partial x_j}\right)\frac{\partial}{\partial x_{i}}\left(\frac{\partial Y}{\partial x_j}\right),
    \label{PS SDR equation}
\end{equation}
where $S_{ij} = \frac{1}{2}\left(\frac{\partial u_{i}}{\partial x_{j}} + \frac{\partial u_{j}}{\partial x_{i}}\right)$ is the symmetric strain rate tensor. Two terms on the right-hand side of the equation are referred to as the stretching term and the diffusion term, respectively. The stretching term reflects the intrinsic mechanism by which the strain field stretches the magnitude of the scalar gradient $\nabla Y$, thereby driving SDR evolution. This term can be further investigated by its orthogonal decomposition form \citep{ashurst1987alignment,vincent1991spatial,dresselhaus1992kinematics}, given by:
\begin{equation}
    T_{stretch}^{\chi} = -2\frac{\partial Y}{\partial x_{i}}S_{ij}\frac{\partial Y}{\partial x_{j}} = -2\chi s_{i}\lambda_{i}^2 = \chi R_{stretch},
    \label{stretch rate definition}
\end{equation}
where $s_{i}$ is the eigenvalue of the symmetric strain rate tensor $S_{ij}$, referred to as the principal strain rate, and $\boldsymbol{e}_{i}$ is the corresponding eigenvector. The alignment of the scalar gradient with $\boldsymbol{e}_{i}$ is given by $\lambda_{i} = \frac{\nabla Y \cdot e_{i}}{|\nabla Y|}$. Therefore, the suitable definition of stretching rate is expressed as the combination of principal strain rate and alignment of scalar gradient \citep{han2024mixing}:  
\begin{equation}
    R_{stretch}=-2s_{i}\lambda_{i}^2.
    \label{stretching rate definition}
\end{equation}
As demonstrated in various mixing scenarios, such as homogeneous isotropic turbulence \citep{girimaji1990material,vincent1991spatial,danish2016influence}, shock-turbulence interaction \citep{tian2017numerical,gao2020parametric}, and turbulent combustion \citep{swaminathan2006interaction,zhao2018dynamics}, the stretching rate remains a positive value in general cases. Consequently, the stretching term is typically considered as the primary source of SDR growth. 

Given the significance of the stretching term, the stretching mechanism has been primarily investigated from two perspectives. The first approach involves statistical analysis. Based on the persistent straining hypothesis, the stretching of the scalar gradient is treated as a series of independent, sequential stretching processes with short correlation time \citep{townsend1951diffusion,batchelor1952effect,cocke1969turbulent}. Under this assumption, the turbulent stretching rate is modeled as a Gaussian distribution, leading to the well-known exponential stretching observed in isotropic turbulence. This statistical framework has been extended to other types of flows, such as porous media mixing \citep{le2013stretching} and sheared particulate suspensions \citep{souzy2018mixing}. The second perspective examines the influence of local velocity gradients on the stretching mechanism. As expressed in Eq. \ref{stretch rate definition}, the stretching rate is the product of the principal strain rate and alignment, both of which are closely related to local velocity gradients. By decomposing the velocity gradient and scalar gradient equations on the frame composed by the eigenvectors of the symmetric strain rate tensor, \citet{she1991structure} and \citet{dresselhaus1992kinematics} derived the principal strain rate and alignment equations. These governing equations demonstrate that both the principal strain rate and alignment are influenced by self-interaction, pressure Hessian, and viscous effects \citep{nomura1998structure,lapeyre1999does,lapeyre2001dynamics,tom2021exploring}. Since these quantities are critical to characterizing small-scale flow structures, the dynamics of stretching rate within various flow structures, such as flow topologies, have attracted significant attention \citep{brethouwer2003micro,danish2016influence,zhao2018dynamics}. Specifically, for the single-vortex, it has been revealed that the axisymetric distribution of the velocity gradients and pressure Hessian matrix lead to its algebraic stretching characteristic \citep{lapeyre1999does}. 
These comprehensive investigations have deepened our understanding of the stretching mechanism in  various types of flow. It is also noteworthy that the exponential stretching characteristic identified in SBI \citep{yang1993applications,kumar2005stretching} contradicts the algebraic stretching characteristic in a single-vortex \citep{lapeyre1999does}, a contradiction that will be further explored in the subsequent sections of this study.

In contrast to the relatively well-developed understanding of the stretching mechanism, the relationship between SDR and stretching dynamics has been addressed qualitatively. It is generally concluded that flow structures with high stretching rates promote the formation of scalar structures with high values of SDR \citep{boukharfane2018evolution, gao2020parametric}. The primary obstacle to developing a quantitative theory for SDR lies in the insufficient understanding of the non-linear diffusion term in Eq. \ref{PS SDR equation}, specifically $ - 2\mathcal{D}\frac{\partial}{\partial x_{i}}\left(\frac{\partial Y}{\partial x_j}\right)\frac{\partial}{\partial x_{i}}\left(\frac{\partial Y}{\partial x_j}\right)$. This term is strictly negative, representing the reduction in SDR due to the diffusional cancellation of gradients in the scalar field. Since the mathematical expression of the diffusion term involves the scalar gradient $\frac{\partial Y}{\partial x_{i}}$, the diffusion process is also influenced by stretching dynamics. The pioneering work by \citet{ranz1979applications} states that the relationship between the diffusion process and stretching rate in PS mixing can be established analytically. By solving the advection-diffusion equation for deforming scalar strips, the diffusion of the mass fraction $Y$, enhanced by the stretching rate, can be analytically described. This approach has led to key concepts such as mixing time \citep{meunier2003vortices} and the Batchelor mixing scale \citep{souzy2018mixing}. Villermaux's review \citep{villermaux2019mixing} provides a comprehensive overview of the theoretical methodology, which has been widely applied to characterize the distribution of mass fraction $Y$ in ideal vortices \citep{villermaux2003mixing}, turbulent jets \citep{duplat2008mixing, duplat2010nonsequential}, porous media \citep{le2013stretching}, and stratified shear flows \citep{petropoulos2023settling}. \citet{liu2022mixing} firstly extended this framework to describe the diffusion process in VD SBI. They proposed a model for mixing time by quantitatively expressing the VD stretching rate, which is enhanced by secondary baroclinic vorticity. This model offers a reliable prediction for the mixing time, defined as the time required to achieve a steady mixing state in VD SBI, and highlights the significant challenges posed by the inherent density variations in exploring the diffusion process in VD mixing. \textcolor{black}{However, the analysis of mixing time only necessitates the local stretching rate at the characteristic location. Therefore, the stretching rate in this theory was not analyzed from the perspective of stretching dynamics, which limits its predictive capability for mixing time in VD SBI. Consequently, a comprehensive quantitative theory that elucidates the role of the stretching rate across the entire flow field in the evolution of the mixing process, as reflected by SDR within VD mixing, remains to be developed.} 

In summary, the intrinsic mechanism of VD mixing in SBI, particularly from the perspective of stretching dynamics and its role on mixing, remains insufficiently understood. This study aims to elucidate this mechanism by providing a quantitative description of the evolution of the mixing rate SDR and mixedness, as a significant advancement of our previous work on the VD mixing time in SBI \citep{liu2022mixing}. Beginning with the relatively simple case of PS SBI, the stretching rate within a single-vortex is firstly explored from the standpoint of stretching dynamics, leading to the identification of an algebraic stretching characteristic. Based on the analytical expression of this stretching rate, a PS mixing model for SDR is proposed by considering the diffusion process on a series of stretched scalar strips. Within this framework, by integrating two density-gradient effects, referred to as the secondary baroclinic effect on stretching dynamics and the density source effect on diffusion process, the rigorous relationship between the stretching dynamics and the mixing rate SDR (mixedness) in VD mixing is established through a well-developed VD mixing model. Moreover, the critical role of stretching dynamics on the mixing rate can be demonstrated through scaling behavior of time-averaged mixing rate with dimensionless P\'{e}clet number.

The organization of this paper is as follows: \S\ \ref{Numerical method} provides an overview of the numerical method and case setup. In \S\ \ref{Stretching dynamics analysis preliminaries}, we outline the framework of the stretching dynamics, and introduce the simplest case of stretching dynamics, specifically the stretching process within a single-vortex, as the foundation for subsequent analysis. Based on the analytical expression of single-vortex stretching rate, a relatively simplified model of PS mixing is proposed in \S\ \ref{PS mixing theory}. \S\ \ref{VD mixing theory} discusses two density-gradient effects influencing VD mixing, which inspire the development of VD mixing model. In \S\ \ref{Model validation and scaling behavior of the mixing rate}, the VD mixing model is validated through its application to predict the evolution of SDR and mixedness in VD SBI across a wide range of Mach numbers. The crucial role of stretching rate in VD mixing is also demonstrated using a scaling analysis of the time-averaged SDR with respect to the P\'{e}clet number. Finally, our study is summarized and potential future research directions are discussed in \S\ \ref{Conclusions and future work}.

\section{Numerical method and setup}
\label{Numerical method}
\subsection{Numerical method}
In this study, the governing equations used for simulating PS and VD SBI are the compressible Navier-Stokes equations with multiple components, expressed as:
\begin{equation}
  \left\{
\begin{aligned}
& \frac{\partial{\rho}}{\partial  t} + \frac{\partial ({\rho}{u_i})}{\partial {x_i}} = 0, \\
& \frac{\partial\left({\rho}{u_i}\right)}{\partial  t} + \frac{\partial ({\rho}{u_i}{u_j})}{\partial {x_j}} = -\frac{\partial{p}}{\partial{x_i}} + \frac{\partial {\sigma}_{ij}}{\partial {x_j}}, \\
&  \frac{\partial({\rho}{E})}{\partial {t}} + \frac{\partial ({\rho}{u_i}{H})}{\partial {x_i}} = -\frac{\partial {q_i}}{\partial {x_i}} + \frac{\partial({u_j}{\sigma}_{ij})}{\partial {x_i}},\\
&\frac{\partial({\rho}{Y}_m)}{\partial {t}} + \frac{\partial({\rho}{u_i}{Y}_m)}{\partial {x_i}} = \frac{\partial}{\partial {x_i}}\left({\rho}\mathcal{D}\frac{\partial {Y}_m}{\partial {x_i}}\right),\\
& m = 1,2,\cdots,s-1.
\end{aligned} 
  \right.
  \label{NSequation}
\end{equation}
Here, the symbols $\rho$, $p$, $E$, $H = E + \frac{p}{\rho}$ denote the mixture's density, pressure, total energy per unit mass, total enthalpy per unit mass, respectively. $u_i$ is the velocity of the mixture in the $ith$ direction, $Y_{m}$ refers to the mass fraction of species $m$ with $\sum\limits_{m=1}^{s} Y_{m} = 1$, and $s$ is labeled as the total number of species. With the ideal gas hypothesis and Dalton's law of partial pressures, the system of Eq \ref{NSequation} is closed by the ideal gas state equation: 
\begin{equation}
    p = \sum\limits_{m=1}^{s} p_{m} = \sum\limits_{m=1}^{s} \rho_{m}\frac{\mathcal{R}_{0}}{M_{m}}T = \rho \mathcal{R}T = \left(\gamma - 1\right)\rho e,
    \label{ideal gas state equation}
\end{equation}
where $\mathcal{R} = \mathcal{R}_{0}/M$ and $\mathcal{R}_{0} = 8.314\ \rm{J \left(mol\ K\right)^{-1}}$ are the gas constant and universal gas constant, respectively; $M = \left(\sum\limits_{m=1}^{s}Y_{m}/M_{m}\right)^{-1}$ and $T$ are the molar mass and temperature of mixture respectively; $p_{m}, \rho_{m} = \rho Y_{m}$ and $M_{m}$ represent the partial pressure, density and molar mass of the component $m$, respectively. The relation of the total energy per unit mass $E$ and the specific internal energy $e$ is given by:
\begin{equation}
    E = e + \frac{1}{2} u_{i}u_{i} = h - \frac{p}{\rho} + \frac{1}{2} u_{i}u_{i}.
\end{equation}
Here, $h$ is the enthalpy of the gas mixture,
\begin{equation}
    h = \sum\limits_{m=1}^{s} Y_{m}h_{m} = \sum\limits_{m=1}^{s} Y_{m}\left(h_{fm}^{0} + \int_{T_0}^{T}\frac{C_{pm}}{M_{m}}dT\right),
\end{equation}
in which $h_{fm}^{0}$ is the heat generated by the component $m$ at the reference temperature $T_{0}$. The constant-pressure specific heat value $C_{pm}$ is fitted by the following temperature-dependent polynomial function:
\begin{equation}
    C_{pm} = \mathcal{R}_{0}(a_{1m} + a_{2m}T + a_{3m}T^2 + a_{4m}T^3 + a_{5m}T^4),
\end{equation}
where the coefficients $a_{1m},....,a_{5m}$ are obtained from the NASA thermochemical polynomial fit coefficient data \citep{kee1996chemkin}.

With the definition of the constant-volume specific heat:
\begin{equation}
    C_{pm} - C_{vm} = \frac{\mathcal{{R}}_{0}}{M_{m}},
\end{equation}
the constant-pressure specific heat and constant-volume specific heat of gas mixture are $C_p = \sum\limits_{m=1}^{s} Y_{m}C_{pm}$ and  $C_v = \sum\limits_{m=1}^{s} Y_{m}C_{vm}$, and the specific heat ratio of gas mixture is $\gamma = C_{p}/C_{v}$.

For the Newtonian fluid considered in this work, the viscous stress tensor and heat flux are defined as:
\begin{equation}
  \left\{
\begin{aligned}
& {\sigma}_{ij} = \mu\left(\frac{\partial {u_i}}{\partial {x_j}} + \frac{\partial {u_j}}{\partial {x_i}} -  \frac{2}{3}\delta_{ij}\frac{\partial {u_k}}{\partial {x_k}}\right), \\
& {q_i} = -\lambda \frac{\partial {T}}{\partial {x_i}}, \\
& \lambda = C_p \mu/Pr.
\end{aligned} 
  \right.
  \label{Viscous tensor and heat flux}
\end{equation}
Here, $\delta_{ij}$ refers to the Kronecker delta. Based on the observation that the transport coefficients converge to nearly steady values after shock interaction  \citep{liu2022mixing}, constant transport coefficients are employed in the numerical simulations to simplify the expression of the mixing model in the subsequent sections. Specifically, the viscous stress tensor ${\sigma}_{ij}$ is calculated by the constant dynamic viscosity $\mu = 125 \times 10^{-6}\, \rm{Pa\cdot s}$. The mass diffusion is simplified by ignoring pressure and temperature diffusion, and Fickian diffusivity is assumed to be constant for different components, with a value of $\mathcal{D} = 355 \times 10^{-6}\, {\rm m^2\,s^{-1}}$. \textcolor{black}{Setting the diffusivity $\mathcal{D}$ as a constant does not accurately reflect the true physical situation. Assigning a constant value to $\mathcal{D}$ is typically used when designing SBI cases with varying P\'{e}clet numbers, which has also been employed in our previous studies \citep{yu2020scaling, yu2022effects}.} The heat flux $q_{i}$ is calculated by the Fourier's law of heat conduction, and the heat conduction coefficient $\lambda$ is calculated using a constant Prandtl number $Pr = 0.72$. After the mathematical model is constructed, the compressible Navier-Stokes equations are numerically solved using an in-house high-resolution code, $ParNS3D$, which has undergone extensive validation \citep{wang2018scaling,liang2019hidden,yu2022effects,liu2022mixing}. Temporal discretization is performed using the third-order Total Variation Diminishing (TVD) Runge-Kutta method. The fifth-order Weighted Essentially Non-Oscillatory (WENO) scheme \citep{liu1994weighted,jiang1996efficient} is employed for the discretization of convection terms, while the second-order central difference method discretizes the viscous terms. 

\subsection{Initial conditions for VD and PS SBI}

\begin{figure}
  \centering
    \includegraphics[clip=true,width=0.9\textwidth]{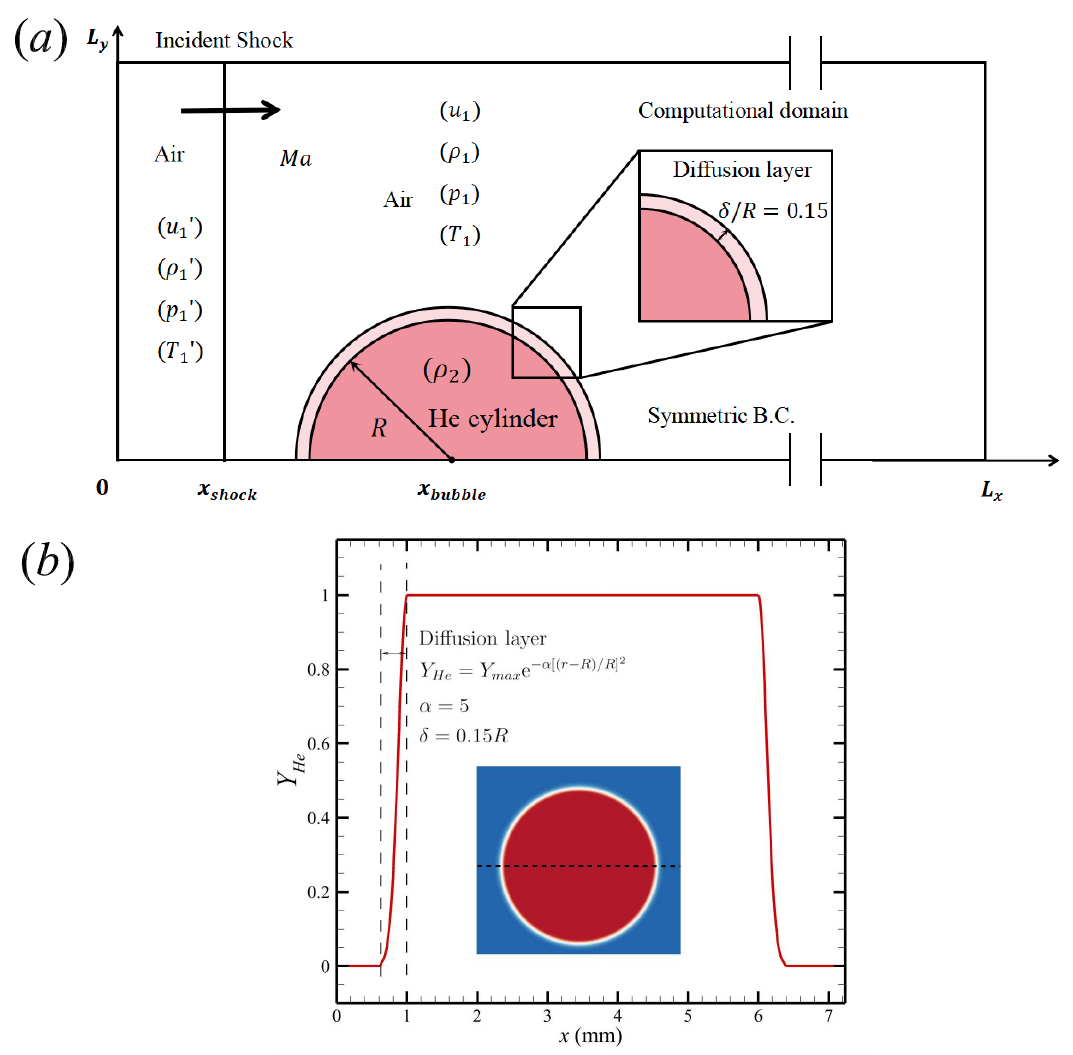}\\
    \caption{\textcolor{black}{$(a)$ Schematic of the initial conditions for VD SBI. $(b)$ Distribution of the initial mass fraction of helium $Y_{He}$ across the diameter of the bubble.}}
    \label{Initial conditions}
\end{figure}

This section presents the initial conditions for the numerical simulation of VD and PS SBI, using a case with a shock strength of $Ma = 2.4$ as an example. The case settings for other simulations, covering a broad range of Mach numbers from $1.22$ to $4$, are provided in Appendix \ref{wide range Ma number VD SBI settings}. The configuration for VD SBI settings is depicted in Fig. \ref{Initial conditions} $(a)$. The computation domain encompasses a two-dimensional space in a Cartesian coordinate system, with $X \times Y = [0, L_x] \times [0, L_y]$, where the length is $L_x = 30 \,\rm{mm}$ and the height is $L_y = 5.5\,\rm{mm}$. The shock is positioned at $x_{shock} = 0.7\,\rm{mm}$. The pre-shock air has the same thermodynamic states of $1\,\rm{atm}$ pressure and $293\,\rm{K}$ temperature prior to being impacted by the shock. The detailed initial conditions for post-shock air ($u_{1}', \rho_{1}', P_{1}'$, and $T_{1}'$) are calculated from the Rankine-Hugoniot equation. To avoid spurious vorticity arisen from the mesh discretisation \citep{niederhaus2008computational}, the helium bubble, centered at the position $x_{bubble} = 4.5\,\rm{mm}$, consists of a core region with a radius of $R=2.5\,{\rm mm}$ and a diffusive layer with a width of $\delta$. The diffusive layer is positioned at the edge of the core region, as illustrated in the inserted figure in Fig. \ref{Initial conditions} $(a)$. \textcolor{black}{Based on this configuration, the distribution of mass fraction of helium is set as: 
\begin{equation}
  Y_{He}(r) = \left\{
\begin{aligned}
& Y_{max}, & &r < R, \\
& Y_{max}\mathrm{e}^{-\alpha[(r-R)/\delta]^2}, & & R \leq r < R + \delta, \\
& 0. & &r \geq R + \delta, 
\end{aligned} 
  \right.
  \label{Mass fraction distribution}
\end{equation}
where $Y_{max} = 1.0$, $\alpha = 5$ and $\delta=5$. Using these parameters, the mass fraction distribution across the bubble is shown in Fig.\ref{Initial conditions} $(b)$. It can be observed that, with the selected parameters in Eq.~\ref{Mass fraction distribution}, the mass fraction within the thin diffusive layer smoothly connects the mass fraction inside and outside the bubble, which is consistent with the results reported in our previous studies \citep{wang2018scaling,liu2022mixing,yu2022effects}. To simplify subsequent expressions, the helium mass fraction $Y_{He}$ is denoted as $Y$.}

Since the influence of the reflected pressure wave on mixing is not considered in this study, the left boundary is set as an inlet, while the upper and right boundaries utilize a fourth-order extrapolation to prevent interference from pseudo-pressure reflection waves affecting the flow structures of interest. A symmetric boundary condition is applied at the bottom of the computational domain. \textcolor{black}{The grid resolution study, detailed in Appendix \ref{grid resolution study}, demonstrates that a uniform grid with a spacing of ${\Delta} = 1.25 \times 10^{-5}\,\rm{m}$ is sufficient to accurately resolve the mixing process for the VD SBI case. Accordingly, the time step for the numerical simulation is selected as $\Delta t = 2\times10^{-9}\ {\rm s}$ to ensure the Courant number satisfies the condition:
\begin{equation}
    {\rm CFL} = \frac{\Delta t}{\Delta/u_{max}} \approx \frac{\Delta t}{\Delta/u_2'} <1.
\end{equation}
Here, $u_2'$ indicates the post-shock velocity of the helium bubble, which approximates the maximum velocity within the computation domain.
}

To investigate the stretching dynamics and its impact on mixing in VD SBI, we begin by developing a quantitative mixing model using a relatively simple PS SBI case. In this scenario, the density difference between the post-shock bubble and the ambient air is eliminated, specifically setting the post-shock Atwood number $At^{+} = \left(\rho_2' - \rho_1'\right)/\left(\rho_2' + \rho_1'\right) \approx 0$, where $\rho_2'$ and $\rho_1'$ represent the post-shock bubble density and ambient air density, respectively. This adjustment allows the density effects to be effectively neglected as the vortex develops, ensuring that the mixing process does not influence the flow dynamics, as expected in PS mixing \citep{dimotakis2005turbulent}. \textcolor{black}{However, the formation of the primary vortex depends on baroclinic vorticity generated by the cylindrical interface of fluids with different densities subjected to shock. If the density of the original bubble matches that of the ambient air, no vorticity is deposited along the bubble interface. Therefore, an artificial density adjustment method is applied to complete the setup of the initial conditions for the PS SBI. On the basis of VD SBI case with the shock Mach number $Ma=2.4$, the cylindrical bubble’s mass fraction is adjusted artificially soon after the shock interaction:
\begin{equation}
    \left\{
    \begin{aligned}
        & Y_{He}^{ps}=0.0001\times Y_{He}^{VD},\\
        & Y_{O_2}^{ps}=0.233\times \left(1 - Y_{He}^{ps}\right),\\
        & Y_{N_2}^{ps}=0.767\times \left(1 - Y_{He}^{ps}\right),
    \end{aligned}
    \right.
\end{equation}
where a small amount of component is maintained to display the mixing process. With the pressure and temperature identical to those in VD SBI:
\begin{equation}
    \left\{
    \begin{aligned}
        & p^{ps}=p^{VD},\\
        & T^{ps}=T^{ps},
    \end{aligned}
    \right.
\end{equation}
the density of the bubble after the shock interaction will increase to the value of post-shock air:
\begin{equation}
    \left\{
    \begin{aligned}
        & \rho^{ps}=\frac{p^{ps}}{\mathcal{R}T^{ps}}\approx\rho^{air},\\
        & \mathcal{R}=\mathcal{R}_{0}\sum_{m=1}^{s}\frac{Y_{m}^{ps}}{M_{m}}.
    \end{aligned}
    \right.
\end{equation}
To ensure comparability of the mixing indicators between PS SBI and VD SBI, the mass fraction in PS SBI, denoted $Y^{ps}$, is normalized by the initial maximum mass fraction $Y_0^{max}$ in the following context, such that $Y = Y^{ps}/Y_0^{max}$. This method increases the bubble's density to match that of the post-shock air, as outlined in our prior research \citep{liu2022mixing}. It is worth noting that this artificial density adjustment, achieved through modification of the mass fraction, is a general technique.  It has been widely employed in the setup of various PS mixing scenarios, including PS SBI \citep{liu2022mixing}, PS shock–turbulence interaction \citep{tian2017numerical,tian2019density}, and PS mixing induced by random weak shock waves \citep{jossy2025mixing}.}

\begin{figure}
    \centering
    \includegraphics[width=1.0\linewidth]{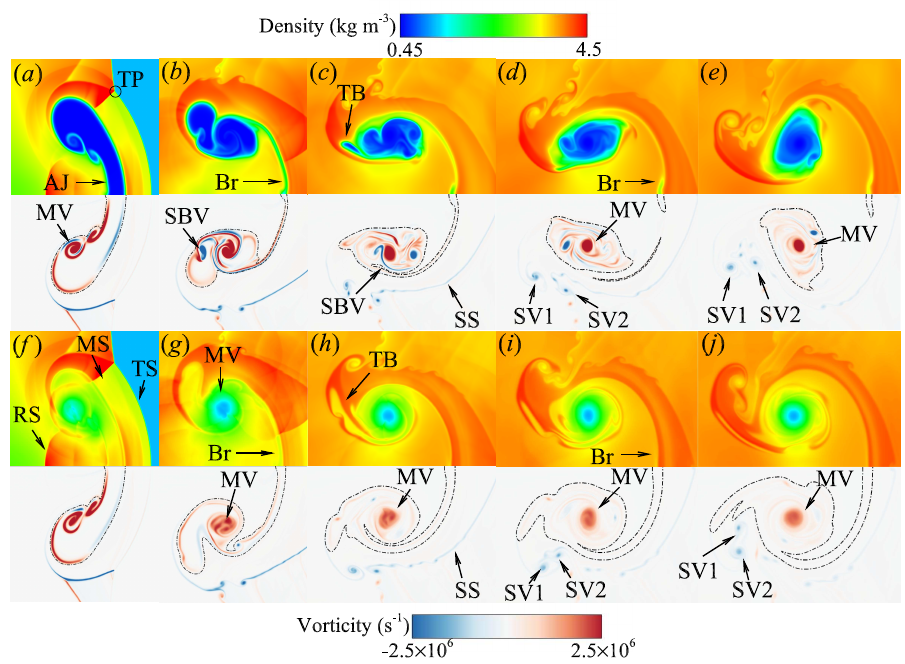}
    \captionsetup{justification=justified, singlelinecheck=false}
    \caption{Time evolution of the density contour (top) and the vorticity contour (bottom) for VD SBI case with shock Mach number 2.4 at $(a)\ t=10.8\ \rm{\upmu s}$; $(b)\ t=21.6\ \rm{\upmu s}$; $(c)\ t=32.4\ \rm{\upmu s}$; $(d)\ t=43.2\ \rm{\upmu s}$; and $(e)\ t=54.0\ \rm{\upmu s}$. The PS SBI case is presented at the same moments as that of VD SBI in the bottom from $(f)$ to $(j)$. Dashed-dot line indicates the isoline of $Y = 0.01Y_{max}$($Y_{max} = 1.0$ for VD SBI and $Y_{max} = 1.0$ for VD SBI). AJ, air jet; TP, triple point; MV, main vortex; MS, Mach stem; TS, transmitted shock; RS, reflected shock; Br, bridge; SBV, secondary baroclinic vorticity; TB, trailing bubble; SS, slip stream; SV1 and SV2, secondary vortex 1 and secondary vortex 2.}
    \label{density contour and vorticity contour for PS and VD SBI}
\end{figure}

\begin{figure}
    \centering
    \includegraphics[width=0.7\linewidth]{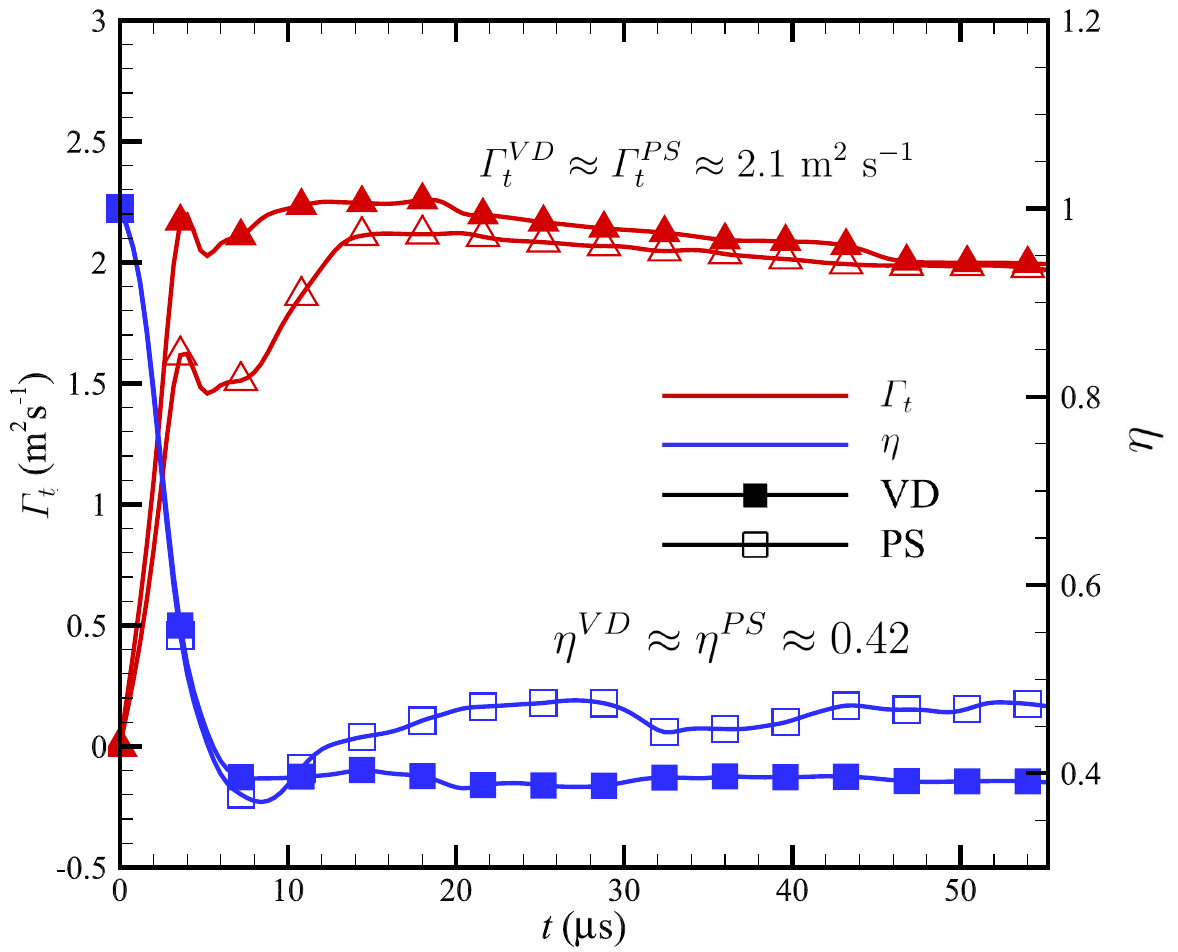}
    \captionsetup{justification=justified, singlelinecheck=false}
    \caption{Comparison of the total circulation $\varGamma_{t}$ and compression rate $\eta$ between the PS and VD SBI cases at a shock Mach number of 2.4.}
    \label{total circulation and compression rate in VD and PS SBI cases}
\end{figure}

The numerical simulation results of the VD and PS SB, based on the setup described, above are presented in Fig. \ref{density contour and vorticity contour for PS and VD SBI}, which depicts the time evolution of both density and vorticity contours. In the VD SBI case, positive primary baroclinic vorticity is generated due to the misalignment between the density gradient of the bubble and the pressure gradient induced by the shock. This primary vorticity drives the formation of a large-scale main vortex, as shown in the lower half of Fig. \ref{density contour and vorticity contour for PS and VD SBI}. As the main vortex develops, a vortex-bilayer structure emerges, characterized by dominant negative vorticity that intensifies and rolls up the positive vorticity. This structure, identified as secondary baroclinic vorticity (SBV), is consistent with previous studies \citep{gupta2003shock, peng2003vortex} and is recognized as a classical phenomenon in VD RMI \citep{peng2021mechanism, peng2021effects}. In contrast, the density contour of the PS SBI case shows that the artificial density adjustment method effectively removes the density difference between the bubble and ambient air. Consequently, the SBV vanishes, and only the main vortex is preserved. In both PS SBI and VD SBI cases, the formation of a nearly steady single-vortex is observed. Accordingly, the total circulation $\varGamma_{t}$ and the compression rate $\eta$ are examined. These two parameters are defined as:
\textcolor{black}{
\begin{equation}
  \left\{
\begin{aligned}
& \varGamma_{t} = \int_{Y > Y_{threshold}} \omega\ \rm{dV}\\
& \eta = \frac{\mathcal{V}_{b}}{\mathcal{V}_0} = \frac{\int_{Y > Y_{threshold}}\ \rm{dV}}{\mathcal{V}_0}, 
\end{aligned} 
  \right.
  \label{total circulation and compression rate definition}
\end{equation}
where $\omega$ is the magnitude of the vorticity vector $\mathbf{\omega} = \nabla \times \boldsymbol{u}$, $\mathcal{V}_{0}$ is the initial volume of the helium bubble, $\mathcal{V}_{b}$ is the volume of the compressed bubble following the shock interaction, and the threshold is chosen as $Y_{threshold}=1 \% Y_{max}$ to obtain the stable values of total circulation $\varGamma_{t}$ and the compression rate $\eta$ in the later stage of VD SBI.} Figure \ref{total circulation and compression rate in VD and PS SBI cases} illustrates the evolution of $\varGamma_{t}$ and $\eta$ in both VD and PS SBI cases. Following the shock interaction with the bubble, both the total circulation $\varGamma_t$ and the compression rate $\eta$ converge to nearly constant values, indicating the formation of a nearly steady single-vortex. \textcolor{black}{It is also observed that the total circulation $\varGamma_t$ in both the PS and VD SBI cases is nearly identical. Although the compression rate in the PS SBI case is about 20$\%$ greater than in the VD SBI case, the impact of this small difference in $\eta$ on the subsequent discussions is negligible. Therefore, the compression rate $\eta$ can be considered a consistent value of 0.42 for both the PS and VD SBI cases. Consequently, both the total circulation $\varGamma_t$ and the compression rate $\eta$ are unaffected by the specific settings of the PS SBI case.} Furthermore, due to the conservation of the total circulation $\varGamma_{t}$ and the alteration in bubble density in the PS SBI, the dynamic viscosity is adjusted to maintain the same dimensionless Reynolds number $Re = \varGamma_{t}/\nu$. This modification is expressed as $\mu^{ps} = \mu^{VD}\left(\frac{2\rho_{2}'}{\rho_{1}' + \rho_{2}'}\right)$, ensuring the kinematic viscosity $\nu = \mu / \overline{\rho'}$ remains constant. The maintenance of  $Pe = \varGamma_{t}/\mathcal{D}$ is achieved by keeping the same Fickian diffusivity $\mathcal{D}$. The setup for the PS SBI case is summarized in Table \ref{PS SBI parameters}.

\begin{table}
\begin{center}
\def~{\hphantom{0}}
  \begin{tabular}{lccccccccccc}
    \hline
    &  &$Ma$ & $p_{1}'\,\rm{(Pa)}$ & $T_{1}'\,\rm{(K)}$ & $u_{1}'\,\rm{(m\,s^{-1})}$ & $W_{t}\,\rm{(m\ s^{-1})}$ & $At^{+}\,\left(-\right)$ & $\mu\,\rm{(Pa\,s)}$ & $\varGamma_{t}\,\rm{(m^{2}\,s^{-1})}$ & $Re\,\left(-\right)$ & $Pe\,\left(-\right)$\\ 
    \hline
    & \textcolor{black}{VD} & 2.4 & 667391.3 & 596.87 & 568.31 & 1678.08 & -0.845 & $125.0 \times 10^{-6}$ & 2.09 & 38446.2 & 5887.3 \\
    & \textcolor{black}{PS} & 2.4 & 667391.3 & 596.87 & 568.31 & 1678.08 & 0.0 & $230.9 \times 10^{-6}$ & 2.09 & 38446.2 & 5887.3 \\
    \hline
  \end{tabular}
  \captionsetup{justification=justified, singlelinecheck=false}
  \caption{The setup for the PS SBI case based on the VD SBI case with a shock Mach number of $Ma = 2.4$. Compared to VD SBI, the post-shock parameters $p_{1}'$, $T_{1}'$, $u_{1}'$ and $T_{1}'$ in PS SBI remain unchanged, while the Atwood number, reflecting the post-shock density difference, is set to zero $\left(At^{+} = 0\right)$. With the total circulation $\varGamma_{t}$ preserved, the P\'{e}clet number is maintained, and the Reynolds number $Re$ is held constant by adjusting the dynamic viscosity $\mu$ in the PS SBI case.}
  \label{PS SBI parameters}
  \end{center}
\end{table}

The reduction of SBV due to the absence of density differences is a primary characteristic of PS mixing. Another feature of PS mixing is that the evolution of the mass fraction $Y$ follows the advection-diffusion equation \citep{warhaft2000passive}, expressed as:

\begin{equation}
    \frac{\partial Y}{\partial t} + \frac{\partial (u_i Y)}{\partial x_i} = \mathcal{D}\frac{\partial^2 Y}{\partial {x_j}^2}.
    \label{advection-diffusion equation}
\end{equation}
Here, the absence of density differences also leads to the reduction of the multi-component transport equation in Eq. \ref{NSequation} to the advection-diffusion equation, indicating a shift from VD mixing characteristic to PS mixing characteristic. As shown in Eq. \ref{NSequation}, the multi-component transport equation for the mass fraction of helium $Y$ is expressed as:

\begin{equation}
    \frac{\partial \left(\rho Y\right)}{\partial t} + \frac{\partial (\rho u_i Y)}{\partial x_i} = \frac{\partial}{\partial x_{j}}\left(\rho \mathcal{D} \frac{\partial Y}{\partial x_j}\right).
    \label{multi-component transport equation}
\end{equation}
Since the Fickian diffusivity $\mathcal{D}$ is treated as constant in our numerical simulations, this equation can be rewritten in the form VD advection-diffusion equation:
\begin{equation}
   \frac{\partial  Y}{\partial t} + \frac{\partial ( u_i Y)}{\partial x_i} = \frac{\mathcal{D}}{\rho}\left(\frac{\partial \rho}{\partial x_j}\frac{\partial Y}{\partial x_j}\right) + \mathcal{D}\frac{\partial^2 Y}{\partial x_j^2}.
    \label{VD advection-diffusion equation}
\end{equation}
When the density gradient $\frac{\partial \rho}{\partial x_j}$ vanishes in the PS SBI, this equation reduces to the standard advection-diffusion equation.

In summary, the PS SBI setup transforms the VD SBI into a system with two key PS mixing characteristics: (1) the disappearance of SBV during the development of the single-vortex, and (2) the reduction of the multi-component transport equation to the advection-diffusion equation. These two features facilitate the development of a quantitative theory for describing the stretching dynamics and its impact on mixing, which will be presented in the following sections.

\section{SDR evolution of SBI: an analogy to single-vortex dynamics}
\label{Stretching dynamics analysis preliminaries}
In this section, we present the preliminaries for analyzing stretching dynamics. By examining the temporal evolution of SDR in both PS and VD SBI, the fundamental role of stretching dynamics in governing mixing is highlighted. Utilizing the precise definition of the stretching rate in Eq. \ref{stretch rate definition}, the framework for stretching dynamics is developed within the eigenframe of the strain rate tensor. This framework comprises the stretching dynamics governing equations for the principal strain (SDGE-$s_i$) and for the scalar gradient alignment (SDGE-$\lambda_i$). 
Considering that the formation of a nearly steady single-vortex is observed in both PS and VD SBI cases, as shown in Fig. \ref{density contour and vorticity contour for PS and VD SBI}, 
the mixing in SBI is reasonable to be assumed to be governed by the large scale single-vortex. Therefore, SDGE-$s_i$ and SDGE-$\lambda_i$ are analytically solved under single-vortex assumptions. The section concludes by exploring the interpretation of the stretching rate from the geometric view, providing the foundation for developing a quantitative model to describe the mixing rate in both PS and VD SBI.

\subsection{Importance of the stretching term to SDR evolution in SBI}

To explore the role of stretching dynamics in mixing for PS and VD SBI, it is essential to define the stretching rate within an appropriate framework. Following previous studies on PS mixing \citep{cetegen1993experiments, buch1996experimental}, the SDR, denoted as $\chi$ and defined in Eq. \ref{SDR definition}, serves as a suitable mixing indicator for this purpose. The definition and physical interpretation of SDR stem from another mixing indicator, the mixedness parameter $f$ defined in Eq. \ref{mixedness definition}. This parameter quantifies the extent of mixing, and reaches its maximum when the mass fractions of the bubble’s relevant component and the surrounding air are equal.
\begin{figure}
    \centering
    \includegraphics[width=0.85\linewidth]{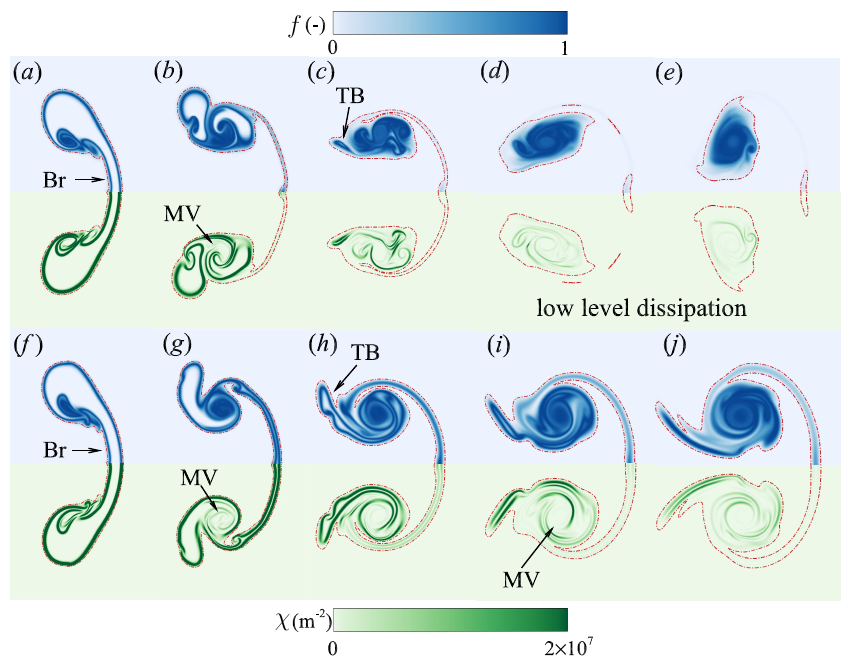}
    \captionsetup{justification=justified, singlelinecheck=false}
    \caption{Time evolution of the mixedness contour (top) and the SDR contour (bottom) for VD SBI case with shock Mach number 2.4 at $(a)\ t=10.8\ \rm{\upmu s}$; $(b)\ t=21.6\ \rm{\upmu s}$; $(c)\ t=32.4\ \rm{\upmu s}$; $(d)\ t=43.2\ \rm{\upmu s}$; and $(e)\ t=54.0\ \rm{\upmu s}$. The PS SBI case is presented at the same moments as that of VD SBI in the bottom from $(f)$ to $(j)$. Red dashed-dot line indicates the isoline of $Y = 0.01$}
    \label{Mixedness and mixing rate contour in PS and VD SBI}
\end{figure}
\begin{figure}
    \centering
    \includegraphics[width=0.7\linewidth]{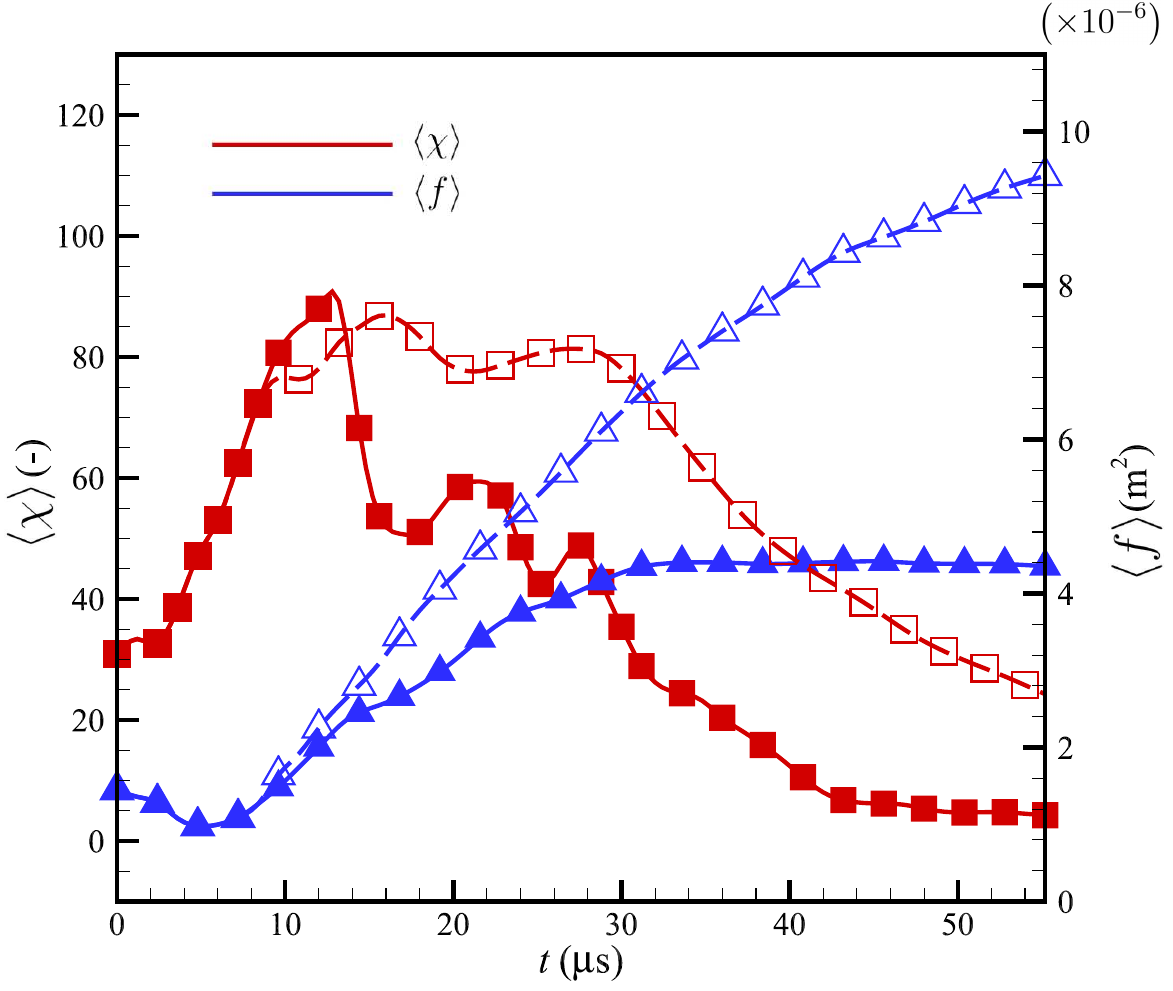}
    \captionsetup{justification=justified, singlelinecheck=false}
    \caption{Time evolution of \textcolor{black}{the total mixedness} (blue) and SDR (red) for PS and VD SBI case demonstrated in Fig.~\ref{Mixedness and mixing rate contour in PS and VD SBI}, respectively. The solid line labels the VD SBI, while the dashed line indicates the PS SBI.}
    \label{Time evolution of mixedness and mixing rate in PS and VD SBI}
\end{figure}
Following the definition of mixedness and based on Eq. \ref{multi-component transport equation}, the behavior of mixedness $f$ is governed by the following equation:
\begin{equation}
\left(\frac{\partial}{\partial t} + u_j \frac{\partial}{\partial x_j} - \mathcal{D}\frac{\partial^2}{\partial x_j^2}\right)f = 8\mathcal{D}\frac{\partial Y}{\partial x_j}\frac{\partial Y}{\partial x_j}+4\left(1-2Y\right)\frac{\mathcal{D}}{\rho}\frac{\partial \rho}{\partial x_j}\frac{\partial Y}{\partial x_j}.
\label{mixedness equation}
\end{equation}
\textcolor{black}{The first term on the right-hand side is strictly positive, representing the growth rate of mixedness. Thus the definition of SDR $\chi = \frac{\partial Y}{\partial x_i}\frac{\partial Y}{\partial x_i}$ reflects the physical meaning of mixing rate}. The second term originates from the density gradient $\frac{\partial \rho}{\partial x_i}$, and vanishes in PS SBI due to the disappearance of density difference.

Figure \ref{Mixedness and mixing rate contour in PS and VD SBI} illustrates the temporal evolution of both mixedness and SDR in PS and VD SBI. As the initially concentrated bubble undergoes the stretching process, the SDR initially increases, subsequently decreases at the later stages, and ultimately dissipates, indicating the achievement of a well-mixed state. Different from the behavior of SDR, the general trend of the mixedness is monotonically increasing. \textcolor{black}{The evolution processes of these two mixing indicators (Eq.~\ref{mixedness definition} and Eq.~\ref{SDR definition}) can be described in detail by examining their total values, which are defined by the corresponding volume integration: 
\begin{equation}
    \left\{
\begin{aligned}
& \left\langle f \right\rangle = \iint_{\mathcal{V}} f \rm{dV},\\
& \left\langle \chi \right\rangle = \iint_{\mathcal{V}} \chi \rm{dV}.
\end{aligned} 
  \right.
\end{equation}}
As shown in Fig. \ref{Time evolution of mixedness and mixing rate in PS and VD SBI}, the mixedness in both PS and VD SBI increases monotonically. In contrast, the SDR initially increases and then decreases after a certain moment. Another notable observation from this figure is that, despite the total circulation $\varGamma_{t}$ and compression rate $\eta$ being consistent between PS SBI and VD SBI, there is a significant difference in \textcolor{black}{the total mixedness $\left\langle f\right\rangle$ and the total SDR $\left\langle \chi\right\rangle$} between the two cases. This phenomenon arises from the discrepancy between the mixing mechanism in PS and VD mixing flow, which will be exhibited further in the following sections.

To gain further insight into the underlying mechanisms driving these two mixing indicators, it is necessary to derive the corresponding governing equations. For the mixedness equation, using the mathematical expression for the time derivative of the volumetric mean value \citep{yu2020scaling}, we have:
\begin{equation}
    \frac{{\rm D } \left\langle f\right\rangle}{{\rm D}t} = \left\langle \frac{{\rm D} f}{{\rm D}t} \right\rangle + \left\langle\left(\frac{\partial u_i}{\partial x_i}\right)f\right\rangle,
    \label{material derivate of volumetric mean value}
\end{equation}
which leads to the governing equation for \textcolor{black}{the total mixedness} $\left\langle f \right\rangle$ as:
\begin{equation}
    \frac{{\rm D } \left\langle f\right\rangle}{{\rm D}t} = \left\langle 8\mathcal{D}\frac{\partial Y}{\partial x_j}\frac{\partial Y}{\partial x_j} \right\rangle + \left\langle 4\left(1-2Y\right)\frac{\mathcal{D}}{\rho}\frac{\partial \rho}{\partial x_j}\frac{\partial Y}{\partial x_j}\right \rangle + \left\langle - \left(\frac{\partial u_i}{\partial x_i}\right)f\right\rangle,
    \label{mean mixedness equation}
\end{equation}

Subsequently, the source of \textcolor{black}{total mixedness growth} $\Delta \left\langle f\right\rangle = \left.\left\langle f\right\rangle\right|_t - \left.\left\langle f\right\rangle\right|_{0}$ can be decomposed by considering the absence of mixedness transport across the boundaries of the integration region, specifically $\left\langle \mathcal{D}\frac{\partial^2 f}{\partial x_j^2} \right\rangle = 0$. This decomposition is described as follows:

\begin{equation}
    \left\{
    \begin{aligned}
        & \Delta \left\langle f\right\rangle = \left\langle T_{SDR}^{f} \right\rangle + \left\langle T_{den}^{f} \right\rangle + \left\langle T_{divU}^{f} \right\rangle,\\
        & \left\langle T_{SDR}^{f} \right\rangle =  \int_{0}^{t}\left\langle 8\mathcal{D}\frac{\partial Y}{\partial x_j}\frac{\partial Y}{\partial x_j} \right\rangle {\rm d}t',\\
        &  \left\langle T_{den}^{f} \right\rangle = \int_{0}^{t} \left\langle 4\left(1-2Y\right)\frac{\mathcal{D}}{\rho}\frac{\partial \rho}{\partial x_j}\frac{\partial Y}{\partial x_j}\right \rangle {\rm d}t', \\
        & \left\langle T_{divU}^{f} \right\rangle = \int_{0}^{t} \left\langle - \left(\frac{\partial u_i}{\partial x_i}\right)f\right\rangle {\rm d}t', 
    \end{aligned}
    \right.
    \label{mean mixedness decomposition}
\end{equation}
where $\left\langle T_{SDR}^{f} \right\rangle$ is the SDR source term, $ \left\langle T_{den}^{f} \right\rangle$ is the density gradient source term, and $\left\langle T_{divU}^{f} \right\rangle$ is the velocity-divergence source term. Similar to \textcolor{black}{the total mixedness} equation in Eq. \ref{mean mixedness equation}, the governing equation for \textcolor{black}{the total SDR} can also be derived based on Eq. \ref{multi-component transport equation} as:
\begin{equation}
    \left\{
     \begin{aligned}
        & \frac{{\rm D}\left\langle \chi \right\rangle}{{\rm D}t} = \left\langle T_{stretch}^{\chi} \right\rangle + \left\langle T_{diff}^{\chi} \right\rangle + \left\langle T_{source}^{\chi} \right\rangle + \left\langle T_{divU}^{\chi} \right\rangle,\\
        & \left\langle T_{stretch}^{\chi} \right\rangle =  \left\langle -2\frac{\partial Y}{\partial x_i} S_{ij} \frac{\partial Y}{\partial x_j}\right\rangle,\\
        &  \left\langle T_{diff}^{\chi} \right\rangle = \left\langle -2\mathcal{D}\frac{\partial}{\partial x_i}\left(\frac{\partial Y}{\partial x_j}\right)\frac{\partial}{\partial x_i}\left(\frac{\partial Y}{\partial x_j}\right)\right\rangle, \\
        & \left\langle T_{source}^{\chi} \right\rangle = \left\langle 2\mathcal{D}\frac{\partial Y}{\partial x_i}\frac{\partial}{\partial x_i}\left(\frac{1}{\rho}\frac{\partial \rho}{\partial x_j}\frac{\partial Y}{\partial x_j}\right) \right\rangle, \\
        & \left\langle T_{divU}^{\chi} \right\rangle = \left\langle -\left(\frac{\partial u_i}{\partial x_i}\right) \chi  \right\rangle,
    \end{aligned}
    \right.
    \label{mean SDR decomposition}
\end{equation}
where $\left\langle T_{stretch}^{\chi} \right\rangle$ is the stretching term expressed as the combination of the scalar gradient $\frac{\partial Y}{\partial x_i}$ and the symmetric strain rate tensor $S_{ij}=\frac{1}{2}\left(\frac{\partial u_{i}}{\partial x_j} + \frac{\partial u_j}{\partial x_i}\right)$, $ \left\langle T_{diff}^{\chi} \right\rangle$ is the diffusion term, $\left\langle T_{source}^{\chi} \right\rangle$ is the density-gradient source term, and $\left\langle T_{divU}^{\chi} \right\rangle$ is the velocity-divergence source term.

\begin{figure}
    \centering
    \includegraphics[width=1.0\linewidth]{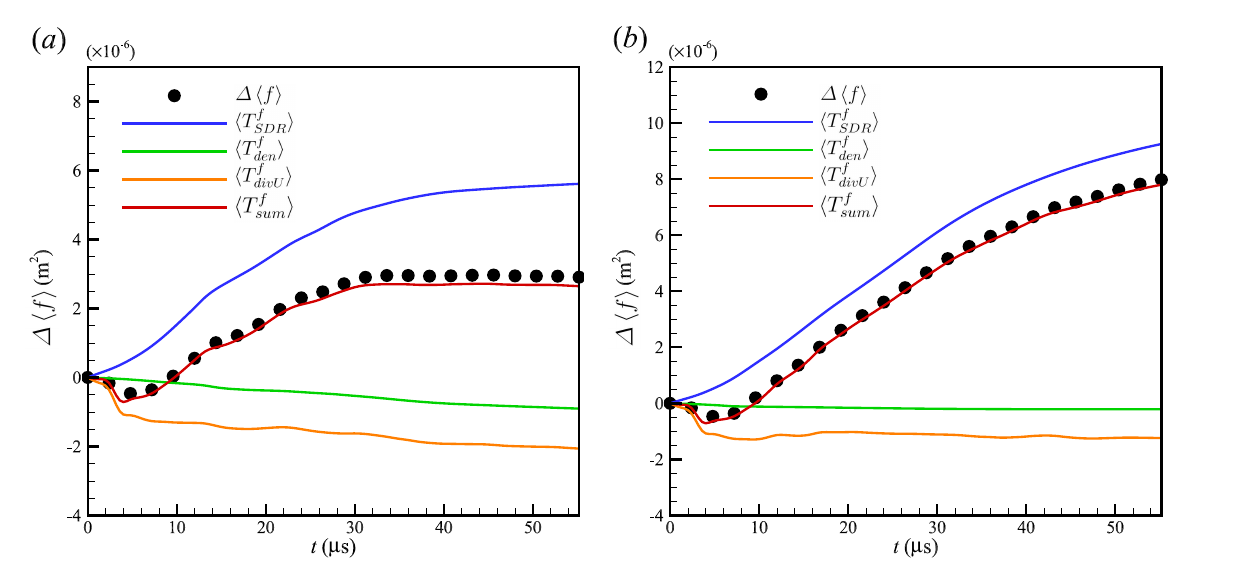}
    \captionsetup{justification=justified, singlelinecheck=false}
    \caption{\textcolor{black}{The decomposition of the total mixedness growth $\Delta\left\langle f \right\rangle$ in  Eq. \ref{mean mixedness decomposition} for $\left(a\right)$ VD SBI and $\left(b\right)$ PS SBI. The black symbols refers to the increment of the total mixedness $\Delta\left\langle f \right\rangle$, which is calculated from the corresponding evolution of $\left\langle f \right\rangle$ in Fig.~\ref{Time evolution of mixedness and mixing rate in PS and VD SBI}. The red solid line indicates the summation of the terms on the right side of Eq.~\ref{mean mixedness decomposition} $\left\langle T_{sum}^{f} \right\rangle$.} }
    \label{mean mixedness decomposition results}
\end{figure}

\begin{figure}
    \centering
    \includegraphics[width=1.0\linewidth]{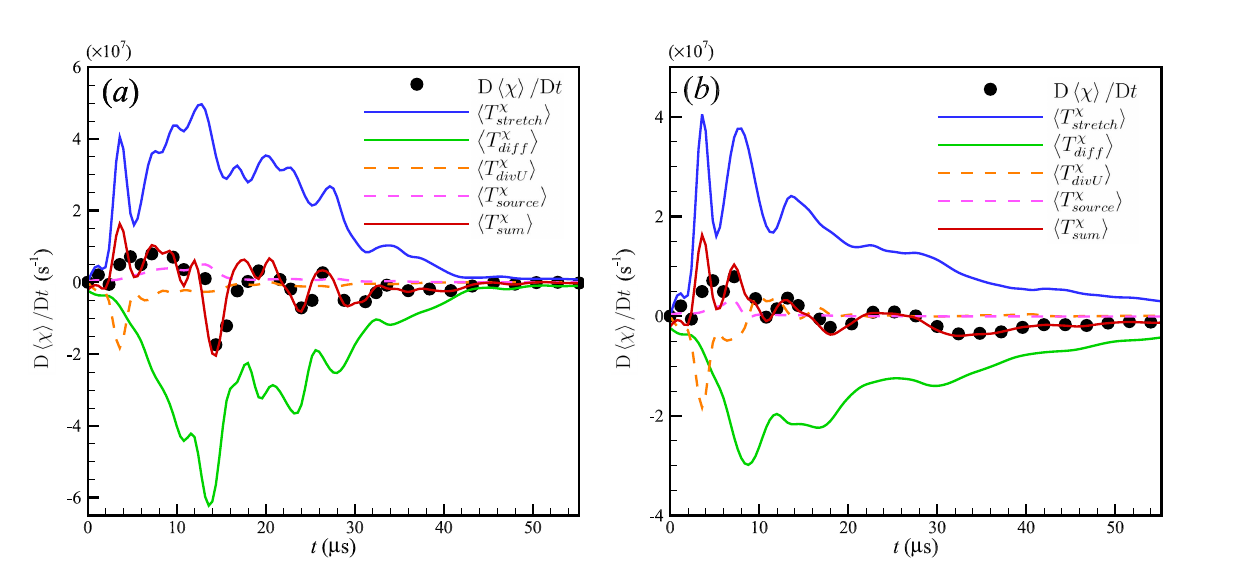}
    \captionsetup{justification=justified, singlelinecheck=false}
    \caption{\textcolor{black}{The decomposition of the material derivative of the total SDR growth $\frac{{\rm D}\left\langle \chi \right\rangle}{{\rm D}t}$ in Eq. \ref{mean SDR decomposition} for $\left(a\right)$ VD SBI and $\left(b\right)$ PS SBI. The black symbols refers to the material derivative of the total SDR growth $\frac{{\rm D}\left\langle \chi \right\rangle}{{\rm D}t}$, which is obtained from the evolution of $\left\langle \chi \right\rangle$ in Fig.~\ref{Time evolution of mixedness and mixing rate in PS and VD SBI}. The red solid line indicates the summation of the terms on the right side of Eq. \ref{mean SDR decomposition} $\left\langle T_{sum}^{\chi} \right\rangle$.}}
    \label{mean SDR decomposition results}
\end{figure}

We begin by analyzing the decomposition of \textcolor{black}{the total mixedness growth $\Delta\left\langle f \right\rangle$} as expressed in Eq. \ref{mean mixedness decomposition}. As shown in Fig. \ref{mean mixedness decomposition results}, the initial decrease in \textcolor{black}{total mixedness} $\Delta\left\langle f \right\rangle$ arises from the velocity-divergence source term $\left\langle T_{divU}^{f} \right\rangle$ in both PS and VD SBI cases, reflecting the effect of shock compression. The primary driver of the monotonic increase in \textcolor{black}{total mixedness}, however, is the SDR source term $\left\langle T_{SDR}^{f} \right\rangle$. These findings from the decomposition of \textcolor{black}{total mixedness growth} confirm that the SDR $\chi$ accurately represents the physical mixing rate and underscore that the growth of mixedness can be almost represented by the SDR source term, which is defined by the time integration of the SDR.

Furthermore, the governing equation for the SDR in Eq. \ref{mean SDR decomposition} is analyzed in Fig. \ref{mean SDR decomposition results}. Consistent with previous studies \citep{danish2016influence, tian2017numerical, gao2020parametric}, the stretching term $\left\langle T_{stretch}^{\chi}\right\rangle$ serves as the primary contributor to SDR growth, while the strictly negative diffusion term $\left\langle T_{diff}^{\chi} \right\rangle$ is the predominant factor in SDR reduction. In comparison, the effects of the density-gradient source term $\left\langle T_{source}^{\chi} \right\rangle$ and the velocity-divergence source term $\left\langle T_{divU}^{\chi} \right\rangle$ on SDR are nearly negligible. This observation highlights that the evolution of SDR is governed by the coupled interaction of stretching dynamics and diffusion processes. Therefore, understanding the SDR necessitates a focus on stretching dynamics, with the stretching term defined rigorously within this framework:
$T_{stretch}^{\chi} = -2 \frac{\partial Y}{\partial x_i} S_{ij} \frac{\partial Y}{\partial x_j}$.

\subsection{Stretching dynamics governing equations (SDGEs)}

To further explore the stretching dynamics, according to the symmetry of the strain rate tensor $S_{ij}$, the stretching term can be decomposed within an orthogonal strain-rate eigenframe, which consists of the principal strain rates $s_i$ and their corresponding eigenvectors $\boldsymbol{e}_i$. The orthogonal expression of the stretching term is provided by Eq. \ref{stretch rate definition}:  $T_{stretch}^{\chi}  = -2\chi s_i \lambda_{i}^2$, indicating that the stretching term $T_{stretch}$ is influenced by both the principal strain rates $s_i$ and the alignment of the scalar gradient with the corresponding eigenvectors $\lambda_i = \nabla Y \cdot \boldsymbol{e}_i/|\nabla Y|$. Consequently, the stretching rate is defined as the product of the principal strain rate $s_i$ and the scalar gradient alignment $\lambda_i$ as described in Eq. \ref{stretching rate definition}: $R_{stretch} = -2 s_i \lambda_i^2$.

Analyzing the stretching rate $R_{stretch}$ from the perspective of stretching dynamics requires the governing equations for the principal strain rate $s_i$ and the alignment of the scalar gradient $\lambda_i$. These two equations can be derived by taking the gradient of the momentum equation in Eq. \ref{NSequation} and the multi-component transport equation in Eq. \ref{multi-component transport equation}, and projecting them onto the strain-rate eigenframe \citep{dresselhaus1992kinematics,ohkitani1995nonlocal,tom2021exploring,han2024mixing}. Details of these derivations are provided in the supplementary materials. We refer to the principal strain rate equations as the stretching dynamics governing equations for $s_i$ (SDGE-$s_i$), which are expressed as follows:
\begin{equation}
    \left\{
    \begin{aligned}
        &\frac{\rm d}{{\rm d} t}s_i = -s_{i}^2 - \frac{1}{4}\left({\widetilde{\omega_i}}^2 - \omega^2\right) + \widetilde{H_{ii}^p} + \widetilde{H_{ii}^b} + \widetilde{H_{ii}^{\nu}},\\
        &\left(s_j - s_i\right)\widetilde{W_{ij}} = -\frac{1}{4}\widetilde{\omega_i}\widetilde{\omega_{j}} + \widetilde{H_{ij}^p} + \widetilde{H_{ij}^b} + \widetilde{H_{ij}^{\nu}},\\
        &\frac{\rm d}{{\rm d}t}\widetilde{\omega_{i}} = s_i \widetilde{\omega_i} - \left(\frac{\partial u_i}{\partial x_i}\right)\widetilde{\omega_i} + \widetilde{B_i} + \widetilde{T_i^{\nu}} - \widetilde{\omega_{j}}\widetilde{W_{ij}},
    \end{aligned}
    \right.
    \label{principal strain rate equation}
\end{equation}
In the first of these equations, $s_i$ represents the principal strain rate, $\widetilde{\omega_{i}} = \boldsymbol{e}_i^{T} \cdot \boldsymbol{\omega}$ are the vorticity component in the strain eigenframe, where $\boldsymbol{\omega} = \nabla \times \boldsymbol{u}$ is the vorticity vector. The norm of the vorticity vector is given by $\omega = ||\boldsymbol{\omega}||$. The terms $H_{ij}^{p} = -\frac{1}{\rho}\frac{\partial^2 p}{\partial x_i \partial x_j}$, $H_{ij}^{b} = \frac{1}{2\rho^2}\left(\frac{\partial \rho}{\partial x_i}\frac{\partial p}{\partial x_j} + \frac{\partial \rho}{\partial x_j}\frac{\partial p}{\partial x_i}\right)$, and $H_{ij}^{\nu}=\frac{1}{2}\left[\frac{\partial}{\partial x_j}\left(\frac{1}{\rho}\frac{\partial \sigma_{ik}}{\partial x_k}\right) + \frac{\partial}{\partial x_i}\left(\frac{1}{\rho}\frac{\partial \sigma_{jk}}{\partial x_k}\right)\right]$ correspond to the pressure Hessian, baroclinic strain, and viscous stress matrix, respectively. The notation $\widetilde{\left(\right)}$ denotes the projection of these matrices onto the strain eigenframe, calculated as:
\begin{equation}
    \widetilde{H_{ij}^{p}} = \boldsymbol{e}_{i}^T \boldsymbol{H}^{p} \boldsymbol{e}_{j},\qquad \widetilde{H_{ij}^{b}} = \boldsymbol{e}_{i}^T \boldsymbol{H}^{b} \boldsymbol{e}_{j},\qquad
    \widetilde{H_{ij}^{\nu}} = \boldsymbol{e}_{i}^T \boldsymbol{H}^{\nu} \boldsymbol{e}_{j}.
\end{equation}
In the second equation, $\widetilde{W_{ij}}$ characterizes the rate of rotation within the plane formed by two eigenvectors about the axis of the third one, that is $\widetilde{W_{ij}} = \frac{\rm d}{{\rm d}t}\boldsymbol{e}_j \cdot \boldsymbol{e}_{i}$. The third equation pertains to the vorticity dynamics, where $\widetilde{B_i}$ and $\widetilde{T_i^{\nu}}$ represent the projections of the baroclinic vorticity $B_{i} = \epsilon_{ijk}\frac{1}{\rho^2}\frac{\partial \rho}{\partial x_j}\frac{\partial p}{\partial x_k}$ and the viscous contribution $T_{i}^{\nu} = \epsilon_{ijk}\frac{\partial}{\partial x_j}\left(\frac{1}{\rho}\frac{\partial \sigma_{kn}}{\partial x_n}\right)$ onto the strain eigenframe, expressed as:
\begin{equation}
    \widetilde{B_i} = \boldsymbol{e}_{i}^{T}\boldsymbol{B},\qquad \widetilde{T_i^{\nu}} = \boldsymbol{e}_{i}^{T}\boldsymbol{T^{\nu}},
\end{equation}
here, $\epsilon_{ijk}$ is the permutation symbol.

The stretching dynamics governing equations for $\lambda_i$ (SDGE-$\lambda_i$) pertains to the alignment of the scalar gradient $\lambda_{i}$, and is formulated as follows:
\begin{equation}
    \frac{\rm d}{{\rm d}t}\lambda_i = -s_i\lambda_i + \xi_{stretch}\lambda_{i} - \widetilde{W_{ij}}\lambda_{j} + \frac{1}{2}\epsilon_{ijk}\widetilde{\omega_{j}}\lambda_{k} + \frac{1}{\sqrt{\chi}}T_{i}^{others},
    \label{alignment equation}
\end{equation}
where $\xi_{stretch} = s_{i}\lambda_{i}^2$ represents the stretching contribution, and $\frac{1}{\sqrt{\chi}}T_{i}^{others}$ characterizes the diffusion effect. Due to the presence of \textcolor{black}{the square root of SDR} on the denominator of the diffusion effect, $\frac{1}{\sqrt{\chi}}T_{i}^{others}$ can generally be neglected in the context of flow structures with a large SDR value. This simplification is justified and demonstrated in Appendix \ref{Validation of the stretching dynamics framework}.

\textcolor{black}{By focusing on the principal strain $s_i$ and the scalar gradient alignment $\lambda_i$, the framework for the stretching dynamics, which centers on the stretching rate $R_{stretch} = -2 s_{i} \lambda_{i}^2$, can be comprehensively described through the combination of SDGE-$s_i$ and SDGE-$\lambda_i$. The validity of the framework is corroborated through Lagrangian methods, as detailed in Appendix \ref{Validation of the stretching dynamics framework}.}

\subsection{Analytical solution of SDGEs in a single-vortex}

Motivated by the formation of a nearly steady single-vortex following the shock's interaction with the bubble, as depicted in Fig. \ref{density contour and vorticity contour for PS and VD SBI}, this subsection investigates the stretching dynamics within a single-vortex. This research is conducted by solving the SDGEs analytically in a cylindrical coordinate system $\left(r, \varphi\right)$, where the center of the main vortex serves as the origin. In this context, $r$ represents the radial direction, while $\varphi$ indicates the azimuthal direction. \textcolor{black}{To facilitate the simplification of the SDGEs, several assumptions and observations regarding vortex morphology for PS mixing are introduced:}

\textcolor{black}{\textbf{Assumption PS.(i)} The single-vortex is axisymmetric in the azimuthal direction $\varphi$, with negligible radial velocity, implying ${\partial}/{\partial \varphi} = 0 $ and $u_r = 0$. This assumption simplifies the momentum equation in Eq. \ref{NSequation} to $-\frac{u_{\varphi}^2}{r} = -\frac{1}{\rho}\frac{\partial p}{\partial r}$.}

\textcolor{black}{\textbf{Assumption PS.(ii)} Following the shock interaction, the single-vortex is considered incompressible, aligning with previous research on RMI \citep{peng2021mechanism}.}

\textcolor{black}{\textbf{Assumption PS.(iii)} The density gradient within the single-vortex is negligible, i.e., $\nabla \rho = 0$.}

\textcolor{black}{\textbf{Assumption PS.(iv)} The mixing of PS SBI can be approximated as the mixing of a compressed semicircle composed of a series of scalar strips, each undergoing deformation due to the continuous stretching effect of the single-vortex.}

\textcolor{black}{The validation of these assumptions is provided in the supplementary materials. Under assumptions PS.(i) and two-dimensional flow condition, the SDGE-$s_i$ presented in Eq. \ref{principal strain rate equation} can be simplified by expressing these equations in cylindrical coordinates as follows:}

\begin{equation}
    \left\{
    \begin{aligned}
        &\frac{\rm d}{{\rm d}t}s_1 = -\frac{1}{2\rho^2}\frac{1}{r}\frac{\partial \rho}{\partial \varphi}\frac{\partial p }{\partial r} + \nu\left(\frac{\partial^2 s_1}{\partial r^2} + \frac{1}{r}\frac{\partial s_1}{\partial r}\right),\\
        &\frac{\rm d}{{\rm d}t}s_2 = \frac{1}{2\rho^2}\frac{1}{r}\frac{\partial \rho}{\partial \varphi}\frac{\partial p }{\partial r} + \nu\left(\frac{\partial^2 s_2}{\partial r^2} + \frac{1}{r}\frac{\partial s_2}{\partial r}\right),\\
        &\widetilde{W_{12}}\approx \frac{\widetilde{H_{12}^p} + \widetilde{H_{12}^b}}{s_2 - s_1} = -\frac{u_{\varphi}}{r},\\
        & \frac{\rm d}{{\rm d}t}\omega = -\frac{1}{2\rho^2}\frac{1}{r}\frac{\partial \rho}{\partial \varphi}\frac{\partial p }{\partial r} + \nu\left(\frac{\partial^2 s_1}{\partial r^2} + \frac{1}{r}\frac{\partial s_1}{\partial r}\right).
    \end{aligned}
    \right.
    \label{simplified SDGE1 2}
\end{equation}
\textcolor{black}{In these equations, with assumption PS.(ii), $s_1$ and $s_2$, representing the positive and negative principal strain rate respectively, satisfy the relation $s_1 + s_2 = 0$. Additionally, $\omega$ denotes the vorticity magnitude perpendicular to the plane of the two-dimensional single-vortex. Furthermore, these equations can be further simplified under assumption PS.(iii) by ignoring the density gradient $\nabla \rho$: }
\begin{equation}
    \left\{
    \begin{aligned}
        &\frac{\rm d}{{\rm d}t}s_1 = \nu\left(\frac{\partial^2 s_1}{\partial r^2} + \frac{1}{r}\frac{\partial s_1}{\partial r}\right),\\
        &\frac{\rm d}{{\rm d}t}s_2 = \nu\left(\frac{\partial^2 s_2}{\partial r^2} + \frac{1}{r}\frac{\partial s_2}{\partial r}\right),\\
        &\widetilde{W_{12}}\approx \frac{\widetilde{H_{12}^p} + \widetilde{H_{12}^b}}{s_2 - s_1} = -\frac{u_{\varphi}}{r},\\
        & \frac{\rm d}{{\rm d}t}\omega = \nu\left(\frac{\partial^2 s_1}{\partial r^2} + \frac{1}{r}\frac{\partial s_1}{\partial r}\right),
    \end{aligned}
    \right.
    \label{simplified SDGE1 3}
\end{equation}
These equations suggest that the principal strain rate $s_i$ and the associated parameters are primarily governed by the viscous effects.

Given the constant total circulation $\varGamma_{t}$ depicted in Fig. \ref{total circulation and compression rate in VD and PS SBI cases}, the analytical solution of Eq.~\ref{simplified SDGE1 3} can be expressed as:
\begin{equation}
    \left\{
    \begin{aligned}
        & s_1 = \frac{1}{2}\left(-\frac{\partial u_{\varphi}}{\partial r} + \frac{ u_{\varphi}}{r}\right) = \frac{\varGamma_{t}}{2\upi r^2}\left(1-{\rm exp}\left(-\frac{r^2}{4\nu t}\right)\right) - \frac{\varGamma_{t}}{8\upi \nu t}{\rm exp}\left(-\frac{r^2}{4\nu t}\right),\\
        & \widetilde{W_{12}} = -\frac{u_{\varphi}}{r} = -\frac{\varGamma_{t}}{2\upi r^2}\left(1-{\rm exp}\left(-\frac{r^2}{4\nu t}\right)\right),\\
        & \omega = \frac{\partial u_{\varphi}}{\partial r} + \frac{u_{\varphi}}{r} = \frac{\varGamma_t}{4\upi \nu t}{\rm exp}\left(-\frac{r^2}{4\nu t}\right),\\
        & \boldsymbol{e}_1 = \left( \cos\left(-\frac{\upi}{4} + \varphi \right), \sin\left(-\frac{\upi}{4} + \varphi\right)\right).
    \end{aligned}
    \right.
    \label{SDGE1 solution}
\end{equation}
This solution derived for the principal strain rate $s_i$ within a single-vortex aligns well with the principal strain rate obtained from Saffman's viscous vortex ring model \citep{saffman1970velocity}, given by:
\begin{equation}
    u_{\varphi} = \frac{\varGamma_{t}}{2\upi r}\left(1-{\rm exp}\left(-\frac{r^2}{4\nu t}\right)\right).
    \label{Saffman vortex model}
\end{equation}

\textcolor{black}{For the scalar gradient alignment $\lambda_i$, SDGE-$\lambda_i$ as represented in Eq. \ref{alignment equation} can be firstly simplified under two-dimensional flow condition:} 
\begin{equation}
    \left\{
    \begin{aligned}
        & \frac{\rm d}{{\rm d}t}\lambda_1 = -\left(s_1 - s_2\right)\left(\lambda_1\lambda_2^2 + \frac{\widetilde{W_{12}} + \omega/2}{s_1-s_2}\lambda_2\right)+\frac{1}{\sqrt{\chi}}T_1^{others},\\
        & \frac{\rm d}{{\rm d}t}\lambda_2 = \left(s_1 - s_2\right)\left(\lambda_1^2\lambda_2 + \frac{\widetilde{W_{12}} + \omega/2}{s_1-s_2}\lambda_1\right)+\frac{1}{\sqrt{\chi}}T_2^{others}.
    \end{aligned}
    \right.
    \label{simplified SDGE2 1}
\end{equation}
Moreover, in a two-dimensional flow, the alignment of the scalar gradient $\lambda_i$ satisfies the relation $\lambda_1^2 + \lambda_2^2 = 1$. Therefore, using the triangular transformation:
\begin{equation}
    \lambda_1 = \cos\theta,\qquad \lambda_2 = \sin\theta,
\end{equation}
Eq. \ref{simplified SDGE2 1} can be further simplified by neglecting the high-order terms $\frac{1}{\sqrt{\chi}}T_i^{others}$ on the flow structures with a large SDR value, shown as:
\begin{equation}
    \frac{\rm d}{{\rm d}t}\theta = \frac{s_1- s_2}{2}\left(\sin2\theta + \xi\right),
    \label{simplified SDGE2 2}
\end{equation}
where $\xi = \frac{\widetilde{W_{12}} + \omega/2}{\frac{1}{2}\left(s_1-s_2\right)}$ is the essential parameter that determines the stability of this equation \citep{lapeyre1999does}. With the analytical expressions for the principal strain rate $s_i$ and the related parameters given in Eq. \ref{SDGE1 solution}, $\xi$ satisfies the relationship:
\begin{equation}
    \xi = \frac{\widetilde{W_{12}} + \omega/2}{\frac{1}{2}\left(s_1-s_2\right)} = -1.
    \label{xi equals to -1 relationship}
\end{equation}
By transforming the angle $\theta$ and time $t$ with the following relationships:
\begin{equation}
    \left\{
    \begin{aligned}
        & \zeta = -2\theta + \frac{\upi}{2},\\
        & \tau = \int_{0}^{t}2s_1\left(t'\right){\rm d}t',
    \end{aligned}
    \right.
\end{equation}
Eq. \ref{simplified SDGE2 2} can be can be reformulated into a simple ordinary differential equation (ODE):
\begin{equation}
    \frac{\rm d}{{\rm d}\tau}\zeta = 1 - \cos\zeta.
\end{equation}
The solution to this ODE is:
\begin{equation}
    \zeta = \upi + 2\arctan\left(\tau\right).
\end{equation}
Thus, the analytical solution for the alignment of the scalar gradient $\lambda_i$ in Eq. \ref{alignment equation} is given by:
\begin{equation}
    \lambda_1 = \cos\left(-\frac{\upi}{4} + \arctan\left(\tau\right)\right),\qquad \lambda_2 = \sin\left(-\frac{\upi}{4} + \arctan\left(\tau\right)\right)
    \label{SDGE2 solution}
\end{equation}

\begin{figure}
    \centering
    \includegraphics[width=1.0\linewidth]{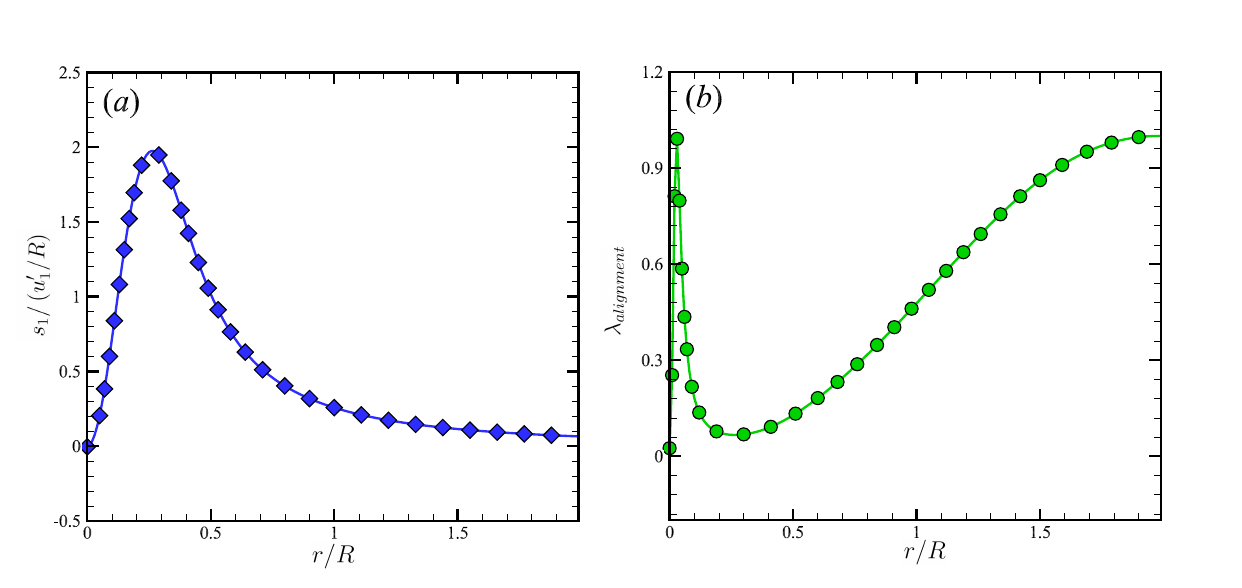}
    \captionsetup{justification=justified, singlelinecheck=false}
    \caption{The examples for the distribution of $\left(a\right)$ the principal strain rate and $\left(b\right)$ the comprehensive alignment effect $\lambda_{alignment}$ along the radial axis in a single-vortex as in Eq. \ref{stretching rate analytical expression}. Here, the principal strain rate is normalized by the ratio between the post-shock air velocity $u_1'$ and the radius of the helium bubble $R$.}
    \label{principal strain rate and alignment distribution}
\end{figure}

With the analytical solution of the SDGEs for the principal strain rate in Eq. \ref{principal strain rate equation} and the alignment of the scalar gradient in Eq. \ref{alignment equation}, the stretching rate, as defined in Eq. \ref{stretch rate definition}, can be determined within a single-vortex as follows:
\begin{equation}
    R_{stretch} = 2s_1\lambda_{alignment} = 4s_1\frac{\tau}{1+\tau^2},
    \label{stretching rate analytical expression}
\end{equation}
where
\begin{equation} 
    \lambda_{alignment} = 1-2\lambda_1^2 =  2\frac{\tau}{1+\tau^2},
    \label{alignment prediction in PS SBI}
\end{equation}
In this expression, $\lambda_{alignment}$ indicates the comprehensive effect of the scalar gradient alignment on the stretching rate. Examples for the distribution of the principal strain $s_1$ and the comprehensive alignment effect $\lambda_{alignment}$ along the radial axis under the condition of PS SBI are illustrated in Fig.~\ref{principal strain rate and alignment distribution}. Notably, if the diffusion term in Eq.~\ref{mean SDR decomposition} is neglected, the analytical expression for the stretching rate in Eq.~\ref{stretching rate analytical expression} corresponds to the algebraic growth of SDR $\chi$:
\begin{equation}
    \frac{\rm d}{{\rm d}t}\chi = T_{stretch}^{\chi} = \chi R_{stretch} \quad \Rightarrow \quad \chi = \chi_0\left(1+\tau^2\right),
\end{equation}
this algebraic stretching characteristic will be further elaborated from a geometric perspective in the subsequent discussion.

\subsection{Physical meaning of the algebraic stretching rate: from the geometric view}

\begin{figure}
    \centering
    \includegraphics[width=0.9\linewidth]{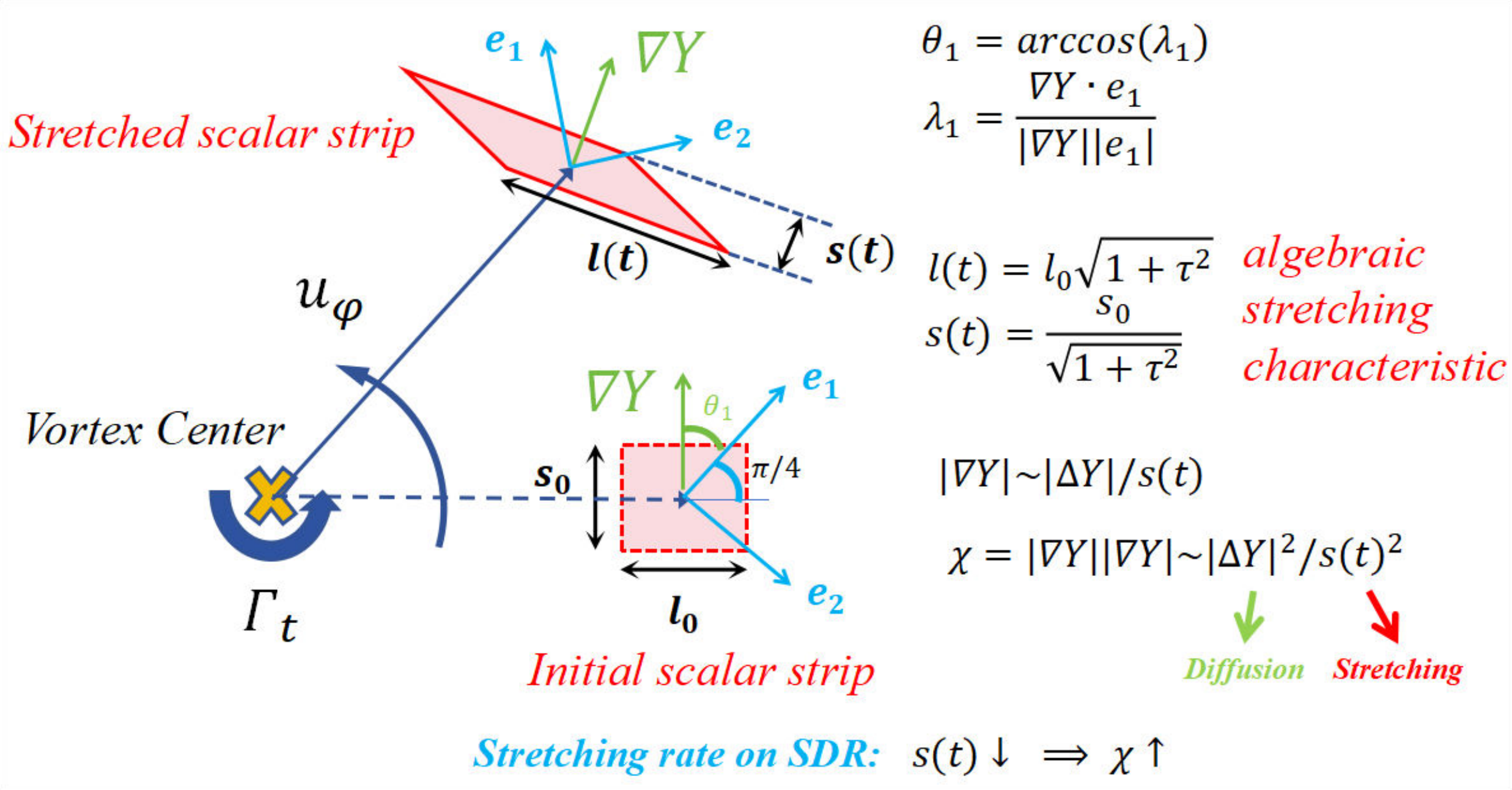}
    \captionsetup{justification=justified, singlelinecheck=false}
    \caption{Schematic illustration of the impact of the stretching rate $R_{stretch}$ on SDR growth in a single-vortex. Within this diagram, the area outlined by the red dashed line represents the initial scalar strip, while the area enclosed by the red solid line denotes the deformed scalar strip as influenced by the stretching effect of the single-vortex. The line perpendicular to the scalar gradient represents the scalar strip length $l(t)$, while the parallel line represents the scalar strip width $s(t)$. According to the estimation for SDR $\chi \sim \left|\Delta Y\right|^2/s(t)^2$, the influence of the stretching rate on SDR growth is reflected by the continuous decrease in the scalar strip width $s(t)$.}
    \label{schematic of stretching rate}
\end{figure}

This subsection elucidates the physical meaning of the analytical expression for the stretching rate $R_{stetch}$ presented in Eq.~\ref{stretching rate analytical expression} from the geometric view. Referring back to the governing equation for the SDR in Eq. \ref{mean SDR decomposition}, the stretching term $T_{stretch}^{\chi}$, from which the definition of the stretching rate $R_{stretch}$ is derived, describes the evolution of the scalar gradient magnitude $\nabla Y$ during a continuous stretching process without molecular diffusion. Figure \ref{schematic of stretching rate} is the schematic illustrating the impact of the stretching rate $R_{stretch}$ on SDR within a single-vortex.

As shown in Fig. \ref{schematic of stretching rate}, the initial scalar strip, with a length of $l_0$ and a width of $s_0$, undergoes deformation due to the continuous stretching effect of the single-vortex. This stretching effect elongates the scalar strip to a new length $l(t)$, while compressing its width to $s(t)$ in accordance with the conservation of the scalar strip's area. The definition of the SDR, $\chi = \lvert \nabla Y\rvert \lvert \nabla Y\rvert$, suggests that the essence of SDR is the square of the magnitude of the scalar gradient. Considering that the SDR for the scalar strip can be approximated by the ratio of the mass fraction difference $\Delta Y$ to the width $s(t)$, expressed as $\chi \sim \lvert \Delta Y \rvert^2/s(t)^2$, an increase in SDR appears to stem from the reduction in the width of the deformed scalar strip $s(t)$ with a constant mass fraction difference $\Delta Y$ when the diffusion process is assumed to be neglected. This observation indicates how the stretching term $T_{stretch}^{\chi}$ drives the growth of SDR.

\begin{figure}
    \centering
    \includegraphics[width=0.9\linewidth]{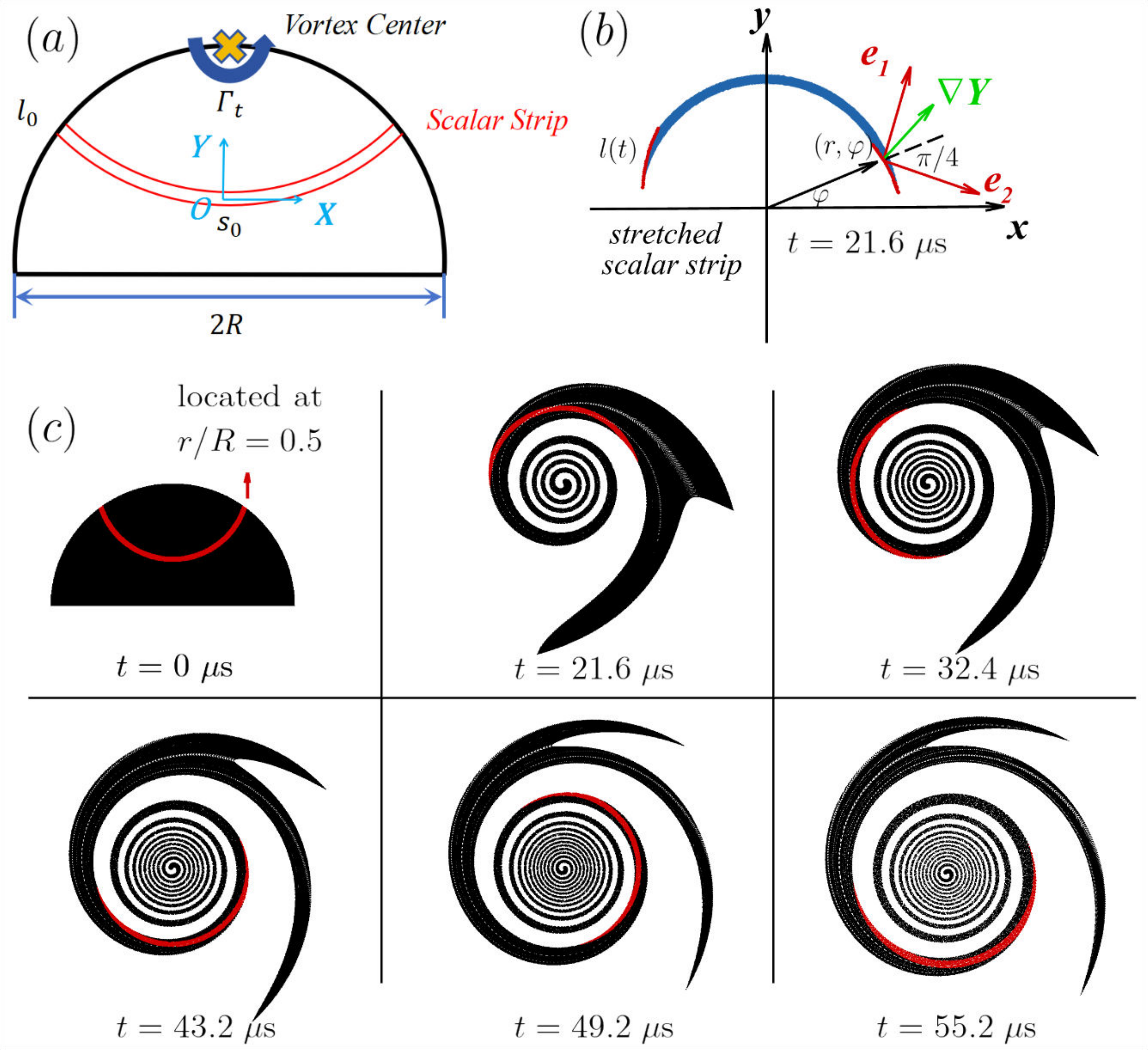}
    \captionsetup{justification=justified, singlelinecheck=false}
    \caption{The schematic illustrating the validation of the algebraic stretching characteristic in a single-vortex using the Lagrangian method. Refer to the previous investigation \citep{liu2022mixing}, the stretching of a post-shock bubble is approximated as a compressed semicircle composed of a series of scalar strips, each undergoing deformation due to the continuous stretching effect of the single-vortex. $\left(a\right)$ The distribution of initial scalar strips used to approximate the stretching process of the post-shock bubble. (b) An example of a stretched scalar strip, along with the corresponding strain eigenframe $\left(\boldsymbol{e}_1,\boldsymbol{e}_2\right)$ and the scalar gradient. $\left(c\right)$ The results of the Lagrangian particle movement for the scalar strip depicted in subfigure $\left(a\right)$.
    }
    \label{validation of the stretching of the scalar strips}
\end{figure}

Under this schematic, the stretching of SDR can be transformed to the reduction in the width of the scalar strip. If the diffusion term $T_{diff}^{\chi}$ is neglected, the governing equation for SDR in Eq. \ref{mean SDR decomposition} can be rewritten based on the estimation $\chi \sim 1/s(t)^2$:
\begin{equation}
    \frac{\rm d}{{\rm d}t}\chi = T_{stretch}^{\chi} = \chi R_{stretch} \quad \Rightarrow \quad \frac{\rm d}{{\rm d}t}\left(\frac{1}{s(t)}\right)^2 = \left(\frac{1}{s(t)}\right)^2 R_{stretch},
\end{equation}
Applying the analytical expression of the stretching rate $R_{stretch}$ in Eq. \ref{stretching rate analytical expression}, this equation becomes:
\begin{equation}
    \frac{\rm d}{{\rm d}t}\left(\frac{1}{s(t)}\right)^2 = \left(\frac{1}{s(t)}\right)^2 4s_1\frac{\tau}{1+\tau^2},
\end{equation}
with the analytical solution:
\begin{equation}
    s\left(t\right) = \frac{s_0}{\sqrt{1+\tau^2}},
    \label{algebraic evolution of s_t}
\end{equation}
and the corresponding evolution of the elongation in the length of the scalar strip $l(t)$ is demonstrated by:
\begin{equation}
    l(t) = l_0\sqrt{1+\tau^2}.
    \label{algebraic stretching of the length}
\end{equation}

These solutions reveal that the length $l(t)$ and the width $s(t)$ evolve in an algebraic form over time, indicating that the stretching within a single-vortex displays an algebraic stretching characteristic.

\begin{figure}
    \centering
    \includegraphics[width=0.65\linewidth]{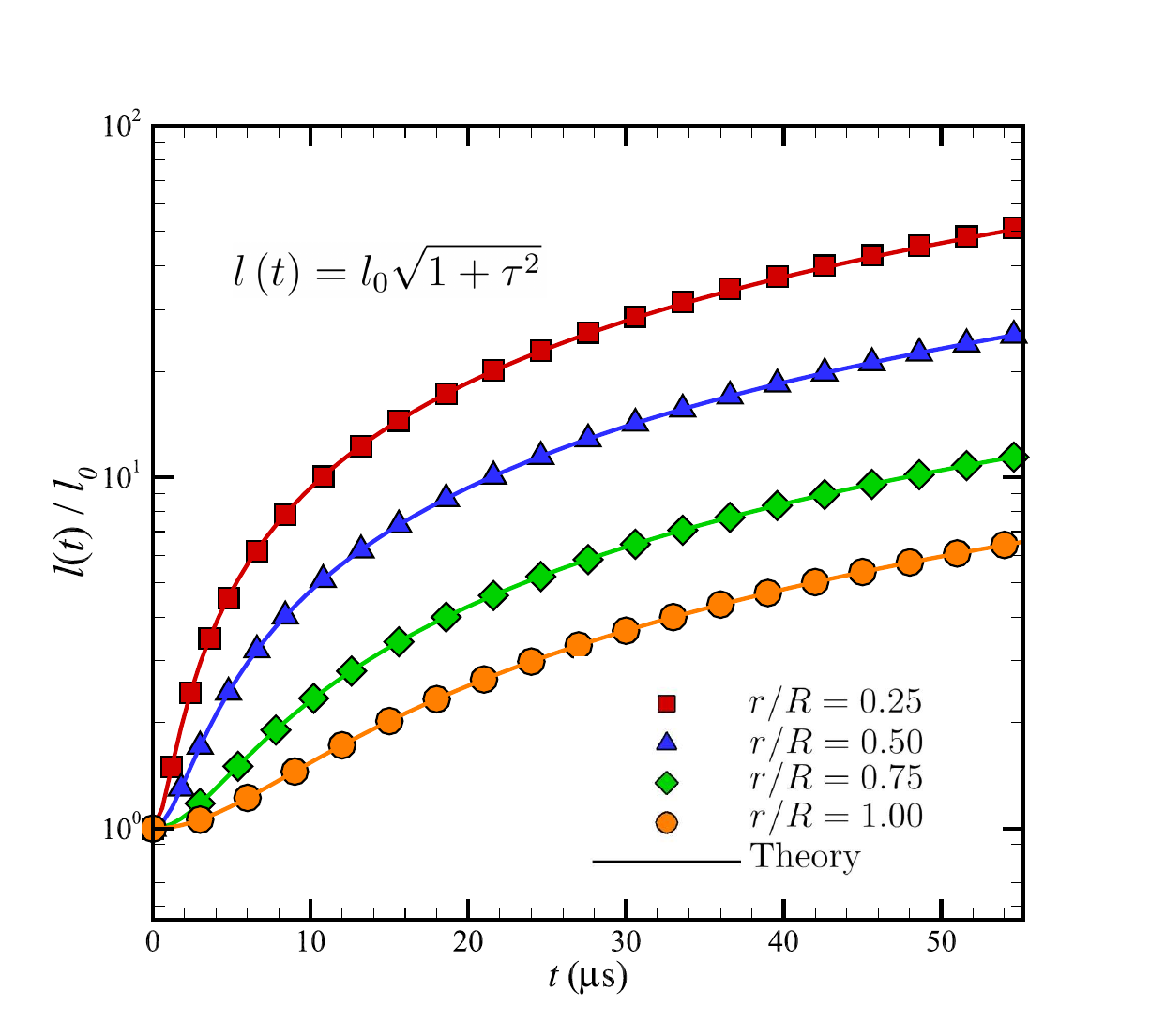}
    \captionsetup{justification=justified, singlelinecheck=false}
    \caption{The comparison of the length of scalar strip at different radius obtained from the Lagrangian particle movement results and  prediction of the stretching dynamics derivation in Eq. \ref{algebraic stretching of the length}.
    }
    \label{comparison of the particle and theory}
\end{figure}

Based on the derivation of the stretching behavior of the scalar strip, the algebraic stretching characteristic is further validated using the Lagrangian method \citep{liang2019hidden,zhang2022numerical}. \textcolor{black}{The morphological changes of the bubble following shock interaction, as shown in Fig.~\ref{Mixedness and mixing rate contour in PS and VD SBI}, support assumption PS.(iv) which approximates the mixing of the post-shock bubble as the mixing of a compressed semicircle composed of a series of scalar strips, as illustrated in Fig.~\ref{validation of the stretching of the scalar strips} $(a)$.} This approximation aligns with our previous investigation \citep{liu2022mixing}. For one of the initial scalar strips with a length $l_0$, perpendicular to the scalar gradient, and a width $s_0$, parallel to the scalar gradient, the Lagrangian particle movement results in Fig.~{\ref{validation of the stretching of the scalar strips} $(c)$} demonstrate the continuous elongation of the length $l(t)$ and reduction of the width $s(t)$. By comparing the lengths of the scalar strips at different radii, obtained according to the trajectories of the Lagrangian particles placed at the edge of semicircle as shown in Fig.~\ref{validation of the stretching of the scalar strips} $(b)$, with the predictions of the stretching dynamics derivation in Eq.~\ref{algebraic stretching of the length}, Fig.~\ref{comparison of the particle and theory} demonstrates the good agreement between the Lagrangian particle movement results and theoretical predictions, thereby the algebraic stretching characteristic within a single-vortex is confirmed.

\section{PS SBI mixing dominated by algebraic stretching dynamics}

In the previous section, the algebraic stretching characteristic within a single-vortex was elucidated using the solution of the SDEGs as the preliminaries for analyzing stretching dynamics. To further investigate the role of stretching dynamics in the context of mixing in SBI, we first examine the relatively simple scenario of mixing in PS SBI as a foundation for understanding VD mixing. In this section, the stretching dynamics of PS SBI are initially assessed. Furthermore, based on the analytical expression of the stretching rate, a PS mixing model for SDR is proposed by incorporating diffusion through the solution of the advection-diffusion equation in Eq. \ref{advection-diffusion equation} along deformed scalar strips subjected to stretching, and this model is validated to accurately capture the evolution of the SDR in PS SBI. The close alignment between the SDR evolution and the model's predictions indicates that mixing in PS SBI is predominantly governed by algebraic stretching dynamics within a single-vortex.
\label{PS mixing theory}

\subsection{Verification of algebraic stretching in PS SBI}

\begin{figure}
    \centering
    \includegraphics[width=0.6\linewidth]{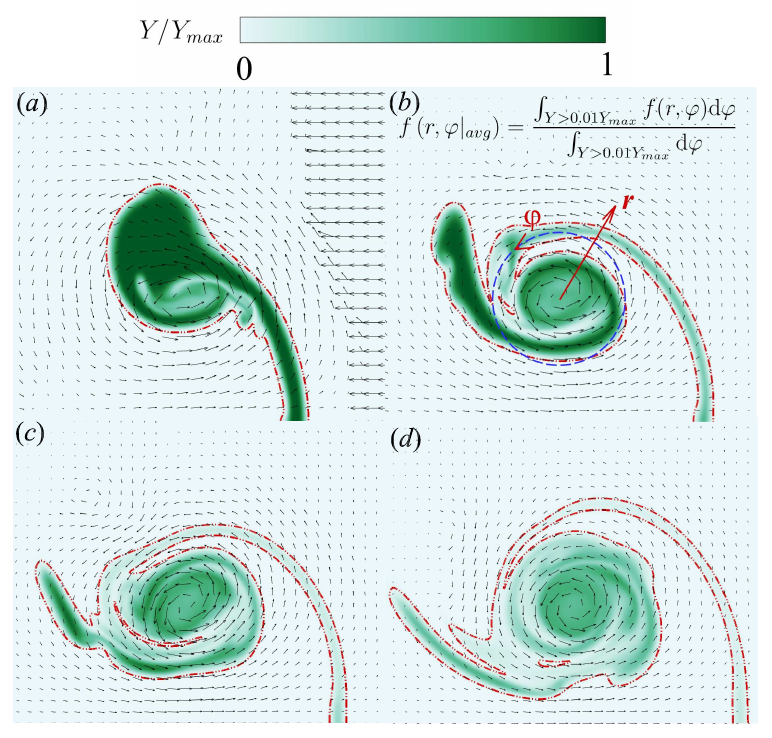}
    \captionsetup{justification=justified, singlelinecheck=false}
    \caption{The distribution of the normalized mass fraction $Y/Y_{max}$ and the relative azimuthal velocity around the vortex center $\left(u-V_{v},v\right)$ in PS SBI at $\left(a\right)$ $t = 13.2\ {\rm \mu s}$, $\left(b\right)$ $t = 26.4\ {\rm \mu s}$, $\left(b\right)$ $t = 39.6\ {\rm \mu s}$, $\left(d\right)$ $t = 52.8\ {\rm \mu s}$. The red dash dot dot line represents the isoline where $Y=0.01Y_{max}$.}
    \label{azimuthal velocity distribution in PS SBI}
\end{figure}

After the shock interaction, the PS bubble attains a translational velocity $V_v$ that is approximately equal to the post-shock velocity $u_1'$. By setting the origin of the cylindrical coordinates at the center of the moving vortex \citep{shariff1992vortex}, the azimuthal velocity around the vortex can be expressed as $\left(u - V_{v},v\right)$, where $u$ represents the velocity in the $x$-direction and $v$ denotes the velocity in the $y$-direction. In Fig.~\ref{azimuthal velocity distribution in PS SBI}, mixing in PS SBI is passively governed by the large-scale main vortex. Given the reduction of the density difference, it is reasonable to assert that the stretching dynamics in PS SBI align with the single-vortex algebraic stretching discussed in Section \ref{Stretching dynamics analysis preliminaries}. This assumption is further validated by analyzing the azimuthal averaged parameters along the radius:
\begin{equation}
    f\left(r,\left.\varphi\right|_{avg}\right)=\frac{\int_{Y>0.01Y_{max}}f(r,\varphi){\rm d}\varphi}{\int_{Y>0.01Y_{max}}{\rm d}\varphi},
\end{equation}
as illustrated by the blue circle in Fig.~\ref{azimuthal velocity distribution in PS SBI} $(b)$.

\begin{figure}
    \centering
    \includegraphics[width=0.7\linewidth]{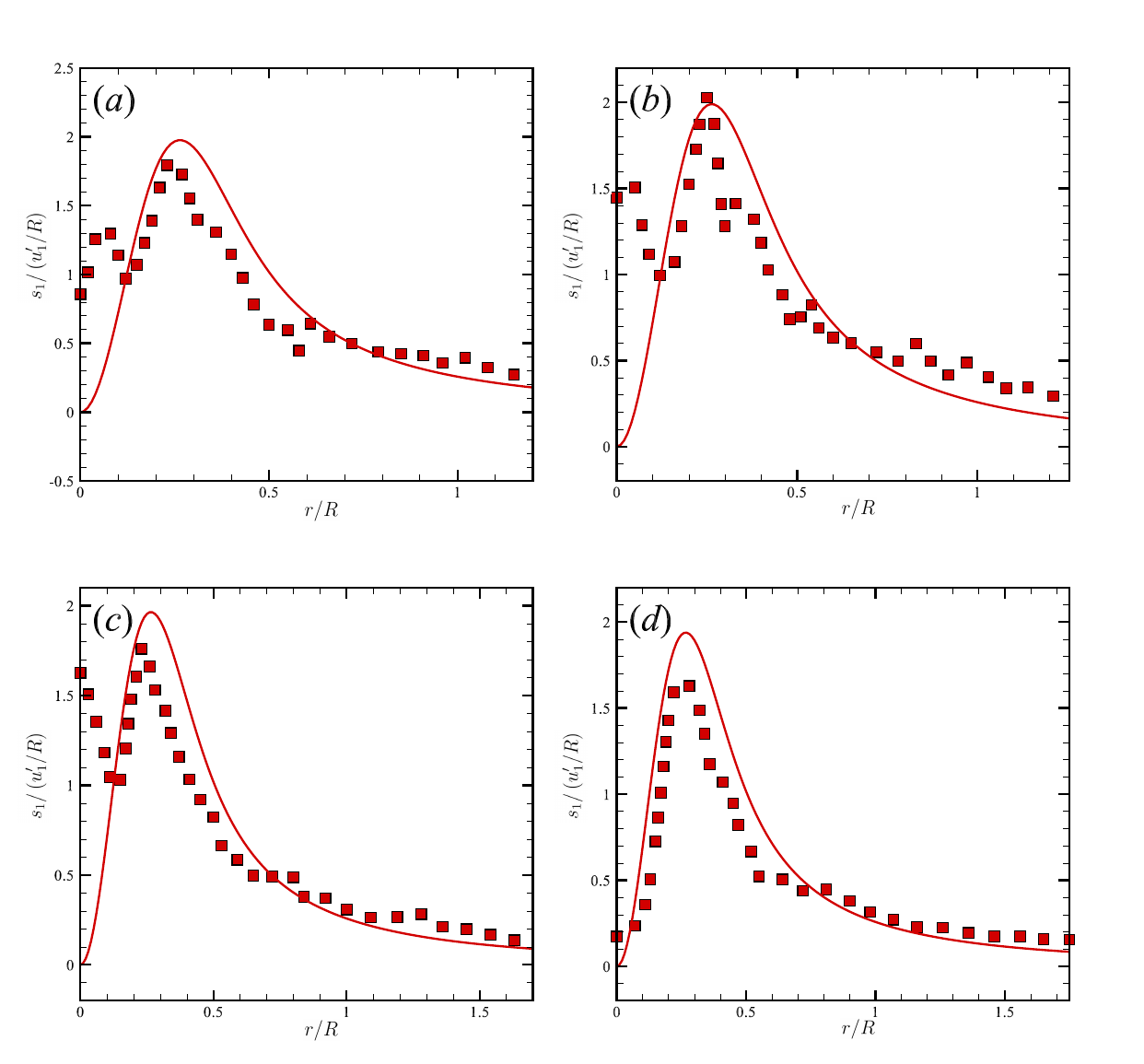}
    \captionsetup{justification=justified, singlelinecheck=false}
    \caption{The distribution of the azimuthal averaged principal strain rate $s_1$ on the radial axis in PS SBI. The moments captured from $\left(a\right)$ to $\left(d\right)$ are the same as those of figure \ref{azimuthal velocity distribution in PS SBI}. The azimuthal averaged principal strain rate is plotted as the red dots, while the prediction of the theory in Eq. \ref{principal strain rate prediction in PS SBI} is plotted as the red solid line.}
    \label{principal strain rate in PS SBI}
\end{figure}

Based on the single-vortex stretching dynamics described in Eq. \ref{stretching rate analytical expression}, we first examine the principal strain rate in PS SBI. Figure \ref{principal strain rate in PS SBI} illustrates the distribution of the azimuthal averaged principal strain rate $s_1$ along the radial axis in PS SBI. With the total circulation of $\varGamma_{t} \approx 2.1\ {\rm m^2\,s^{-1}}$, the analytical expression for the principal strain rate can be expressed as follows:
\begin{equation}
s_{1} = \frac{\varGamma_{t}}{2\upi r^2}\left(1 - \exp\left(-\frac{r^2}{4\nu (t + t_0)}\right)\right) - \frac{\varGamma_{t}}{8\upi \nu t}\exp\left(-\frac{r^2}{4\nu (t + t_0)}\right),
\label{principal strain rate prediction in PS SBI}
\end{equation}
where $t_0$ is estimated based on the scale of the vortex core \citep{saffman1970velocity}:
\begin{equation}
\sqrt{4\nu t_0} = r(s_1^{\text{max}}) \approx 0.27 R, \quad \Rightarrow \quad t_{0} \approx 9.8\times 10^{-4} {\rm s}.
\end{equation}
Notably, since $t_0 \gg t$, the principal strain rate evolves as a nearly steady state, which corresponds with the observation of the nearly steady single-vortex depicted in Fig. \ref{azimuthal velocity distribution in PS SBI}. The comparison shown in Fig. \ref{principal strain rate in PS SBI} demonstrates the principal strain rate in PS SBI aligns closely with the theoretical distribution outlined in Eq. \ref{principal strain rate prediction in PS SBI}. This correspondence indicates that the stretching dynamics in PS SBI exhibits characteristic consistent with single-vortex stretching from the first perspective of the principal strain rate.

Another important parameter for stretching dynamics is the alignment of the scalar gradient $\lambda_{alignment}$. Referring to the derivation of the SDGE-$\lambda_i$ in Eq. \ref{alignment equation}, a significant result for the parameter $\xi = \frac{\widetilde{W_{12}} + \omega/2}{\frac{1}{2}(s_1-s_2)}$, which determines the stability of SDGE-$\lambda_i$, is that $\xi$ satisfies the relationship $\xi = -1$ within a single-vortex (Eq.~\ref{xi equals to -1 relationship}).

\begin{figure}
    \centering
    \includegraphics[width=0.7\linewidth]{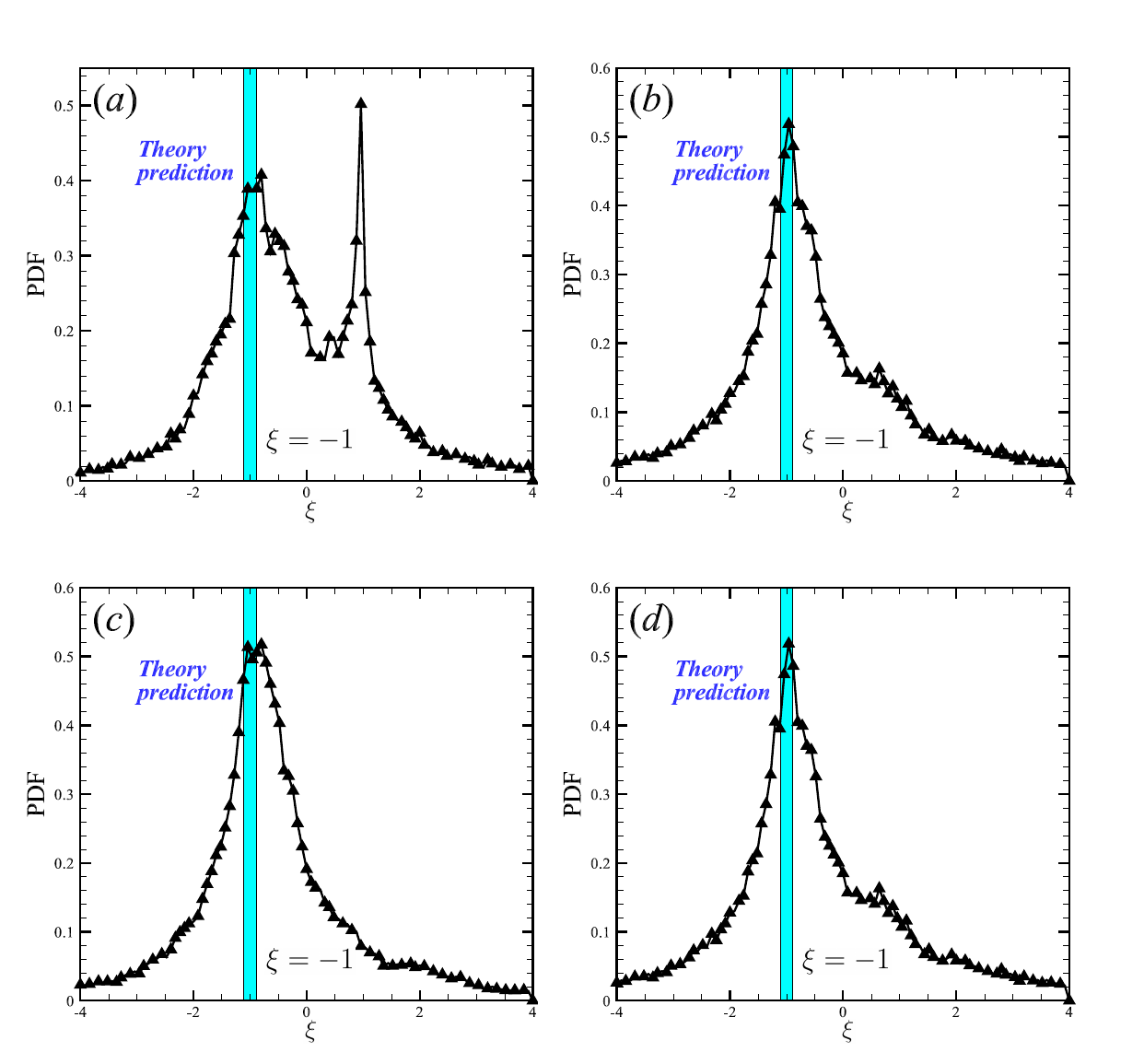}
    \captionsetup{justification=justified, singlelinecheck=false}
    \caption{The probability density function (PDF) of the parameter $\xi = \frac{\widetilde{W_{12}} + \omega/2}{\frac{1}{2}(s_1-s_2)}$ in PS SBI at the same moments as those of figure \ref{azimuthal velocity distribution in PS SBI}.}
    \label{distribution of xi in PS SBI}
\end{figure}

\textcolor{black}{The probability density function (PDF) of $\xi$ in PS SBI is obtained from the region enclosed by the red dash dot dot line in Fig.~\ref{azimuthal velocity distribution in PS SBI}, with the corresponding results shown in Fig.~\ref{distribution of xi in PS SBI}. Notably, the PDF curves exhibit pronounced peaks centered around $\xi = -1$. The deviations from the theoretical value of $\xi = -1$, such as the emergence of $\xi=1$ in the early stages, can be attributed to the small discrepancies between PS SBI and an idealized single-vortex. To evaluate whether the relationship given in Eq.~\ref{xi equals to -1 relationship} remains applicable in PS SBI, we examine if the solution of SDGE-$\lambda_i$ as derived in Section \ref{Stretching dynamics analysis preliminaries} based on Eq.~\ref{xi equals to -1 relationship}, accurately describes the alignment of the scalar gradient $\lambda_{alignment}$. If Eq.~\ref{xi equals to -1 relationship} holds, the distribution of $\lambda_{alignment}$ should follow the theoretical prediction given in Eq.~\ref{alignment prediction in PS SBI}: $\lambda_{alignment} = 2\tau/(1+\tau^2)$, where $\tau$ is defined as $\tau = \int_{0}^{t}2s_1(t'){\rm d}t'$, and the expression for $s_1$ is consistent with that provided in Eq.~\ref{principal strain rate prediction in PS SBI}.}

\begin{figure}
    \centering
    \includegraphics[width=0.7\linewidth]{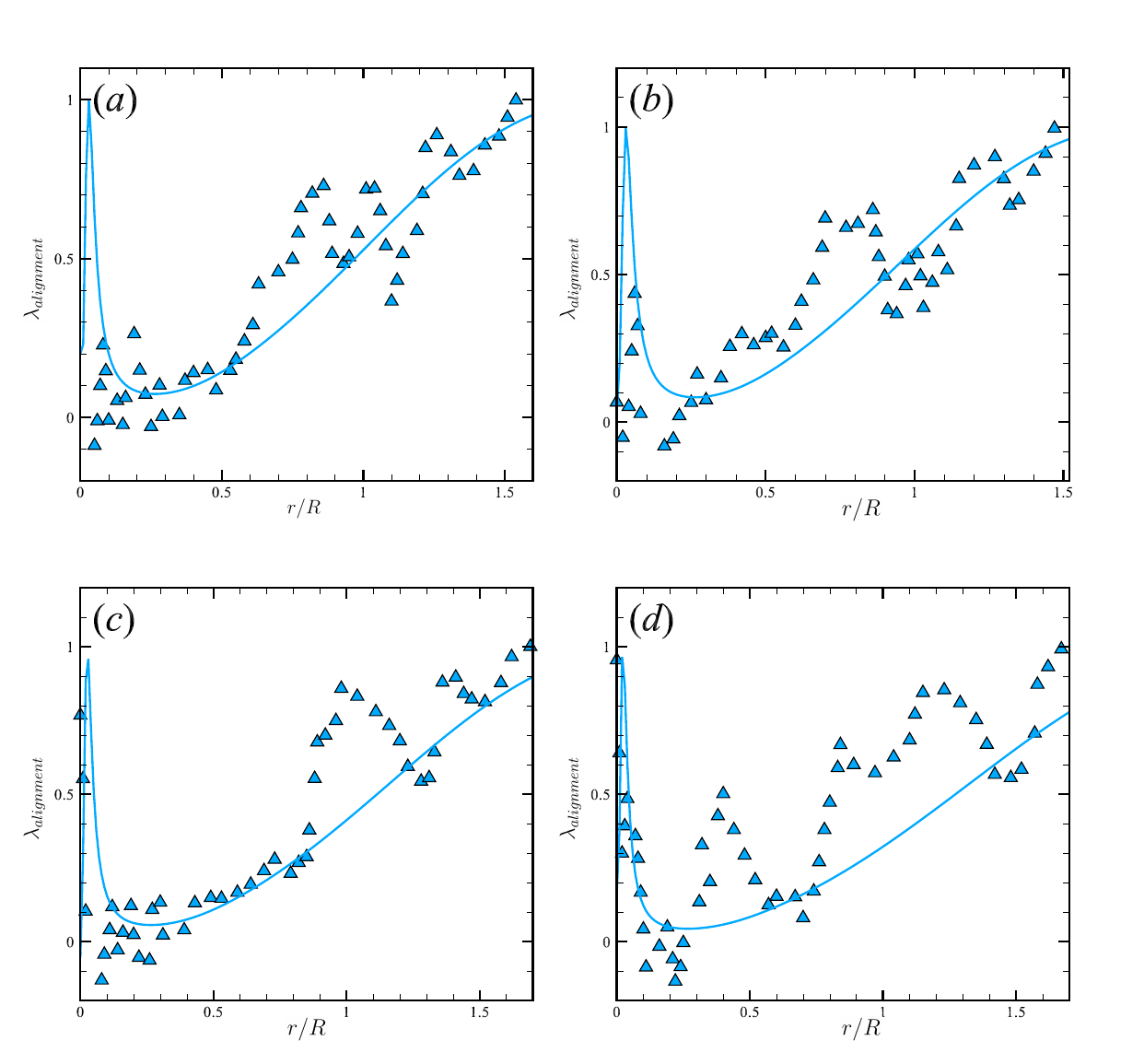}
    \captionsetup{justification=justified, singlelinecheck=false}
    \caption{The distribution of the azimuthal averaged alignment of the scalar gradient $\lambda_{alignment}$ on the radial axis in PS SBI. The moments captured from $\left(a\right)$ to $\left(d\right)$ are the same as those of figure \ref{azimuthal velocity distribution in PS SBI}. The azimuthal averaged alignment is plotted as the blue dots, while the prediction of the theory in Eq. \ref{alignment prediction in PS SBI} is plotted as the blue solid line.}
    \label{alignment in PS SBI}
\end{figure}

\textcolor{black}{Figure \ref{alignment in PS SBI} presents a comparison between the azimuthal averaged $\lambda_{alignment}$ in PS SBI and the theoretical prediction provided in Eq. \ref{alignment prediction in PS SBI}. The distribution of $\lambda_{alignment}$ is accurately captured by the theory described in Eq. \ref{alignment prediction in PS SBI}. The minor discrepancies observed are likely due to the intrinsic complexity of alignment dynamics described by SDGE-$\lambda_i$ in Eq. \ref{alignment equation}, which incorporates the coupled effects of pressure and velocity gradients. These results support the validity of Eq.~\ref{xi equals to -1 relationship} in PS SBI, suggesting that the stretching dynamics within this flow exhibit characteristic consistent with single-vortex stretching from the second perspective of scalar gradient alignment.}

\subsection{Incorporating diffusion process based on algebraic stretching rate}

The stretching dynamics in PS SBI have been shown to exhibit characteristic of single-vortex stretching from two perspectives: the principal strain rate $s_1$ and the alignment of the scalar gradient $\lambda_{alignment}$. However, as indicated in the analysis of the decomposition of the material derivative of \textcolor{black}{the total SD}R $\frac{{\rm D}\left\langle \chi\right\rangle}{{\rm D}t}$ presented in Fig. \ref{mean SDR decomposition results}, the diffusion term $\left\langle T_{diff}^{\chi} \right\rangle$ in Eq. \ref{mean SDR decomposition} serves as another significant factor influencing the evolution of the SDR. Therefore, incorporating the diffusion process based on the algebraic stretching rate is essential for developing the PS mixing model for SDR $\left\langle \chi \right\rangle$.

\begin{figure}
    \centering
    \includegraphics[width=0.6\linewidth]{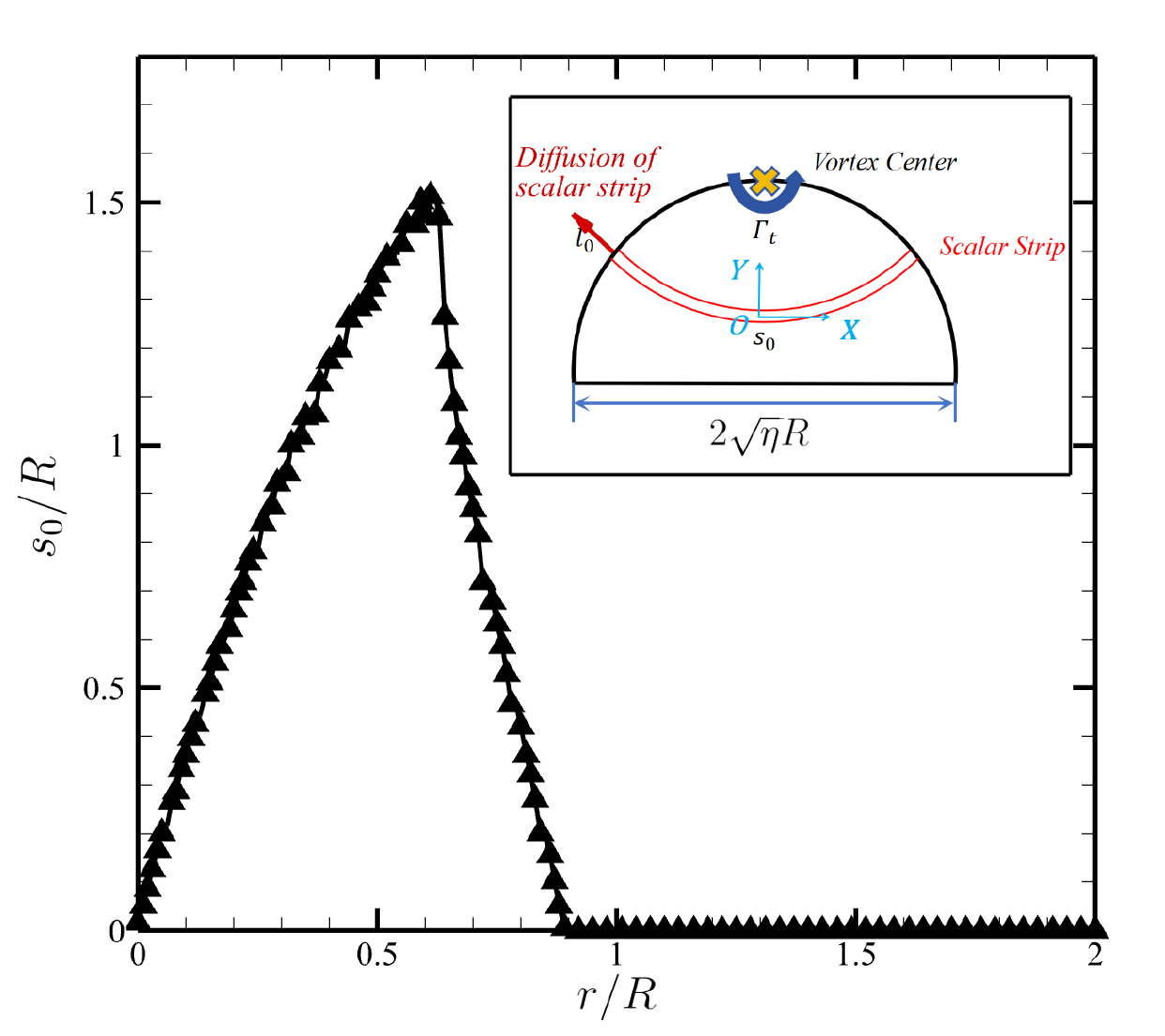}
    \captionsetup{justification=justified, singlelinecheck=false}
    \caption{The schematic of the moving coordinate $(O,X,Y)$ on the deformed scalar strips and the distribution of the related width $s_0$ of the initial scalar strips in the compressed semicircle depicted in Fig. \ref{validation of the stretching of the scalar strips}.}
    \label{distribution of s0}
\end{figure}

As presented in Fig. \ref{schematic of stretching rate}, the estimation $\chi \sim \lvert \Delta Y \rvert^2/s(t)^2$ suggests that the diffusion effect reduces the SDR $\chi$ by altering the magnitude of the mass fraction difference $\lvert \Delta Y \rvert$. \textcolor{black}{To model this diffusion effect within a post-shock cylindrical bubble, it is convenient to apply assumption PS.(iv) to assume that the mixing occurs on a compressed semicircle illustrated in Fig. \ref{validation of the stretching of the scalar strips}. }The radius of this semicircle is $\sqrt{\eta}R$, where $\eta \approx 0.42$ represents the compression rate of PS SBI. By transforming the advection-diffusion equation (Eq. \ref{advection-diffusion equation}) into the coordinate system $\left(O,X,Y\right)$ shown in Fig.~\ref{distribution of s0}, the equation can be reformulated as:
\begin{equation}
\frac{\partial Y}{\partial t} + U\frac{\partial Y}{\partial x} + V\frac{\partial Y}{\partial y} = \mathcal{D}\left(\frac{\partial^2 Y}{\partial x^2} + \frac{\partial^2 Y}{\partial y^2}\right),
\label{ade on the moving coordinate}
\end{equation}
where the origin of the $\left(O,X,Y\right)$ system is a Lagrangian frame set on a moving scalar strip. The direction of this frame changes over time in response to the stretching motion.

\textcolor{black}{On this moving coordinate, the relative velocity $\left(U,V\right)$ can be expressed in terms of the deformation rate of the scalar strip, as described by \citep{villermaux2019mixing}:
\begin{equation}
    \left\{
    \begin{aligned}
        & U = \frac{{\rm d}x}{{\rm d}t} = \frac{{\rm d}x}{{\rm d}s(t)}\frac{{\rm d}s(t)}{{\rm d}t} = \frac{x}{s(t)}\frac{{\rm d}s}{{\rm d}t},\\
        & V = \frac{{\rm d}y}{{\rm d}t} = \frac{{\rm d}y}{{\rm d}s(t)}\frac{{\rm d}s(t)}{{\rm d}t} = -\frac{y}{s(t)}\frac{{\rm d}s(t)}{{\rm d}t},
    \end{aligned}
    \right.
    \label{relative velocity on scalar strips}
\end{equation}
where $s(t)$ denotes the instantaneous width of the scalar strip. Considering the mass fraction gradient $\nabla Y$ is predominantly  aligned with the X-direction, i.e. $\lvert \frac{\partial Y}{\partial x} \rvert \gg \lvert \frac{\partial Y}{\partial y}\rvert$, Eq. \ref{ade on the moving coordinate} becomes:
\begin{equation}
    \frac{\partial Y}{\partial t} + \frac{x}{s(t)}\frac{{\rm d}s(t)}{{\rm d}t}\frac{\partial Y}{\partial x}  = \mathcal{D}\frac{\partial^2 Y}{\partial x^2}.
    \label{ade simplify 1}
\end{equation}
}Since the stretching dynamics is PS SBI is proved to exhibit the single-vortex stretching characteristic, the width $s(t)$ evolves algebraically as in Eq. \ref{algebraic evolution of s_t}.
Using the canonical transformation introduced by Ranz \citep{ranz1979applications}:
\begin{equation}
    \left\{
    \begin{aligned}
        & \varsigma = \frac{x}{s(t)},\\
        & \tau^{\mathcal{D}} = \int_{0}^{t}\frac{\mathcal{D}{\rm d}t'}{s(t')^2} ,
    \end{aligned}
    \right.
    \label{ranz transform}
\end{equation}
Eq. \ref{ade simplify 1} is transformed into a simple Fourier equation :
\begin{equation}
    \frac{\partial Y}{\partial \tau^{\mathcal{D}}} = \frac{\partial^2 Y}{\partial \varsigma^2},
\end{equation}
with the initial condition specified as:
\begin{equation}
    Y(\tau^{\mathcal{D}} = 0) = \left\{
    \begin{aligned}
        & 1 \quad \lvert \varsigma \rvert \leq \frac{1}{2},\\
        & 0 \quad \lvert \varsigma \rvert > \frac{1}{2}.
    \end{aligned}
    \right.
\end{equation}

\begin{figure}
    \centering
    \includegraphics[width=0.6\linewidth]{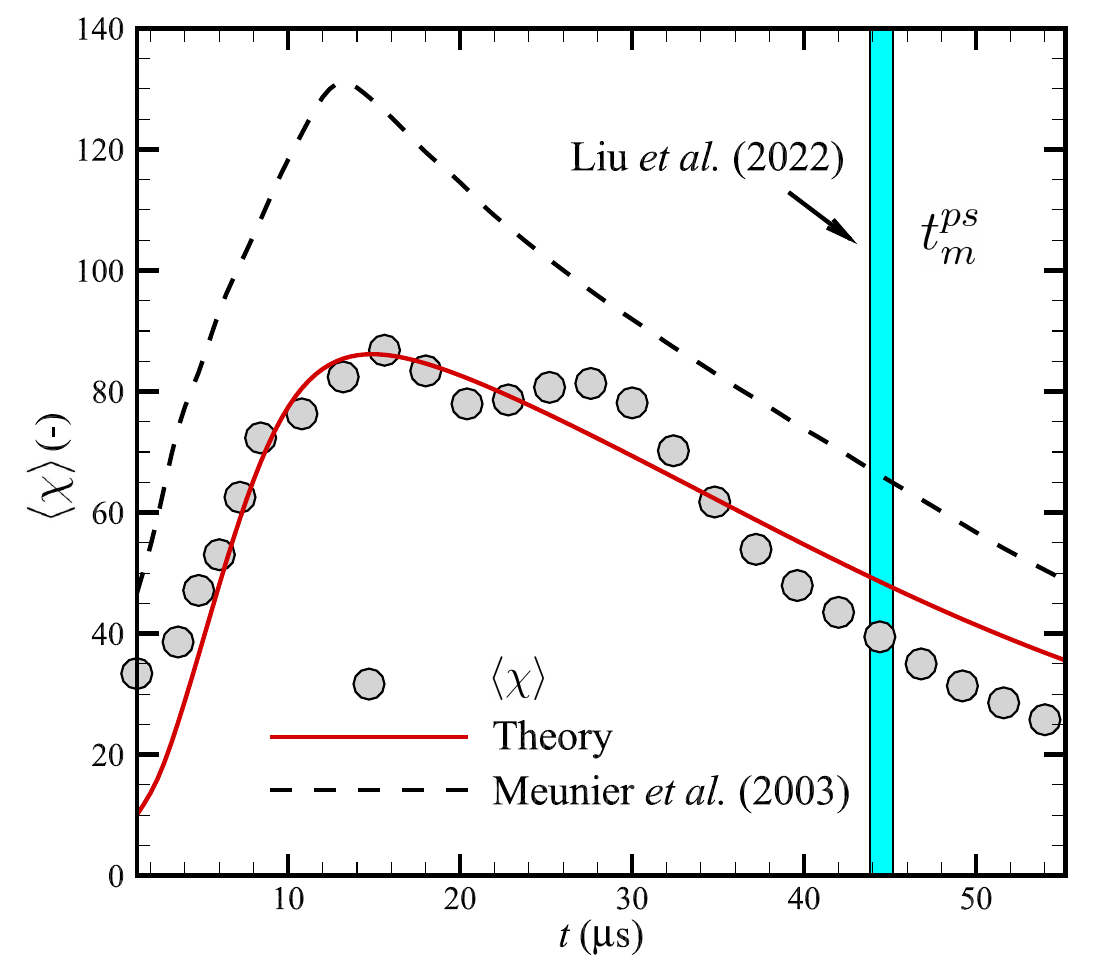}
    \captionsetup{justification=justified, singlelinecheck=false}
    \caption{Comparison of the evolution of \textcolor{black}{total SDR} $\left\langle \chi\right\rangle$ in PS SBI with theoretical predictions: the red solid line represents the prediction from the PS mixing model in Eq. \ref{PS mixing model expression}, and the black dashed line denotes the two-stage approximation from Eq.~\ref{PS mixing model simplification 1} and Eq.~\ref{PS mixing model simplification 2}, which aligns with the approximation of SDR by  \cite{meunier2003vortices}. The blue line indicates the PS mixing time $t_{m}^{ps}$ obtained from the previous research \citep{liu2022mixing}.}
    \label{comparision of PS model and SDR}
\end{figure}

Subsequently, the mass fraction in the moving coordinate frame will diffuse and spread according to the following solution:
\begin{equation}
    Y\left(\varsigma,\tau^{\mathcal{D}}\right) = \frac{1}{2}\left[{\rm erf}\left(\frac{\varsigma + 1/2}{2\sqrt{\tau^{\mathcal{D}}}}\right) - {\rm erf}\left(\frac{\varsigma - 1/2}{2\sqrt{\tau^{\mathcal{D}}}}\right) \right].
    \label{ade solution}
\end{equation}
By considering the mass fraction difference $\Delta Y$ alongside the mass fraction distribution from Eq. \ref{ade solution}, the PS mixing model for the \textcolor{black}{total SDR} $\left\langle \chi\right\rangle $ can be proposed as:
\begin{equation}
    \left\langle \chi\right\rangle = \int \left(\int_{-\infty}^{+\infty}\left(\frac{{\rm d}Y}{{\rm d}x}\right)^2{\rm d}x\right) {\rm d}y = \int_{0}^{\sqrt{2\eta}R}\frac{1+\tau^2}{s_0}\frac{1}{\sqrt{2\upi\tau^{\mathcal{D}}}}\left(1-\mathrm{e}^{-\frac{1}{8\tau^{\mathcal{D}}}}\right) {\rm d}r.
    \label{PS mixing model expression}
\end{equation}
In this model, $\tau = \int_{0}^{t}2s_1(t'){\rm d}t'$ represents the transformed time corresponding to stretching dynamics, while $\tau^{\mathcal{D}} = \int_{0}^{t}\frac{\mathcal{D}{\rm d}t'}{s(t')^2} = \int_{0}^{t}\frac{\mathcal{D}\left(1+\tau^2\right){\rm d}t'}{s_0^2}$ indicates the transformed time according to the diffusion process.

The comparison shown in Fig. \ref{comparision of PS model and SDR} indicates that the PS mixing model presented in Eq. \ref{PS mixing model expression} accurately predicts the evolution of \textcolor{black}{the total SDR} $\left\langle \chi\right\rangle$ during mixing in PS SBI, which proves that the mixing in PS SBI is predominantly governed by the algebraic stretching rate within a single-vortex. Furthermore, We can notice the presence of the characteristic time $t_{charac}$, which delineates the mixing process into two distinct stages: when $t < t_{charac}$, \textcolor{black}{the total SDR} $\left\langle \chi\right\rangle$ increases to its peak value; subsequently, for $t \geq t_{charac}$, it transitions into the decay stage. The existence of this characteristic time and the two distinct mixing stages can be explained through two-stage approximation of Eq. \ref{PS mixing model expression}.

In relation to the transformed time associated with stretching dynamics, defined as $\tau = \int_{0}^{t}2s_1(t'){\rm d}t'$, and given the nearly steady principal strain rate $s_1$ illustrated in Fig. \ref{principal strain rate prediction in PS SBI}, $\tau$ can be approximated as
\begin{equation}
    \tau \approx 2s_1 t.
\end{equation}
Consequently, the width of the scalar strips evolves according to:
\begin{equation}
    s(t) = \frac{s_0}{\sqrt{1+\tau^2}} =\frac{s_0}{\sqrt{1+4s_1^2t^2}},
\end{equation}
which simplifies the expression for the transformed time related to diffusion $\tau^{\mathcal{D}}$ as follows:
\begin{equation}
    \tau^{\mathcal{D}} = \int_{0}^{t}\frac{\mathcal{D}{\rm d}t'}{s(t')^2} = \frac{\mathcal{D}}{s_0^2}\left(t + \frac{4}{3}s_1^2t^3\right) \approx \frac{4}{3}\frac{\mathcal{D}}{s_0^2}s_1^2t^3.
\end{equation}
In the context of Eq. \ref{PS mixing model expression}, if $\tau^{\mathcal{D}} < \frac{1}{8}$, i.e. $t < t_{charac} = \left(\frac{3}{32}\frac{s_0^2}{s_1^2\mathcal{D}}\right)^{\frac{1}{3}}$, Eq. \ref{PS mixing model expression} can be modified to:
\begin{equation}
    \begin{aligned}
        \left\langle \chi\right\rangle & = \int_{0}^{\sqrt{2\eta}R}\frac{1+\tau^2}{s_0}\frac{1}{\sqrt{2\upi\tau^{\mathcal{D}}}}\left(1-\mathrm{e}^{-\frac{1}{8\tau^{\mathcal{D}}}}\right) {\rm d}r \\
        & \approx \int_{0}^{\sqrt{2\eta}R}\frac{1+\tau^2}{s_0}\frac{1}{\sqrt{2\upi\tau^{\mathcal{D}}}}{\rm d}r \approx \int_{0}^{\sqrt{2\eta}R}\sqrt{\frac{6}{\upi \mathcal{D}}}s_1t^{\frac{1}{2}}{\rm d}r,
    \end{aligned}
    \label{PS mixing model simplification 1}
\end{equation}
indicating the growth stage of $\left\langle \chi\right\rangle$. During this phase, the growth of $\left\langle \chi\right\rangle$ arises from the dominance of the stretching effect $\frac{1+\tau^2}{s_0}$ in the PS mixing process. Conversely, if $\tau^{\mathcal{D}} \geq \frac{1}{8}$, i.e. $t \geq t_{charac} = \left(\frac{3}{32}\frac{s_0^2}{s_1^2\mathcal{D}}\right)^{\frac{1}{3}}$, the equation transforms to:
\begin{equation}
    \begin{aligned}
        \left\langle \chi\right\rangle & = \int_{0}^{\sqrt{2\eta}R}\frac{1+\tau^2}{s_0}\frac{1}{\sqrt{2\upi\tau^{\mathcal{D}}}}\left(1-\mathrm{e}^{-\frac{1}{8\tau^{\mathcal{D}}}}\right) {\rm d}r \\
        & \approx \int_{0}^{\sqrt{2\eta}R}\frac{1+\tau^2}{s_0}\frac{1}{\sqrt{2\upi\tau^{\mathcal{D}}}}\frac{1}{8\tau^{\mathcal{D}}} {\rm d}r\approx \int_{0}^{\sqrt{2\eta}R}\frac{3}{16}\sqrt{\frac{3}{2\upi }}\frac{s_0^2}{\mathcal{D}^{\frac{3}{2}}s_1 t^{\frac{5}{2}}}{\rm d}r,
    \end{aligned}
    \label{PS mixing model simplification 2}
\end{equation}
which characterizes the decay stage of $\left\langle \chi\right\rangle$. In this phase, the decay of $\left\langle \chi\right\rangle$ stems from the predominance of the diffusion effect 
$\frac{1}{\sqrt{2\upi\tau^{\mathcal{D}}}}\left(1-\mathrm{e}^{-\frac{1}{8\tau^{\mathcal{D}}}}\right)$ in the PS mixing process. \textcolor{black}{The reasonability of these equation simplifications are provided in the supplementary material.} Notably, the two-stage approximations of the PS mixing model are consistent with the approximations for $\left\langle\chi\right\rangle$ in the previous study on single-vortex PS mixing \citep{meunier2003vortices}. As shown in Fig.~\ref{comparision of PS model and SDR}, the two-stage approximation of the PS mixing model in Eq.~\ref{PS mixing model simplification 1} and Eq.~\ref{PS mixing model simplification 2} accurately captures the trends of SDR evolution during both the growth and decay stages, while still overestimating the evolution of SDR. Compared to this previous estimation of SDR, the PS mixing model effectively synthesizes the stretching dynamics and diffusion effects in Eq. \ref{mean SDR decomposition}, providing a robust predictive framework for the evolution of $\left\langle \chi \right\rangle$. Furthermore, compared with the mixing time for PS SBI predicted by the mixing time model in our earlier work \citep{liu2022mixing}, indicated by the blue line in Fig.~\ref{comparision of PS model and SDR}, the PS mixing model offers a more detailed description of the evolution of mixing indicator, representing a significant advancement over the theory for mixing time prediction.

\section{VD SBI mixing modified by VD effects from perspectives of stretching and diffusion}

\begin{figure}
    \centering
    \includegraphics[width=0.6\linewidth]{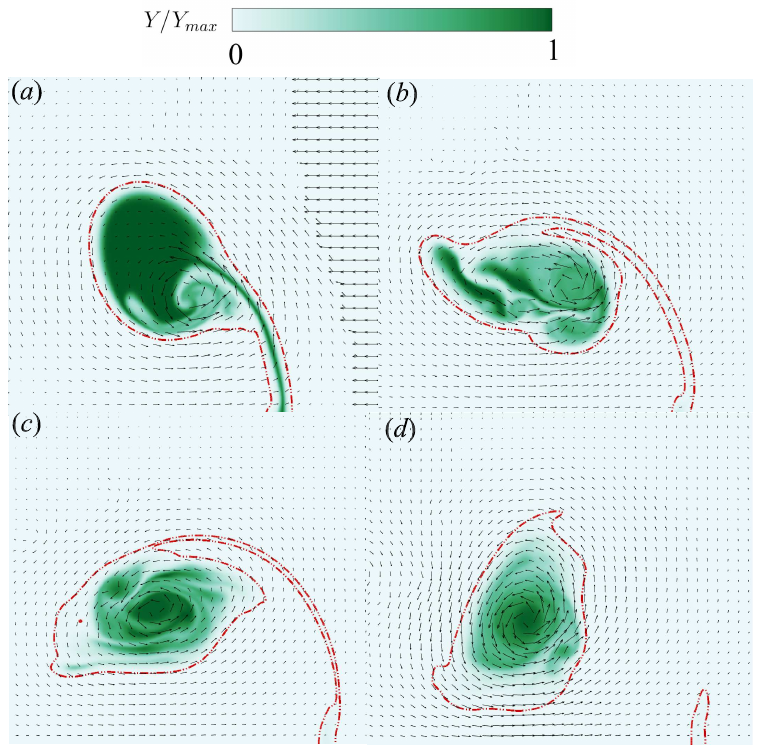}
    \captionsetup{justification=justified, singlelinecheck=false}
    \caption{The distribution of the normalized mass fraction $Y/Y_{max}$ and the relative azimuthal velocity around the vortex center $\left(u-V_{v},v\right)$ in VD SBI at same moments as that of PS SBI in Fig.~\ref{azimuthal velocity distribution in PS SBI}.}
    \label{azimuthal velocity distribution in VD SBI}
\end{figure}

According to the distribution of the azimuthal velocity around the vortex center, Fig.~\ref{azimuthal velocity distribution in VD SBI} illustrates that the mixing in VD SBI is also predominantly governed by the large-scale primary vortex as in PS SBI. However, even with the nearly identical total circulation $\varGamma_{t}$ and the compression rate $\eta$ between VD and PS SBI shown in Fig.~\ref{total circulation and compression rate in VD and PS SBI cases}, significant difference remains in the evolution of the mean scalar dissipation rate SDR $\left\langle \chi \right\rangle$
, as depicted in Fig.~\ref{Time evolution of mixedness and mixing rate in PS and VD SBI}. This suggests that, in comparison to PS mixing based on single-vortex stretching dynamics, additional VD effects challenge the extension of the PS mixing model to VD mixing. The primary objective of this section is to investigate these VD effects from two perspectives of stretching and diffusion, and propose a VD mixing model capable of accurately capturing the evolution of SDR in VD SBI. Through the development of this VD mixing model, the quantitative relationship between the stretching dynamics and single-vortex VD mixing is established.

\label{VD mixing theory}
\subsection{Secondary baroclinic effect on stretching dynamics in VD SBI}
\begin{figure}
    \centering
    \includegraphics[width=0.7\linewidth]{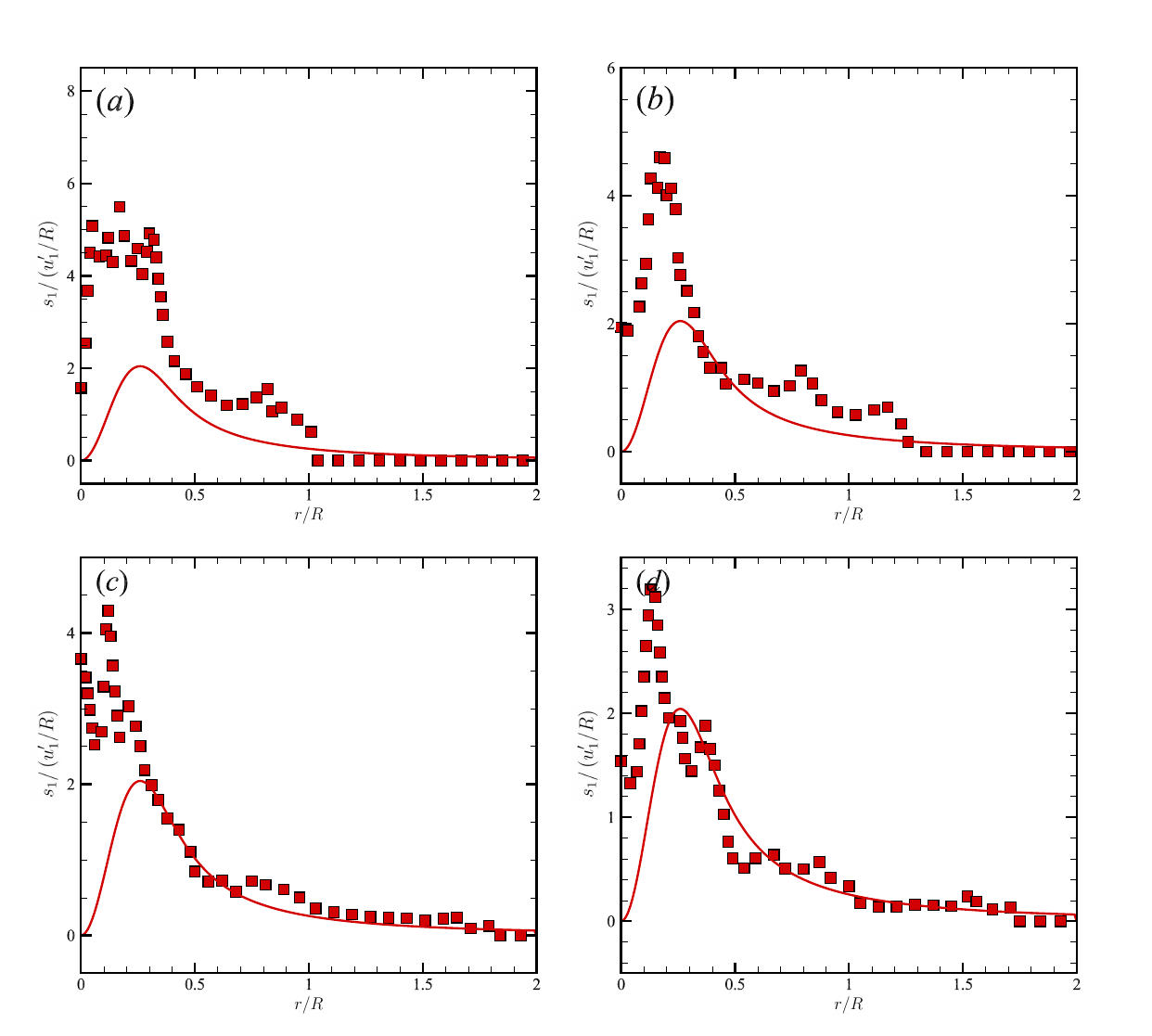}
    \captionsetup{justification=justified, singlelinecheck=false}
    \caption{The distribution of the azimuthal averaged principal strain rate $s_1$ on the radial axis in VD SBI. The moments captured from $\left(a\right)$ to $\left(d\right)$ are the same as those of Fig.~\ref{azimuthal velocity distribution in VD SBI}. The azimuthal averaged principal strain rate is plotted as the red dots, while the prediction of the single-vortex stretching theory in Eq. \ref{principal strain rate prediction in PS SBI} is plotted as the red solid line.}
    \label{principal strain rate in VD SBI 1}
\end{figure}

\begin{figure}
    \centering
    \includegraphics[width=1.0\linewidth]{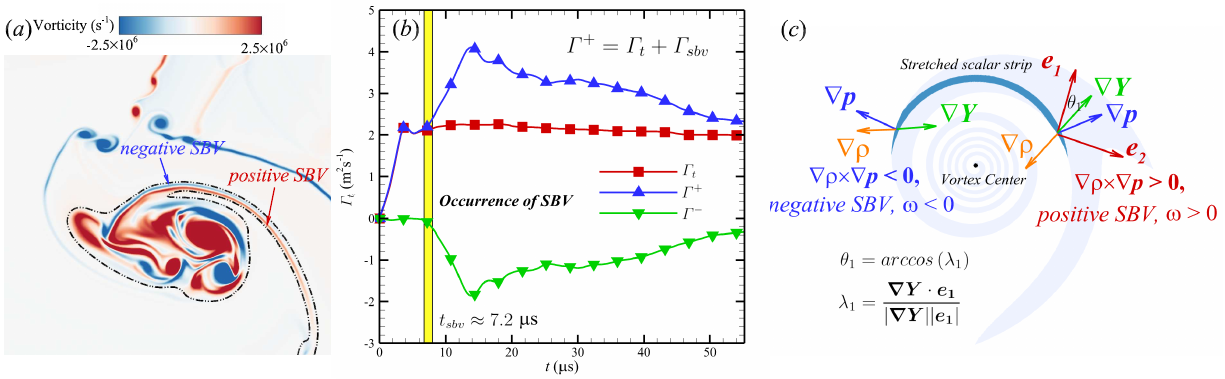}
    \captionsetup{justification=justified, singlelinecheck=false}
    \caption{Schematic of the secondary baroclinic vorticity (SBV) in VD SBI, including (a) an example of the SBV distribution in VD SBI at at $t=21.6\ \rm{ \upmu}s$; (b) the evolution of the total circulation $\varGamma_t$, the positive circulation $\varGamma^{+}$, and negative circulation $\varGamma^{-}$; and (c) the geometric relationship related to the generation of SBV.
    }
    \label{SBV schematic}
\end{figure}

This section examines whether the stretching dynamics in VD SBI manifest the same single-vortex stretching characteristic as observed in PS SBI. Specifically, the azimuthal averaged principal strain rate $s_1$ in VD SBI is compared with the predictions of single-vortex stretching model as outlined in Eq. \ref{principal strain rate prediction in PS SBI}. As illustrated in Fig.~\ref{principal strain rate in VD SBI 1}, the current single-vortex stretching model underestimates $s_1$ in VD SBI. \textcolor{black}{In deriving the distribution of $s_1$ for the single-vortex, the application of assumption PS.(iii) neglected the density gradient $\nabla \rho$, leading to an oversimplified SDGE-$s_i$ in Eq. \ref{simplified SDGE1 3}, where only viscous effect is considered.} Therefore, the inclusion of the baroclinic term, $-\frac{1}{2\rho}\frac{1}{r}\frac{\partial \rho}{\partial \varphi}\frac{\partial p}{\partial r}$, as outlined in Eq. \ref{simplified SDGE1 2}, should be reconsidered in the analysis of VD SBI to provide a more accurate depiction of the principal strain rate $s_1$. Notably, the governing equation for $s_1$ shares the same mathematical structure as that for vorticity $\omega$.
Thus, the influence of the baroclinic term can be reflected by secondary baroclinic vorticity (SBV), a canonical phenomenon in VD mixing, as reported in several studies \citep{peng2021mechanism, peng2021effects, liu2022mixing}. As shown in Fig. \ref{SBV schematic},
SBV manifests as a vortex-bilayer structure characterized by dominant negative vorticity and intensified positive vorticity. This structure is quantitatively described using the positive and negative circulation, defined as follows, consistent with prior investigations \citep{peng2003vortex,niederhaus2008computational}:
\begin{equation}
    \left\{
        \begin{aligned}
            &\varGamma^{+} = \iint_{\omega>0} \omega {\rm dV}, \\
            &\varGamma^{-} = \iint_{\omega<0} \omega {\rm dV}.
        \end{aligned}
    \right.
\end{equation}
Fig.~\ref{SBV schematic} $(b)$ describes the evolution of positive and negative circulation: prior to the time $t_{sbv} \approx 7.2 \rm{\ \upmu s}$, the shock interaction with the bubble triggers the formation of the primary vortex. After $t > t_{sbv}$, the growth of positive and negative circulation $\varGamma^{+}$ and $\varGamma^{-}$ signifies the emergence of SBV. According that the positive circulation is the sum of the total circulation and the baroclinic circulation composed by SBV:
\begin{equation}
     \varGamma^{+} = \varGamma_{t} + \varGamma_{sbv},
\end{equation}
the principal strain rate $s_1$ in VD SBI can be characterized by the combination of the strain rate arising from the primary vortex $s_1^{ps}$ and that induced by the secondary baroclinic effect $s_1^{b,\nu}$:
\begin{equation}
    s_1 = s_1^{ps} + s_1^{b,\nu},
\end{equation}
where $s_1^{ps}$ is determined by the single-vortex principal strain rate as described in Eq. \ref{principal strain rate prediction in PS SBI}. Therefore, accurate modeling of $s_1^{b,\nu}$ is essential to provide a precise description of the principal strain rate $s_1$ in VD SBI.

\textcolor{black}{Considering the linearity of the governing equation for the principal strain rate as specified in Eq. \ref{simplified SDGE1 2}, $s_1^{b,\nu}$ satisfies the following equation:
\begin{equation}
    \frac{\rm d}{{\rm d}t}s_1^{b,\nu} = -\frac{1}{2\rho^2}\frac{1}{r}\frac{\partial \rho}{\partial \varphi}\frac{\partial p }{\partial r} + \nu\left(\frac{\partial^2 s_1^{b,\nu}}{\partial r^2} + \frac{1}{r}\frac{\partial s_1^{b,\nu}}{\partial r}\right),
    \label{baroclinic viscous equation}
\end{equation}
indicating that $s_1^{b,\nu}$ is governed by the baroclinic term $-\frac{1}{2\rho^2}\frac{1}{r}\frac{\partial \rho}{\partial \varphi}\frac{\partial p }{\partial r}$ and viscous term $\nu\left(\frac{\partial^2 s_1^{b,\nu}}{\partial r^2} + \frac{1}{r}\frac{\partial s_1^{b,\nu}}{\partial r}\right)$. Eq.~\ref{baroclinic viscous equation} is relatively complex, and the modeling of the secondary baroclinic strain $s_1^{b,\nu}$ can be outlined as follows:} 

\textcolor{black}{Firstly, to simplify the analysis, the derivation initially focuses solely on the baroclinic term $-\frac{1}{2\rho^2}\frac{1}{r}\frac{\partial \rho}{\partial \varphi}\frac{\partial p }{\partial r}$:
\begin{equation}
    \frac{\rm d}{{\rm d}t}s_1^{b} = -\frac{1}{2\rho^2}\frac{1}{r}\frac{\partial \rho}{\partial \varphi}\frac{\partial p }{\partial r} = \frac{1}{2\rho}\left(\frac{1}{\rho}\frac{\partial p}{\partial r}\right)\left(-\frac{1}{r}\frac{\partial \rho}{\partial \varphi}\right),
    \label{baroclinic equation}
\end{equation}
and the solution derived from this simplification is referred to as $s_1^b$. }

\textcolor{black}{Secondly, based on the expression for $s_1^{b}$, the effect of the viscous term $\nu\left(\frac{\partial^2 s_1^{b,\nu}}{\partial r^2} + \frac{1}{r}\frac{\partial s_1^{b,\nu}}{\partial r}\right)$ is incorporated through a viscous correction.
}

\textcolor{black}{In the process of modeling of the secondary baroclinic strain $s_1^{b,\nu}$, three assumptions for VD mixing are introduced to simplify the equation. These include assumptions VD.(i) and VD.(ii) for modeling the density gradient $\frac{1}{r}\frac{\partial \rho}{\partial \varphi}$ in baroclinic term $-\frac{1}{2\rho^2}\frac{1}{r}\frac{\partial \rho}{\partial \varphi}\frac{\partial p }{\partial r}$, and assumption VD.(iii) for the viscous correction:}

\textcolor{black}{\textbf{Assumption VD.(i):} Similar to previous studies of SBI \citep{yu2020scaling} and VD buoyancy-driven turbulence \citep{aslangil2020variable}, the relationship between the post-shock density and mass fraction in VD SBI can be expressed as:
\begin{equation}
    \frac{1}{\rho} = \frac{1-Y}{\rho_1'} +  \frac{Y}{\rho_2'},
    \label{VD.i}
\end{equation}
where $\rho_1'$ represents the post-shock air density, and $\rho_2'$ denotes the density of helium following the shock interaction.}

\textcolor{black}{\textbf{Assumption VD.(ii):}  The distribution of the mass fraction along the scalar strips, $Y\left(\varsigma,\tau^{\mathcal{D}}\right) = \frac{1}{2}\left[{\rm erf}\left(\frac{\varsigma + 1/2}{2\sqrt{\tau^{\mathcal{D}}}}\right) - {\rm erf}\left(\frac{\varsigma - 1/2}{2\sqrt{\tau^{\mathcal{D}}}}\right) \right]$, informs the evolution of the diffusion scale $\lambda_{\mathcal{D}}$ \citep{souzy2018mixing}:
\begin{equation}
    \lambda_{\mathcal{D}}(t) = \left\{
    \begin{aligned}
       & s(t) = \frac{s_0}{\sqrt{1+\tau^2}}, \qquad \quad\ \ {\rm if} \ \tau^{\mathcal{D}} < \frac{1}{16},\\
       & 4s(t)\sqrt{\tau^{\mathcal{D}}} = \frac{4s_0\sqrt{\tau^{\mathcal{D}}}}{\sqrt{1+\tau^2}},\quad {\rm if} \  \tau^{\mathcal{D}} \geq \frac{1}{16}.
    \end{aligned}
    \right.
    \label{diffusion scale}
\end{equation}}

\textcolor{black}{\textbf{Assumption VD.(iii):} Based on the solution for $s_1^b$, the viscous effect on the secondary baroclinic strain $s_1^{b,\nu}$ can be incorporated through a viscous correction \citep{petropoulos2023settling}, which is given by:
\begin{equation}
    \left\{
    \begin{aligned}
        &s_1^{b,\nu} = s_1^{b}{\rm erf}\left(\frac{1}{4\sqrt{\tau^{\nu}}}\right), \\
        &\tau^{\nu} \approx \frac{\nu}{{s_0}^2}\left[\frac{4}{3}{s_1^{ps}}^2\left(t - t_{sbv}\right)^3\right].
    \end{aligned}
    \right.
    \label{s1 baro nu expression}
\end{equation}}
\textcolor{black}{The validation of the assumptions for VD mixing is provided in the supplementary materials.} 

\textcolor{black}{Here, we present the derivation of $s_1^{b}$. As shown in Eq.~\ref{baroclinic equation}, the baroclinic term is the product of the pressure gradient $\frac{1}{\rho}\frac{\partial p}{\partial r}$ and the density gradient $-\frac{1}{r}\frac{\partial \rho}{\partial \varphi}$.} The pressure gradient, which arises from the primary vortex, can be modeled by the azimuthal velocity as specified in Eq. \ref{Saffman vortex model}:
\begin{equation}
    \frac{1}{\rho}\frac{\partial p}{\partial r} = \frac{u_{\varphi}^2}{r} = \frac{\varGamma_{t}^2}{4\upi^2 r^3}\left(1-{\rm exp}{\left(-\frac{r^2}{4\nu(t+t_{0})}\right)}\right)^2 \approx \frac{\varGamma_{t}^2}{4\upi^2 r^3}\left(1-{\rm exp}{\left(-\frac{r^2}{4\nu t_{0}}\right)}\right)^2.
    \label{pressure gradient}
\end{equation}
This expression presents the steady pressure gradient due to the relationship $t_{0} \gg t$.

\textcolor{black}{The modeling of the azimuthal component of the density gradient $-\frac{1}{r}\frac{\partial \rho}{\partial \varphi}$ requires Eq.~\ref{VD.i} from assumption VD.(i).} Using this relationship, the density gradient $\nabla \rho$ is expressed in terms of the scalar gradient $\nabla Y$:
\begin{equation}
    \nabla \rho = -\rho^2\frac{\rho_1' - \rho_2'}{\rho_1' \rho_2'} \nabla Y,
\end{equation}
Owing to the distribution of $\nabla \rho$ across scalar strips as depicted in Fig.~\ref{SBV schematic} $(c)$, this relationship can be simplified to:
\begin{equation}
    \nabla \rho \approx -\rho_2'^2\frac{\rho_1' - \rho_2'}{\rho_1' \rho_2'} \nabla Y = -\rho_2'\left(1-\sigma\right)\nabla Y,
    \label{relation between density gradient and scalar gradient}
\end{equation}
where $\sigma = \rho_2'/\rho_1'$ denotes the density ratio between post-shock helium and post-shock air. By applying this expression of the density gradient and the geometric schematic depicted in Fig.~\ref{SBV schematic} $(c)$, the azimuthal component of the density gradient 
$-\frac{1}{r}\frac{\partial \rho}{\partial \varphi}$ is derived by projecting the scalar gradient $\nabla Y$ onto the azimuthal direction:
\begin{equation}
    -\frac{1}{r}\frac{\partial \rho}{\partial \varphi} = \rho_2'\left(1-\sigma\right)\lvert\nabla Y\rvert \cos\left(\theta_1 + \frac{\upi}{4}\right).
\end{equation}

Based on the solution of SDGE-$\lambda_i$ detailed in Eq. \ref{SDGE2 solution}, specified as  $\theta_1 = -\frac{\upi}{4} + \arctan\left(\tau\right)$, the azimuthal density gradient component can be refined further to:
\begin{equation}
    -\frac{1}{r}\frac{\partial \rho}{\partial \varphi} = \rho_2'\left(1-\sigma\right)\lvert\nabla Y\rvert \frac{1}{\sqrt{1+\tau^2}}.
    \label{density gradient}
\end{equation}
Moreover, the magnitude of the scalar gradient $\nabla Y$ in this formulation is expressed by the averaged SDR over the scalar strips:
\begin{equation}
    \lvert \nabla Y \rvert \approx \sqrt{\overline{\chi}},\qquad  \overline{\chi} = \int_{-\infty}^{+\infty}\left(\frac{{\rm d}Y}{{\rm d}x}\right)^2 {\rm d}x/\lambda_{\mathcal{D}} = \frac{1}{s(t)}\int_{-\infty}^{+\infty}\left(\frac{{\rm d}Y}{{\rm d}\varsigma}\right)^2 {\rm d}\varsigma / \lambda_{\mathcal{D}}.
\end{equation}
\textcolor{black}{Here, the mass fraction is presumed to follow the distribution described in Eq. \ref{ade solution} on the scalar strips, specified as: $Y\left(\varsigma,\tau^{\mathcal{D}}\right) = \frac{1}{2}\left[{\rm erf}\left(\frac{\varsigma + 1/2}{2\sqrt{\tau^{\mathcal{D}}}}\right) - {\rm erf}\left(\frac{\varsigma - 1/2}{2\sqrt{\tau^{\mathcal{D}}}}\right) \right]$, and the diffusion scale $\lambda_{\mathcal{D}}$ is determined using Eq.~\ref{diffusion scale} from assumption VD.(ii). With the expression of $\lambda_{\mathcal{D}}$ in Eq.~\ref{diffusion scale}, the averaged SDR on scalar strips can be rewritten as:}
\begin{equation}
    \overline{\chi} = \left\{
    \begin{aligned}
        &\frac{1+\tau^2}{s_0^2}\frac{1}{\sqrt{2\upi \tau^{\mathcal{D}}}}\left(1 - \mathrm{e}^{-\frac{1}{8\tau^{\mathcal{D}}}}\right), \qquad {\rm if} \ \tau^{\mathcal{D}} < \frac{1}{16},\\
        &\frac{1+\tau^2}{4s_0^2\tau^{\mathcal{D}}}\frac{1}{\sqrt{2\upi}}\left(1 - \mathrm{e}^{-\frac{1}{8\tau^{\mathcal{D}}}}\right), \qquad \quad \ {\rm if} \ \tau^{\mathcal{D}} \geq \frac{1}{16},
    \end{aligned}
    \label{averaged SDR}
    \right.
\end{equation}
which can be further simplified considering the distinct stages of SDR evolution, following approximations similar to Eq.~\ref{PS mixing model simplification 1} and Eq.~\ref{PS mixing model simplification 2}. The simplified form is given by:
\begin{equation}
    \overline{\chi} = \left\{
    \begin{aligned}
        &\frac{1+\tau^2}{s_0^2}\frac{1}{\sqrt{2\upi \tau^{\mathcal{D}}}}, \qquad {\rm if} \ \tau^{\mathcal{D}} < \frac{1}{8},\\
        &\frac{1+\tau^2}{4s_0^2\tau^{\mathcal{D}}}\frac{1}{\sqrt{2\upi}}\frac{1}{8\tau^{\mathcal{D}}}, \quad  \ {\rm if} \ \tau^{\mathcal{D}} \geq \frac{1}{8}.
    \end{aligned}
    \right.
\end{equation}

Combining the expressions for the pressure gradient from Eq. \ref{pressure gradient} and density gradient from Eq. \ref{density gradient}, the baroclinic term specified in Eq. \ref{baroclinic equation} can be reformulated using the approximation $\tau^{\mathcal{D}}\approx \frac{\mathcal{D}}{s_0^2}\left(\frac{4}{3}{s_1^{ps}}^2 t^3\right)$, yielding: 
\begin{equation}
    -\frac{1}{2\rho^2}\frac{1}{r}\frac{\partial \rho}{\partial \varphi}\frac{\partial p}{\partial r} = \left\{
    \begin{aligned}
        &\frac{\varGamma_{t}^2}{16\upi^2 r^3}\left(1-{\rm exp}{\left(-\frac{r^2}{4\nu t_{0}}\right)}\right)^2\left(1-\sigma\right)\frac{1}{\left(\frac{8\upi \mathcal{D}}{3}\right)^{\frac{1}{4}}\left(s_0 s_1^{ps}\right)^{\frac{1}{2}}}t^{-\frac{3}{4}}, \quad {\rm if}\ t < t_{charac},\\
        & \frac{3\varGamma_{t}^2}{256\sqrt{2}\upi^2 r^3}\left(1-{\rm exp}{\left(-\frac{r^2}{4\nu t_{0}}\right)}\right)^2\left(1-\sigma\right)\frac{s_0}{\left(2\upi\right)^{\frac{1}{4}}\mathcal{D}{s_1^{ps}}^2 t^3}, \quad\ \ {\rm if}\ t \geq t_{charac},
    \end{aligned}
    \right.
\end{equation}
where $t_{charac} = \left(\frac{3s_0}{32\mathcal{D}{s_1^{ps}}^2}\right)^{\frac{1}{3}}$ is determined from $\tau^{\mathcal{D}} = \frac{1}{8}$. 

Noticing that Fig.~\ref{SBV schematic} $\left(b\right)$ illustrates the presence of SBV after $t > t_{sbv}$, the solution to the governing equation for $s_1^b$ in Eq. \ref{baroclinic equation} should consider the emergence time $t_{sbv}$:

$\quad {\rm if}\ t_{charac} <\  t_{sbv}:$
\begin{equation}
    s_1^{b} = \frac{3\varGamma_{t}^2}{512\sqrt{2}\upi^2 r^3}\left(1-{\rm exp}{\left(-\frac{r^2}{4\nu t_{0}}\right)}\right)^2\left(1-\sigma\right)\frac{s_0}{\left(2\upi\right)^{\frac{1}{4}}\mathcal{D}{s_1^{ps}}^2 }\left(\frac{1}{t_{sbv}^2} - \frac{1}{t^2}\right)
\end{equation}
$\qquad {\rm if}\ t_{charac} \geq\  t_{sbv}:$
\begin{equation}
    s_1^b = \left\{
    \begin{aligned}
        &\frac{\varGamma_{t}^2}{4\upi^2r^3}\left(1-{\rm exp}{\left(-\frac{r^2}{4\nu t_{0}}\right)}\right)^2\frac{1}{\left(\frac{8\upi \mathcal{D}}{3}\right)^{\frac{1}{4}}\left(s_0 s_1^{ps}\right)^{\frac{1}{2}}}\left(t^{\frac{1}{4}} - {t_{sbv}}^{\frac{1}{4}}\right),  \quad\quad \  {\rm if}\ t < t_{charac}\\
        &s_1^{b}|_{0}-\frac{3\varGamma_{t}^2}{512\sqrt{2}\upi^2 r^3}\left(1-{\rm exp}{\left(-\frac{r^2}{4\nu t_{0}}\right)}\right)^2\left(1-\sigma\right)\frac{s_0}{\left(2\upi\right)^{\frac{1}{4}}\mathcal{D}{s_1^{ps}}^2 }\frac{1}{t^2},  \ {\rm if}\ t \geq t_{charac}
    \end{aligned}
    \right.
\end{equation}
where $s_1^b|_{0}$ is proposed to ensure the continuity of this solution. \textcolor{black}{Following this solution for $s_1^b$, the viscous effect described in Eq. \ref{baroclinic viscous equation} is incorporated using the viscous correction from assumption VD.(iii).} \textcolor{black}{The reasonability of the equation simplifications applied for deriving the secondary baroclinic principal strain $s_1^{b,\nu}$ are provided in the supplementary material.}

Based on the analytical expressions for the principal strain induced by the secondary baroclinic effect $s_1^{b,\nu}$ and the principal strain induced by the primary vortex $s_1^{ps}$, the principal strain rate in VD SBI $s_1$ can be articulated as follows:
\begin{equation}
    \left\{
    \begin{aligned}
        &s_1 = s_1^{ps} + s_1^{b,\nu},\\
        &s_{1}^{ps} = \frac{\varGamma_{t}}{2\upi r^2}\left(1 - \exp\left(-\frac{r^2}{4\nu (t + t_0)}\right)\right) - \frac{\varGamma_{t}}{8\upi \nu t}\exp\left(-\frac{r^2}{4\nu (t + t_0)}\right),\\
        &s_1^{b,\nu} = s_1^{b} {\rm erf}\left(\frac{1}{4\sqrt{\tau^{\nu}}}\right).
    \end{aligned}
    \right.
    \label{s1 result for VD SBI}
\end{equation}
\begin{figure}
    \centering
    \includegraphics[width=0.7\linewidth]{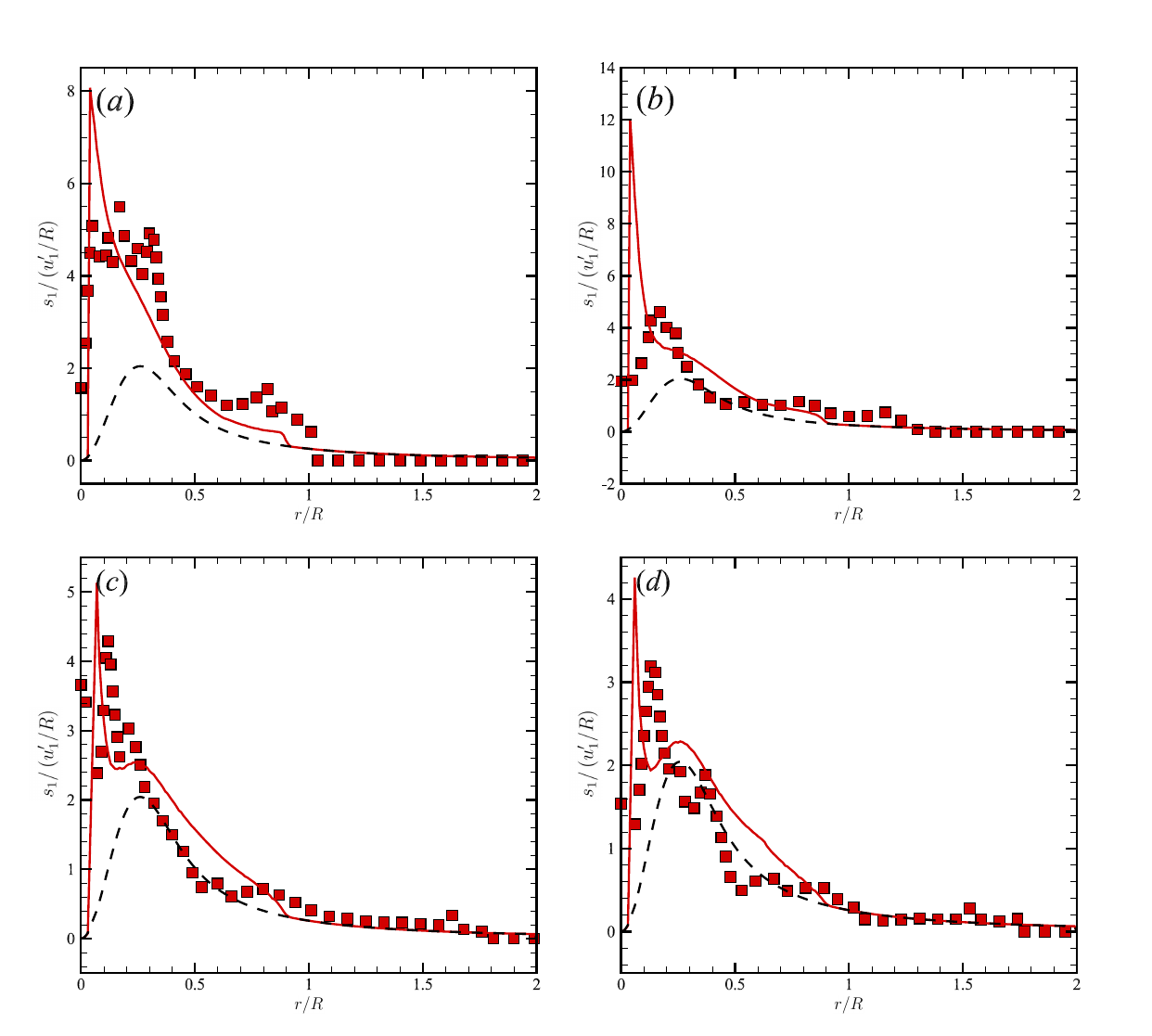}
    \captionsetup{justification=justified, singlelinecheck=false}
    \caption{The distribution of the azimuthal averaged principal strain rate $s_1$ on the radial axis in VD SBI. The moments captured from $\left(a\right)$ to $\left(d\right)$ are the same as those of Fig.~\ref{azimuthal velocity distribution in VD SBI}. The azimuthal averaged principal strain rate is plotted as the red dots, while the prediction of the single-vortex theory for $s_1^{ps}$ in Eq. \ref{principal strain rate prediction in PS SBI} is plotted as the black dashed line, and the prediction for $s_1$ in Eq. \ref{s1 result for VD SBI} is plotted as the red solid line.}
    \label{principal strain rate in VD SBI 2}
\end{figure}
Figure \ref{principal strain rate in VD SBI 2} compares the azimuthal averaged principal strain rate $s_1$ in VD SBI with the prediction of single-vortex theory for $s_1^{ps}$ from Eq. \ref{principal strain rate prediction in PS SBI}, alongside the prediction for $s_1$ from Eq. \ref{s1 result for VD SBI}. This comparison underscores that Eq. \ref{s1 result for VD SBI} effectively captures the distribution of $s_1$ in VD SBI.

\begin{figure}
    \centering
    \includegraphics[width=0.7\linewidth]{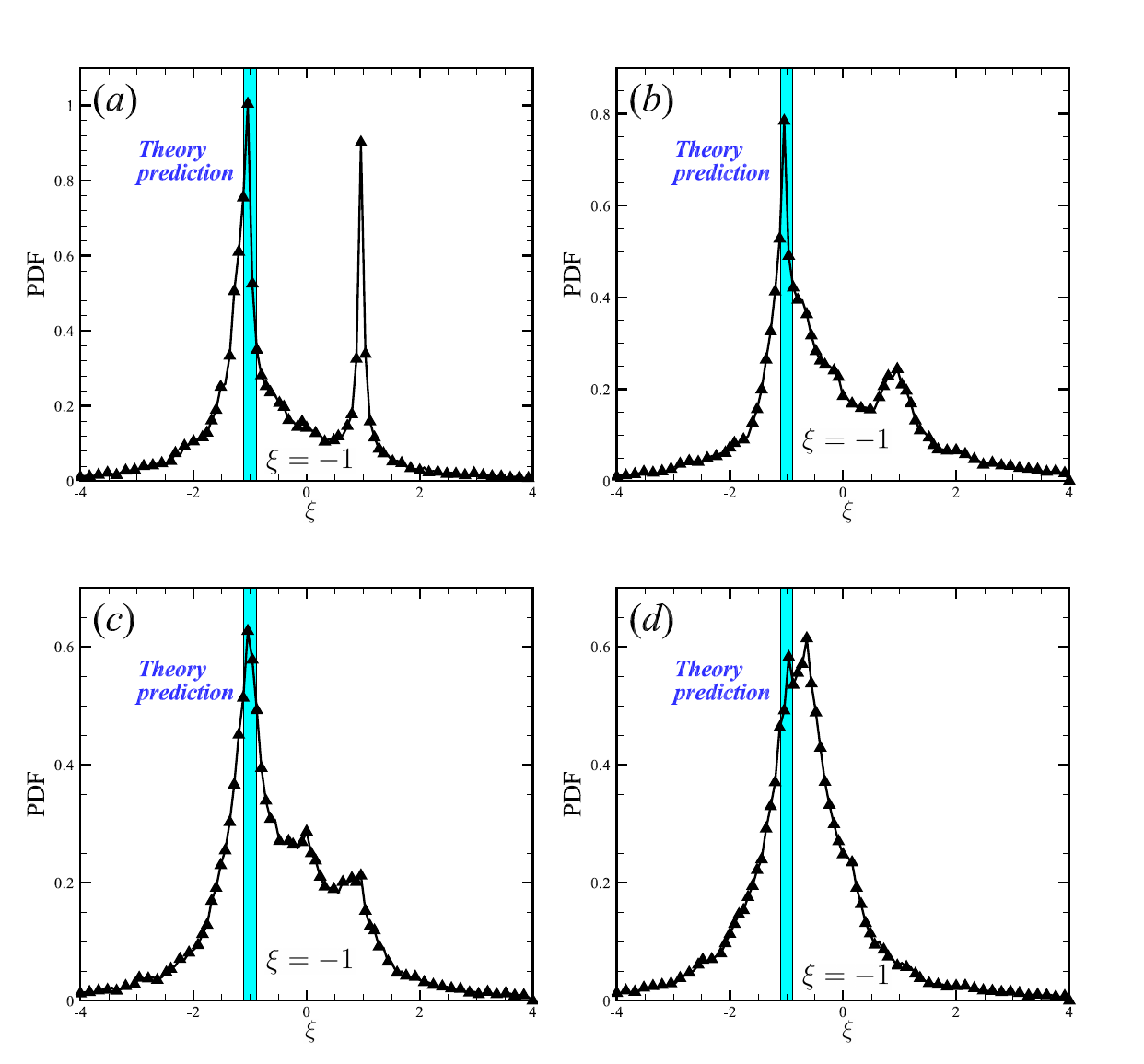}
    \captionsetup{justification=justified, singlelinecheck=false}
    \caption{The PDF of the parameter $\xi = \frac{\widetilde{W_{12}} + \omega/2}{\frac{1}{2}(s_1-s_2)}$ in VD SBI at the same moments as those of Fig.~\ref{azimuthal velocity distribution in VD SBI}.}
    \label{distribution of xi in VD SBI}
\end{figure}

\begin{figure}
    \centering
    \includegraphics[width=0.7\linewidth]{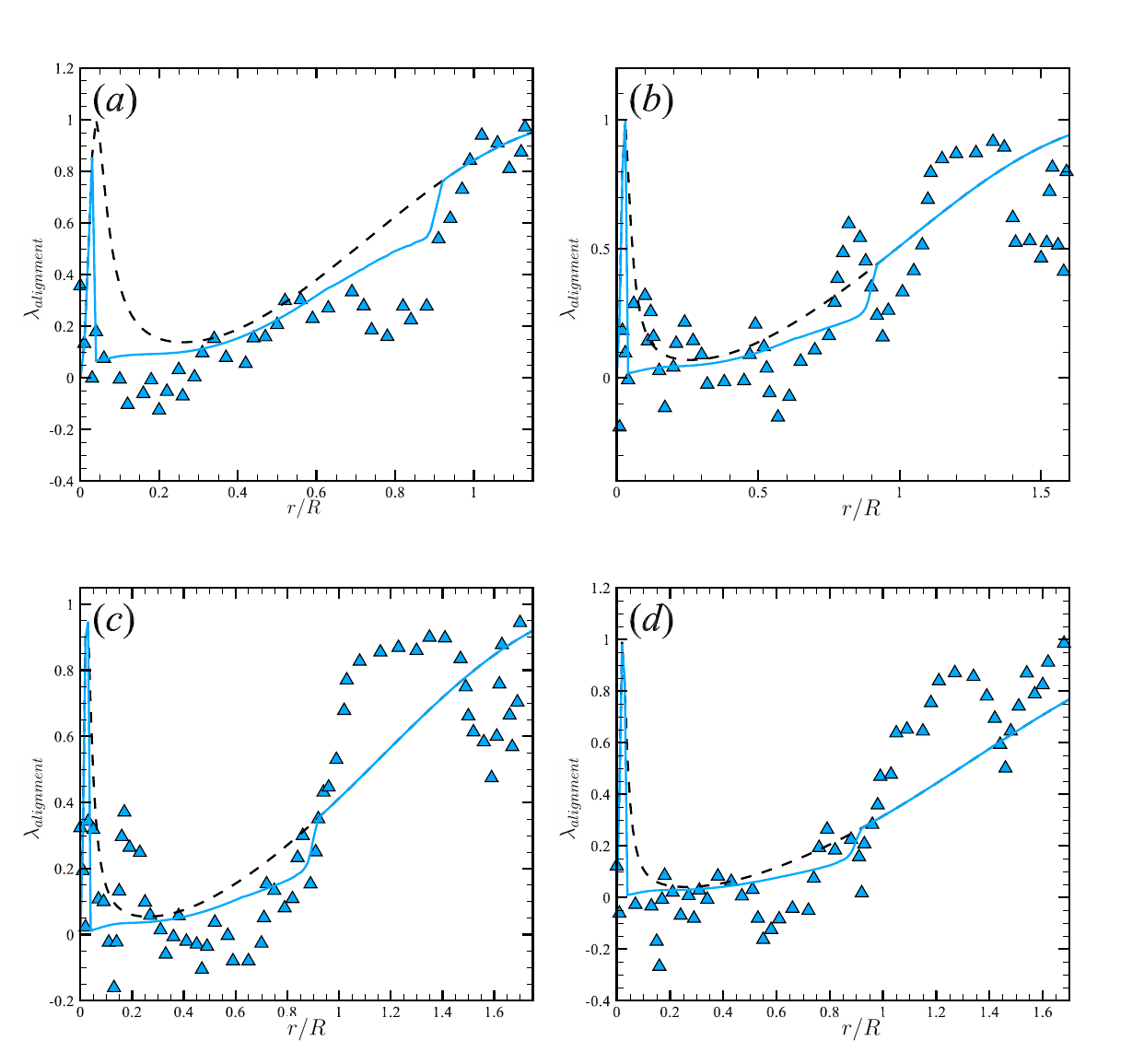}
    \captionsetup{justification=justified, singlelinecheck=false}
    \caption{The distribution of the azimuthal averaged alignment of the scalar gradient $\lambda_{alignment}$ on the radial axis in VD SBI. The moments captured from $\left(a\right)$ to $\left(d\right)$ are the same as those of Fig.~\ref{azimuthal velocity distribution in VD SBI}. The azimuthal averaged alignment is plotted as the blue dots, while the prediction of the single-vortex theory in Eq. \ref{alignment prediction in PS SBI} is plotted as the black dashed line, and the prediction for $\lambda_{alignment}$ in Eq. \ref{alignment prediction in VD SBI} is plotted as blue solid line.}
    \label{alignment in VD SBI}
\end{figure}

Following the examination of the principal strain rate $s_1$, the alignment of the scalar gradient $\lambda_{alignment}$ is further analyzed with respect to the stretching dynamics in VD SBI. A critical observation in PS SBI is the behavior of the parameter $\xi$ (Eq.~\ref{xi equals to -1 relationship}): the distribution of $\xi$ typically centers around $-1$, as illustrated in Fig.~\ref{distribution of xi in PS SBI}. This behavior is similarly investigated in the context of VD SBI. The PDF of $\xi$ in VD SBI is displayed in Fig.~\ref{distribution of xi in VD SBI}, it is evident that the peaks of the PDF curves cluster around $\xi = -1$. This consistency suggests that the $\xi = -1$ feature is not significantly altered by the baroclinic effect and remains applicable in VD SBI scenarios. Consequently, the derivation of SDGE-$\lambda_i$ outlined in Section \ref{Stretching dynamics analysis preliminaries} remains appropriate for analyzing the alignment of the scalar gradient $\lambda_{alignment}$ in VD SBI. The distribution of $\lambda_{alignment}$ is expected to follow Eq. \ref{alignment prediction in PS SBI}: $\lambda_{alignment} = 2{\tau}/\left({1+\tau^2}\right)$, where the expression for $s_1$ is altered from $s_1^{ps}$ in PS SBI to $s_1^{ps} + s_1^{b,\nu}$ in VD SBI:
\begin{equation}
    \left\{
    \begin{aligned}
        & \tau = \int_{0}^{t} 2s_1(t'){\rm d}t',\\
        & s_1 = s_1^{ps} + s_1^{b,\nu}.
    \end{aligned}
    \right.
    \label{alignment prediction in VD SBI}
\end{equation}
Therefore, the secondary baroclinic effect on $\lambda_{alignment}$ is captured by the additional principal strain $ s_1^{b,\nu}$ in VD SBI. Figure \ref{alignment in VD SBI} compares the azimuthal averaged $\lambda_{alignment}$ in VD SBI with the predictions from the single-vortex theory in Eq.~\ref{alignment prediction in PS SBI}, as well as those outlined the refined predictions in Eq.~\ref{alignment prediction in VD SBI}. The results confirm that the distribution of $\lambda_{alignment}$ in VD SBI aligns closely with the theoretical prediction by Eq.~\ref{alignment prediction in VD SBI}.

By examining the stretching dynamics from two perspectives: the principal strain $s_1$ and the alignment of the scalar gradient $\lambda_{alignment}$. It is evident that the effects of the baroclinic term in VD SBI are comprehensively captured by the combination of Eq.~\ref{s1 result for VD SBI} and Eq.~\ref{alignment prediction in VD SBI}. These equations for $s_1$ and $\lambda_{alignment}$ confirm the preservation of the same algebraic stretching characteristic in VD SBI with that observed in PS SBI, which means that the width of the scalar strip $s(t)$ in VD SBI evolves according to Eq.~\ref{algebraic evolution of s_t}: $s(t) = {s_0}/{\sqrt{1 + \tau^2}}$. 
The only difference between PS SBI and VD SBI lies in the expression for the principal strain $s_1$, as it appears in the definition of $\tau = \int_{0}^{t} 2s_1(t'){\rm d}t'$. 
\textcolor{black}{Based on the description of the scalar strip width in VD SBI, the mixing model incorporating only the secondary baroclinic effect (SBE) on the stretching dynamics is modified as:
\begin{equation}
     \left\langle \chi\right\rangle  = \int_{0}^{\sqrt{2\eta}R} \frac{1+\tau^2}{s_0}\frac{1}{\sqrt{2\upi\tau^{\mathcal{D}}}}\left(1-\mathrm{e}^{-\frac{1}{8\tau^{\mathcal{D}}}}\right) {\rm d}r,
     \label{VD mixing model modified by SBE}
\end{equation}
where $\tau = \int_{0}^{t}2s_1(t'){\rm d}t',\ s_1 = s_1^{ps}+s_1^{b,\nu}$.
However, the prediction of the SBE model still does not align with the observed SDR evolution in VD SBI, as will be further discussed in Fig.~\ref{comparision of VD model and SDR}.}

\subsection{Density source effect on diffusion process in VD SBI}

The stretching dynamics in VD SBI have been shown to be affected by the baroclinic effect, while the algebraic stretching characteristic is still retained. The derivation of the VD mixing model for \textcolor{black}{total SDR} $\left\langle \chi \right\rangle$ should consider another VD mixing feature, specifically transitioning from the advection-diffusion equation used in PS SBI (Eq. \ref{advection-diffusion equation}), to the multi-component transport equation applicable to VD SBI (Eq. \ref{multi-component transport equation}). In Section \ref{Numerical method}, the density variation in VD SBI introduces a density source term $\frac{\mathcal{D}}{\rho}\left(\frac{\partial \rho}{\partial x_j}\frac{\partial Y}{\partial x_j}\right)$ in Eq. \ref{VD advection-diffusion equation}. This section investigates the impact of this term on the diffusion process.

For the VD advection diffusion equation (Eq. \ref{VD advection-diffusion equation}), using the relation between the density gradient $\nabla \rho$ and scalar gradient $\nabla Y$ in Eq. \ref{relation between density gradient and scalar gradient}, this equation can be reformulated as:
\begin{equation}
    \frac{\partial Y}{\partial t} + u_j \frac{\partial Y}{\partial x_j} = -\mathcal{D}\left(1-\sigma\right)\frac{\partial Y}{\partial x_j}\frac{\partial Y}{\partial x_j} + \mathcal{D}\frac{\partial^2 Y}{\partial x_j^2}.
    \label{VD advection-diffusion equation 2} 
\end{equation}
\textcolor{black}{Similar to the approach used in the derivation of the PS mixing model described in Section \ref{PS mixing theory}, assumption PS.(iv) can also be applied to assume the mixing in VD SBI occurs within a compressed semicircle depicted in Fig.~\ref{distribution of s0}}. Consequently, Eq.~\ref{VD advection-diffusion equation 2} is transformed into the coordinate system $\left(O,X,Y\right)$ that aligns with the deforming scalar strips:
\begin{equation}
    \frac{\partial Y}{\partial t} + U\frac{\partial Y}{\partial x} = -\mathcal{D}\left(1-\sigma\right)\frac{\partial Y}{\partial x}\frac{\partial Y}{\partial x} + \mathcal{D}\frac{\partial^2 Y}{\partial x^2},
\end{equation}
where the relatively small component of the scalar gradient $\frac{\partial Y}{\partial y}$ is neglected. Employing the same expression for the relative velocity $(U,V)$ as in Eq. \ref{relative velocity on scalar strips} and utilizing the Ranz transformation described in Eq. \ref{ranz transform} as applied in PS SBI, this equation is further simplified into a quasi-Fourier equation:
\begin{equation}
    \frac{\partial Y}{\partial \tau^{\mathcal{D}}} = -\left(1-\sigma\right)\frac{\partial Y}{\partial \varsigma}\frac{\partial Y}{\partial \varsigma} + \frac{\partial^2 Y}{\partial \varsigma^2},
    \label{quasi-Fourier}
\end{equation}
the transformed time $\tau^{\mathcal{D}}$ in this equation is given by:
\begin{equation}
    \left\{
    \begin{aligned}
         & \tau^{\mathcal{D}} = \int_{0}^{t} \frac{\mathcal{D}}{s_0^2}\left(1+\tau^2\right){\rm d}t',\\
         & \tau = \int_{0}^{t} 2s_1(t'){\rm d}t',\\
         & s_1 = s_1^{ps} + s_1^{b,\nu}.
    \end{aligned}
    \right.
    \label{transformed tau in VD mixing}
\end{equation}

The solution to Eq. \ref{quasi-Fourier} is assumed to take the form $g(\tau^{\mathcal{D}})Y^{ps}(\varsigma,\tau^{\mathcal{D}})$, where $Y^{ps}(\varsigma,\tau^{\mathcal{D}})$ denotes the solution to the Fourier equation without the density source term, thus exhibiting the same expression with PS mixing as described in Eq. \ref{ade solution}. The term $g(\tau^{\mathcal{D}})$ corresponds to the modified function arising from the inclusion of the density source term. Substituting $g(\tau^{\mathcal{D}})Y^{ps}(\varsigma,\tau^{\mathcal{D}})$ into Eq.~\ref{quasi-Fourier}, the governing equation for the modified function $g(\tau^{\mathcal{D}})$ is derived as:
\begin{equation}
    \frac{1}{{g(\tau^{\mathcal{D}})}^2}\frac{\partial g(\tau^{\mathcal{D}})}{\partial \tau^{\mathcal{D}}} = -\left(1-\sigma\right)\frac{1}{Y^{ps}}\frac{\partial Y^{ps}}{\partial \varsigma}\frac{\partial Y^{ps}}{\partial \varsigma}.
    \label{equation for g_tau}
\end{equation}
In this equation, $\frac{\partial Y^{ps}}{\partial \varsigma}\frac{\partial Y^{ps}}{\partial \varsigma}$ can be approximated by its averaged value across scalar strips, using the approach analogous to the averaged SDR in Eq. \ref{averaged SDR}:
\begin{equation}
    \left.\frac{\partial Y^{ps}}{\partial \varsigma}\frac{\partial Y^{ps}}{\partial \varsigma}\right|_{\rm avg} = \left\{
    \begin{aligned}
        &\int_{-\infty}^{+\infty}\left(\frac{\partial Y^{ps}}{\partial \varsigma}\right)^2 {\rm d}\varsigma, \qquad \qquad \quad{\rm if}\ \tau^{\mathcal{D}} < \frac{1}{16},\\
        &\int_{-\infty}^{+\infty}\left(\frac{\partial Y^{ps}}{\partial \varsigma}\right)^2 {\rm d}\varsigma/\left(4\sqrt{\tau^{\mathcal{D}}}\right), \quad {\rm if}\ \tau^{\mathcal{D}} \geq \frac{1}{16}.
    \end{aligned}
    \right.
\end{equation}
Simultaneously, $Y^{ps}$ is approximated as half of the maximum value on the scalar strips:
\begin{equation}
    \left.Y^{ps}\right|_{\rm avg} = \frac{1}{2}{\rm erf}\left(\frac{1}{4\sqrt{\tau^{\mathcal{D}}}}\right) = \left\{
    \begin{aligned}
        &\frac{1}{2}, \qquad \qquad {\rm if}\ \tau^{\mathcal{D}} < \frac{1}{16},\\
        &\frac{1}{8\sqrt{\tau^{\mathcal{D}}}}, \qquad {\rm if}\ \tau^{\mathcal{D}} \geq \frac{1}{16}.
    \end{aligned}
    \right.
\end{equation}
Using these approximations, the right-hand side of Eq. \ref{equation for g_tau} is simplified:
\begin{equation}
    \left.\frac{1}{Y^{ps}}\frac{\partial Y^{ps}}{\partial \varsigma}\frac{\partial Y^{ps}}{\partial \varsigma}\right|_{\rm avg} = \left\{
    \begin{aligned}
        & \sqrt{\frac{2}{{\upi}}}{\tau^{\mathcal{D}}}^{\frac{1}{2}}, \qquad {\rm if}\ \tau^{\mathcal{D}} < \frac{1}{8},\\
        & \frac{1}{4\sqrt{2\upi}}{\tau^{\mathcal{D}}}^{-\frac{3}{2}}, \quad {\rm if}\ \tau^{\mathcal{D}} \geq \frac{1}{8}.
    \end{aligned}
    \right.
\end{equation}
\textcolor{black}{This simplification is validated to be reasonable in the supplementary material.} Substituting this result into Eq. \ref{equation for g_tau}, we obtain the 
expression for the modified function $g(\tau^{\mathcal{D}})$:
\begin{equation}
    g(\tau^{\mathcal{D}}) = \left\{
    \begin{aligned}
        & \frac{1}{1+2\sqrt{\frac{2}{\upi}}\left(1-\sigma\right){\tau^{\mathcal{D}}}^{\frac{1}{2}}}, \qquad \quad {\rm if}\ \tau^{\mathcal{D}} < \frac{1}{8},\\
        & \frac{1}{1 + 2\left(1-\sigma\right)\left(\frac{1}{\sqrt{\upi}}-\frac{{\tau^{\mathcal{D}}}^{-\frac{1}{2}}}{4\sqrt{2\upi}}\right)}, \quad {\rm if}\ \tau^{\mathcal{D}} \geq \frac{1}{8}.
    \end{aligned}
    \right.
    \label{g_tau form}
\end{equation}
\begin{figure}
    \centering
    \includegraphics[width=0.6\linewidth]{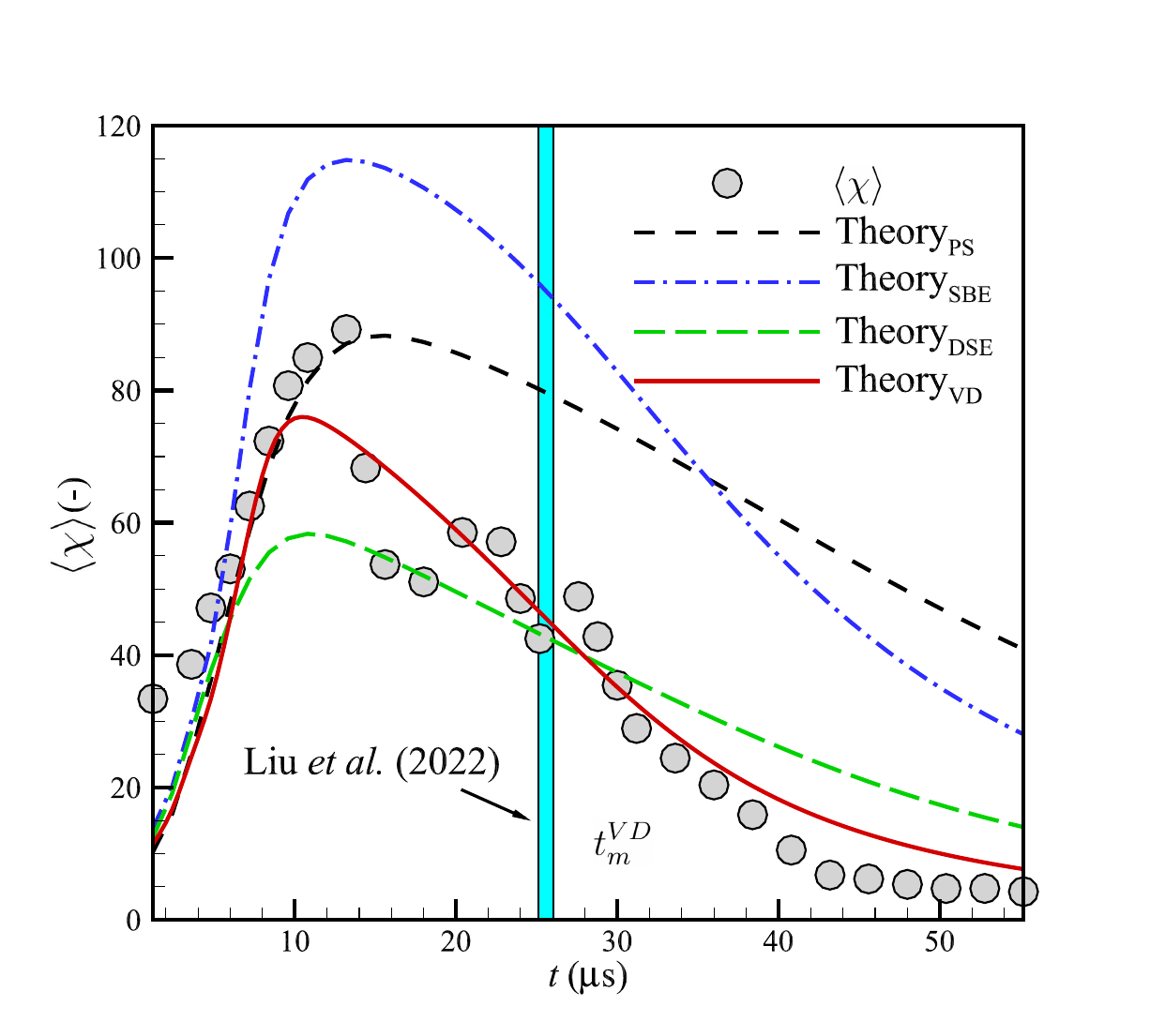}
    \captionsetup{justification=justified, singlelinecheck=false}
    \caption{\textcolor{black}{The comparison between the evolution of the total SDR $\left\langle \chi\right\rangle$ in VD SBI and the predictions of the mixing models. The prediction of PS mixing model as detailed in Eq. \ref{PS mixing model expression} is plotted as black dotted line, the prediction of VD mixing model as detailed in Eq. \ref{VD mixing model expression} is plotted as red solid line, the prediction of mixing model incorporating only the secondary baroclinic effect (SBE) on the stretching dynamics (Eq.~\ref{VD mixing model modified by SBE}) is plotted as blue dash dot line, and the prediction of mixing model incorporating only the density source effect (DSE) on the diffusion process (Eq.~\ref{VD mixing model modified by SBE}) is plotted as green long dash line. The blue line indicates the VD mixing time $t_{m}^{VD}$ obtained from the previous research \citep{liu2022mixing}.}}
    \label{comparision of VD model and SDR}
\end{figure}
The mass fraction $Y$ on scalar strips in VD SBI evolves as follows:
\begin{equation}
    \left\{
        \begin{aligned}
            &Y = g(\tau^{\mathcal{D}})Y^{ps},\\
            & Y^{ps} = \frac{1}{2}\left[{\rm erf}\left(\frac{\varsigma + 1/2}{2\sqrt{\tau^{\mathcal{D}}}}\right) - {\rm erf}\left(\frac{\varsigma - 1/2}{2\sqrt{\tau^{\mathcal{D}}}}\right) \right],
        \end{aligned}
    \right.
\end{equation}
\textcolor{black}{Based on this mass fraction distribution, the VD mixing model for the total SDR $\left\langle \chi \right\rangle$ can be proposed as:
\begin{equation}
    \left\langle \chi\right\rangle  = \int_{0}^{\sqrt{2\eta}R} {g(\tau^{\mathcal{D}})}^2\frac{1+\tau^2}{s_0}\frac{1}{\sqrt{2\upi\tau^{\mathcal{D}}}}\left(1-\mathrm{e}^{-\frac{1}{8\tau^{\mathcal{D}}}}\right) {\rm d}r,
    \label{VD mixing model expression}
\end{equation}
where 
\begin{equation}
    \left\{
    \begin{aligned}
        & s_1 = s_1^{ps} + s_1^{b,\nu},\\
        &\tau = \int_{0}^{t}2s_1{\rm d}t',\\
        &\tau^{\mathcal{D}} = \int_{0}^{t}\frac{\mathcal{D}{\rm d}t'}{s(t)^2} = \int_{0}^{t}\frac{\mathcal{D}(1+\tau^2){\rm d}t'}{s_0^2}.
    \end{aligned}
    \right.
\end{equation}}

\textcolor{black}{A comparison of the mathematical form of the VD mixing model with the PS mixing model presented in Eq. \ref{PS mixing model expression} reveals that the VD mixing model synthesizes two key VD effects on mixing. First, the baroclinic effect on stretching dynamics is reflected in the inclusion of the additional principal strain: $s_1^{b,\nu}$, which modifies the transformed time $\tau$ and $\tau^{\mathcal{D}}$, as defined in Eq. \ref{transformed tau in VD mixing}. Second, the influence of the density source on the diffusion process is accounted for by the introduction of the modified function $g(\tau^{\mathcal{D}})$. Similar to the SBE model (Eq.~\ref{VD mixing model modified by SBE}), which considers only the secondary baroclinic effect on VD mixing, the DSE model is proposed, where only the modified function $g(\tau^{\mathcal{D}})$ representing the density source effect (DSE) is included. The DSE model has the same form as Eq.~\ref{VD mixing model expression}, but with the secondary baroclinic strain absent: $s_1 = s_1^{ps}$. As with the SBE model (Eq.~\ref{VD mixing model modified by SBE}), the absence of the secondary baroclinic strain $s_1^{b,\nu}$ in the DSE model also leads to an inaccurate prediction of the SDR evolution in VD SBI, as discussed in Fig.~\ref{comparision of VD model and SDR}.}

\textcolor{black}{Figure~\ref{comparision of VD model and SDR} compares the evolution of the total SDR $\left\langle \chi \right\rangle$ with theoretical predictions from the PS mixing model (Eq.~\ref{PS mixing model expression}), the mixing model incorporating only the secondary baroclinic effect (SBE) on stretching dynamics (Eq.~\ref{VD mixing model modified by SBE}), the mixing model incorporating only the density source effect (DSE) on diffusion process, and the VD mixing model (Eq.~\ref{VD mixing model expression}). It is evident that the PS mixing model overestimates the total SDR $\left\langle \chi \right\rangle$. In comparison, the inclusion of the additional principal strain rate $s_1^{b,\nu}$ in Eq.~\ref{VD mixing model modified by SBE} enhances the growth of $\left\langle \chi \right\rangle$ during the early stage and accelerates its decay in the later stage. However, the prediction of Eq.~\ref{VD mixing model modified by SBE} still overestimates the observed evolution of $\left\langle \chi \right\rangle$ due to the absence of the modified function $g(\tau^{\mathcal{D}})$. The DSE model predicts the approximate value of total SDR $\left\langle\chi\right\rangle$ by including the modified function $g(\tau^{\mathcal{D}})$. However, in the absence of secondary baroclinic effect, this model significantly underestimates the principal strain $s_1$, which leads to a slower growth rate of $\left\langle\chi\right\rangle$ in the early stage and a smaller decay rate in the later stage. In contrast, the close agreement between the observed evolution of $\left\langle \chi \right\rangle$ and the predictions of the VD mixing model demonstrates that this model, which incorporates both secondary baroclinic effect on stretching and density source effect on diffusion, accurately captures the underlying mechanisms governing $\left\langle \chi \right\rangle$ in VD mixing. Therefore, compared to previous research on the VD mixing time $t_m^{VD}$ \citep{liu2022mixing}, the VD mixing model establishes a quantitative relationship between stretching dynamics and the evolution of the mixing rate in VD SBI.}

\section{Model validation and scaling behavior of the SBI mixing rate}
In Section \ref{VD mixing theory}, the VD mixing model for the evolution of SDR detailed in Eq. \ref{VD mixing model expression} is developed based on the characterization of stretching dynamics and the mixing process in VD SBI. Since SDR quantifies the mixing rate, this section demonstrates that the increase of mixedness, a measure of the extent of mixing, can also be theoretically predicted using this VD mixing model.
The validity of the VD mixing model is further established by its successful application in predicting the evolution of both SDR and mixedness in VD SBI across a broad range of Mach numbers. Additionally, this section presents a scaling analysis of the VD mixing model, which derives the dependence of the time-averaged mixing rate on the dimensionless P\'{e}clet number.
\label{Model validation and scaling behavior of the mixing rate}
\subsection{VD mixing within a wide range of Mach number}
In Fig. \ref{mean mixedness decomposition results}, the initial decrease in \textcolor{black}{the total mixedness} $\left\langle f\right\rangle$ is attributed to the effect of shock compression, while the monotonic increase in $\left\langle f\right\rangle$ is primarily driven by the SDR source term $\left\langle T_{SDR}^{f}\right\rangle$. Consequently, the increase in $\left\langle f\right\rangle$ can be theoretically predicted using the time integral of the VD mixing model described in Eq. \ref{VD mixing model expression}.

This section focuses on the increase in $\left\langle f\right\rangle$, where the effect of shock compression is ignored by redefining the increase in \textcolor{black}{total mixedness} through the removal of the velocity-divergence term 
$\left\langle T_{divU}^{f} \right\rangle$, expressed as:
\begin{equation}
    \Delta\left\langle f^{*} \right\rangle = \Delta\left\langle f \right\rangle - \left\langle T_{divU}^{f} \right\rangle = \left\langle T_{SDR}^{f} \right\rangle +  \left\langle T_{den}^{f} \right\rangle,
    \label{mixedness growth model}
\end{equation}
here the increase in \textcolor{black}{total mixedness} $\Delta\left\langle f^{*} \right\rangle$ are decomposed into the SDR source term $\left\langle T_{SDR}^{f} \right\rangle$ and density source term $\left\langle T_{den}^{f} \right\rangle$.

According to the definition of these two terms in Eq. \ref{mean mixedness decomposition}, the SDR source term driving the evolution of $\Delta\left\langle f^{*} \right\rangle$ can be modeled based on the VD mixing model for SDR as:
\begin{equation}
    \begin{aligned}
        \left\langle T_{SDR}^{f} \right\rangle = 8\mathcal{D}\int_{0}^{t} \left(\int_{0}^{\sqrt{2\eta}R} {g(\tau^{\mathcal{D}})}^2\frac{1+\tau^2}{s_0}\frac{1}{\sqrt{2\upi\tau^{\mathcal{D}}}}\left(1-\mathrm{e}^{-\frac{1}{8\tau^{\mathcal{D}}}}\right) {\rm d}r\right) {\rm d}t'.
    \end{aligned}
    \label{SDR source term model}
\end{equation}
For the density source term, the relationship between the density gradient $\nabla \rho$ and the scalar gradient $\nabla Y$ (Eq.~\ref{relation between density gradient and scalar gradient}) allows its transformation into the following form:
\begin{equation}
    \begin{aligned}
        \left\langle T_{den}^{f} \right\rangle &=  \int_{0}^{t} \left\langle 4\left(1-2Y\right)\frac{\mathcal{D}}{\rho}\frac{\partial \rho}{\partial x_j}\frac{\partial Y}{\partial x_j}\right \rangle {\rm d}t' \\
        & = \int_{0}^{t} \left\langle -4\mathcal{D}\left(1-2Y\right)\left(1-\sigma\right)\chi\right \rangle {\rm d}t',
    \end{aligned}
\end{equation}
indicating that the density source term is the product of the mass fraction and SDR. Using the expressions for mass fraction and SDR on scalar strips, the density source term is further modeled as:
\begin{equation}
    \left\{
    \begin{aligned}
        &\left\langle T_{den}^{f} \right\rangle = \int_{0}^{t} \left( \int_{0}^{\sqrt{2\eta}R} -4\mathcal{D}\left(1-2Y\right)\left(1-\sigma\right)\chi {\rm d}r\right) {\rm d}t',\\
        &\chi = {g(\tau^{\mathcal{D}})}^2\frac{1+\tau^2}{s_0}\frac{1}{\sqrt{2\upi\tau^{\mathcal{D}}}}\left(1-\mathrm{e}^{-\frac{1}{8\tau^{\mathcal{D}}}}\right),\\
        &Y \approx \frac{1}{2}Y|_{max} = \frac{1}{2} g(\tau^{\mathcal{D}}){\rm erf}\left(\frac{1}{4\sqrt{\tau^{\mathcal{D}}}}\right).
    \end{aligned}
    \right.
    \label{density source term model}
\end{equation}

\begin{figure}
    \centering
    \includegraphics[width=0.55\linewidth]{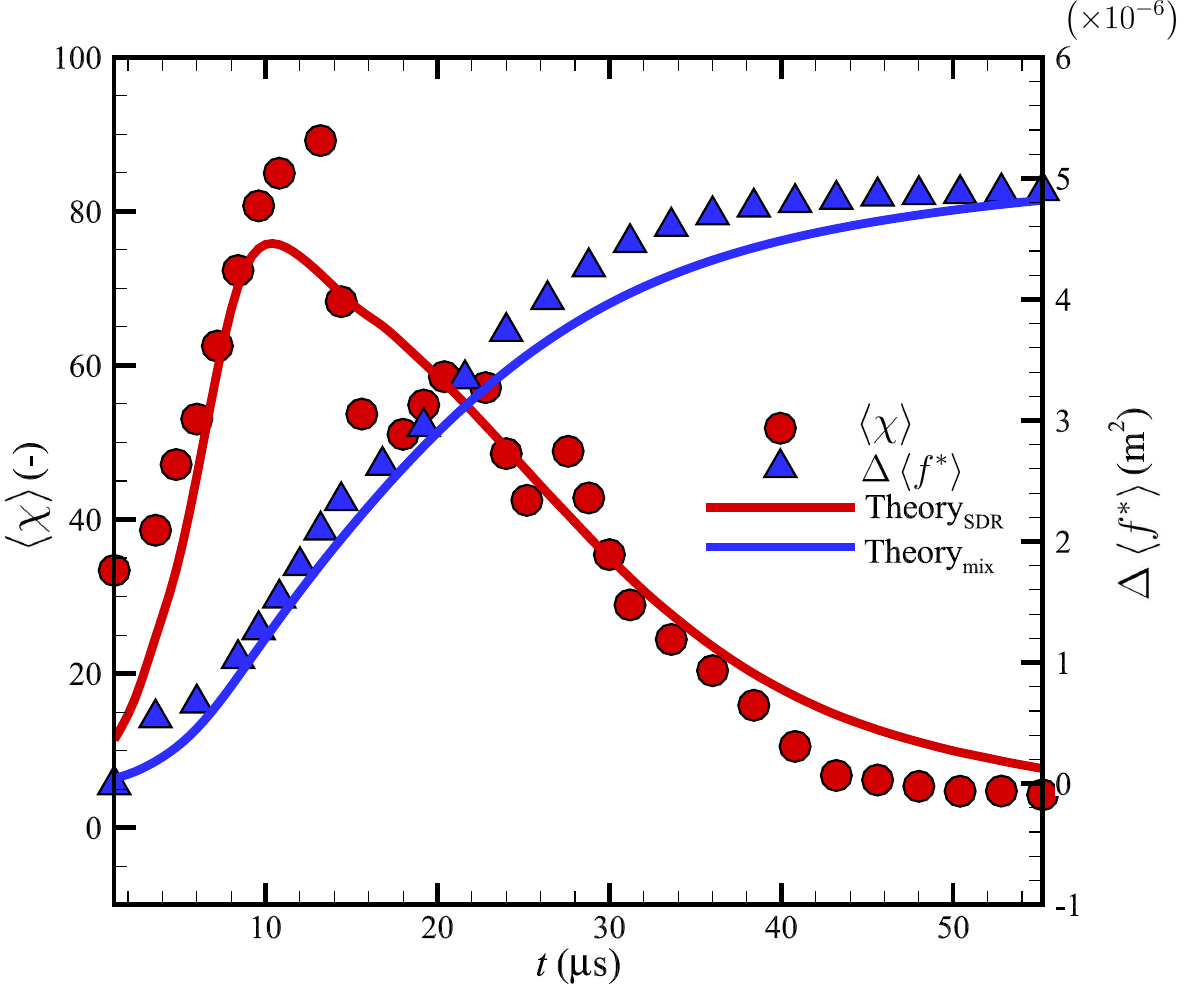}
    \captionsetup{justification=justified, singlelinecheck=false}
    \caption{The comparison between the evolution of \textcolor{black}{the total SDR} $\left\langle \chi\right\rangle$ and increase of \textcolor{black}{the total mixedness} $\Delta \left\langle f^{*} \right\rangle$ in VD SBI and the predictions of the mixing models. The prediction of VD mixing model for \textcolor{black}{total SDR} $\left\langle \chi\right\rangle$ as detailed in Eq. \ref{VD mixing model expression} is plotted as red solid line, while the prediction of VD mixing model for the increase in \textcolor{black}{the total mixedness} $\Delta \left\langle f^{*} \right\rangle$ as detailed in Eq. \ref{mixedness growth model} is plotted as blue solid line.}
    \label{comparision of VD model and mixedness, SDR}
\end{figure}

Based on the modeled expressions for the SDR source term $\left\langle T_{SDR}^{f}\right\rangle$ (Eq. \ref{SDR source term model}) and density source term $\left\langle T_{SDR}^{f}\right\rangle$ (Eq. \ref{density source term model}), the overall description of the increase in \textcolor{black}{total mixedness} $\Delta\left\langle f^{*} \right\rangle$ can be obtained as presented in Eq. \ref{mixedness growth model}. 
Figure \ref{comparision of VD model and mixedness, SDR} illustrates the small discrepancy between the observed evolution of the increase in \textcolor{black}{total mixedness} $\Delta \left\langle f^{*} \right\rangle$ in VD SBI and the theoretical prediction.

The predictive capability of the VD mixing model for both the SDR and the growth of mixedness is also validated in VD SBI across a Mach number range of $1.22\ \sim\ 4.0$, with the case setup detailed in Appendix \ref{wide range Ma number VD SBI settings}. Referring to the structure of the VD mixing model (Eq. \ref{VD mixing model expression}) and the associated expressions for stretching dynamics (Eq. \ref{s1 result for VD SBI} and \ref{alignment prediction in VD SBI}) and diffusion process (Eq. \ref{transformed tau in VD mixing} and \ref{g_tau form}), the undetermined parameters of the VD mixing model include the total circulation $\varGamma_{t}$, the compression rate $\eta$, the initial time $t_0$ for the principal strain rate arising from the primary vortex $s_{1}^{ps}$, the emergence time for the secondary baroclinic effect $t_{sbv}$, and the post-shock density ratio $\sigma$. For VD SBI cases spanning a Mach number range of $1.22\ \sim\ 4.0$, these parameters $\left(\varGamma_{t},\eta, t_0, t_{sbv},\sigma\right)$ are summarized in the following Table \ref{parameters for VD mixing model}. 
\begin{table}
\begin{center}
\def~{\hphantom{0}}
  \begin{tabular}{lccccc}
    \hline
    $Ma$ & $\varGamma_{t}\,\rm{(m^{2}\,s^{-1})}$ & $\eta$ & $\frac{\sqrt{4\nu t_0}}{\sqrt{\eta}R}$ & $\frac{t_{sbv}u_1'}{R}$ & $\sigma$\\ 
    \hline
    1.22 & 0.74 & 0.81 & 0.58 & 1.64 & 0.118 \\
    1.8 & 1.57 & 0.53 & 0.59 & 1.54 & 0.092 \\
    2.4 & 2.09 & 0.39 & 0.43 & 1.63 & 0.083 \\
    3.0 & 2.58 & 0.31 & 0.48 & 1.47 & 0.080 \\
    4.0 & 3.31 & 0.26 & 0.49 & 1.46 & 0.080 \\
    \hline
  \end{tabular}
  \captionsetup{justification=justified, singlelinecheck=false}
  \caption{Parameters for the VD mixing model applied for the VD SBI across a Mach number range of $1.22\ \sim\ 4.0$, including the total circulation $\varGamma_{t}$, the compression rate $\eta$, the initial time $t_0$ for the principal strain rate arising from the primary vortex $s_{1}^{ps}$, the emergence time for $t_{sbv}$, and the post-shock density ratio $\sigma$.}
  \label{parameters for VD mixing model}
  \end{center}
\end{table}
Using these parameters, the predictions of the VD mixing model for \textcolor{black}{the total SDR} $\left\langle \chi\right\rangle$ and the increase in mixedness $\Delta\left\langle f^{*} \right\rangle$ across the Mach number range are presented in Fig.~\ref{comparision of VD model and mixedness, SDR in a Mach number range}. As reflected in this figure, the capacity of the VD mixing model is validated to accurate capture the evolution of the mixing rate and mixedness in VD SBI across a wide range of Mach numbers.

\begin{figure}
    \centering
    \includegraphics[width=1.1\linewidth]{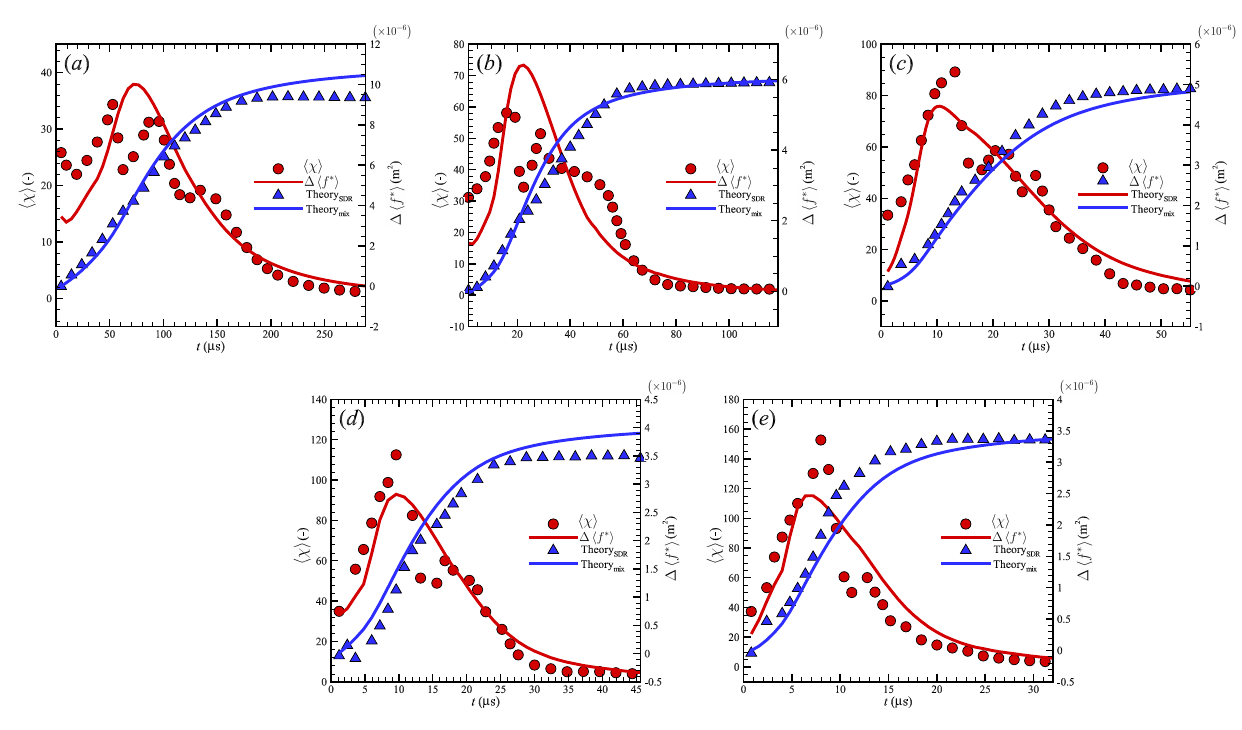}
    \captionsetup{justification=justified, singlelinecheck=false}
    \caption{\textcolor{black}{The comparison between the evolution of the \textcolor{black}{total SDR} $\left\langle \chi\right\rangle$ and increase of the \textcolor{black}{total mixedness} $\Delta \left\langle f^{*} \right\rangle$ in VD SBI across a Mach number range of $1.22\ \sim \ 4.0$ and the predictions of the mixing models, including $\left(a\right)\ Ma=1.22$, $\left(b\right)\ Ma=1.8$, $\left(c\right)\ Ma=2.4$, $\left(d\right)\ Ma=3.0$ and $\left(e\right)\ Ma=4.0$.}}
    \label{comparision of VD model and mixedness, SDR in a Mach number range}
\end{figure}

\begin{figure}
    \centering
    \includegraphics[width=1.0\linewidth]{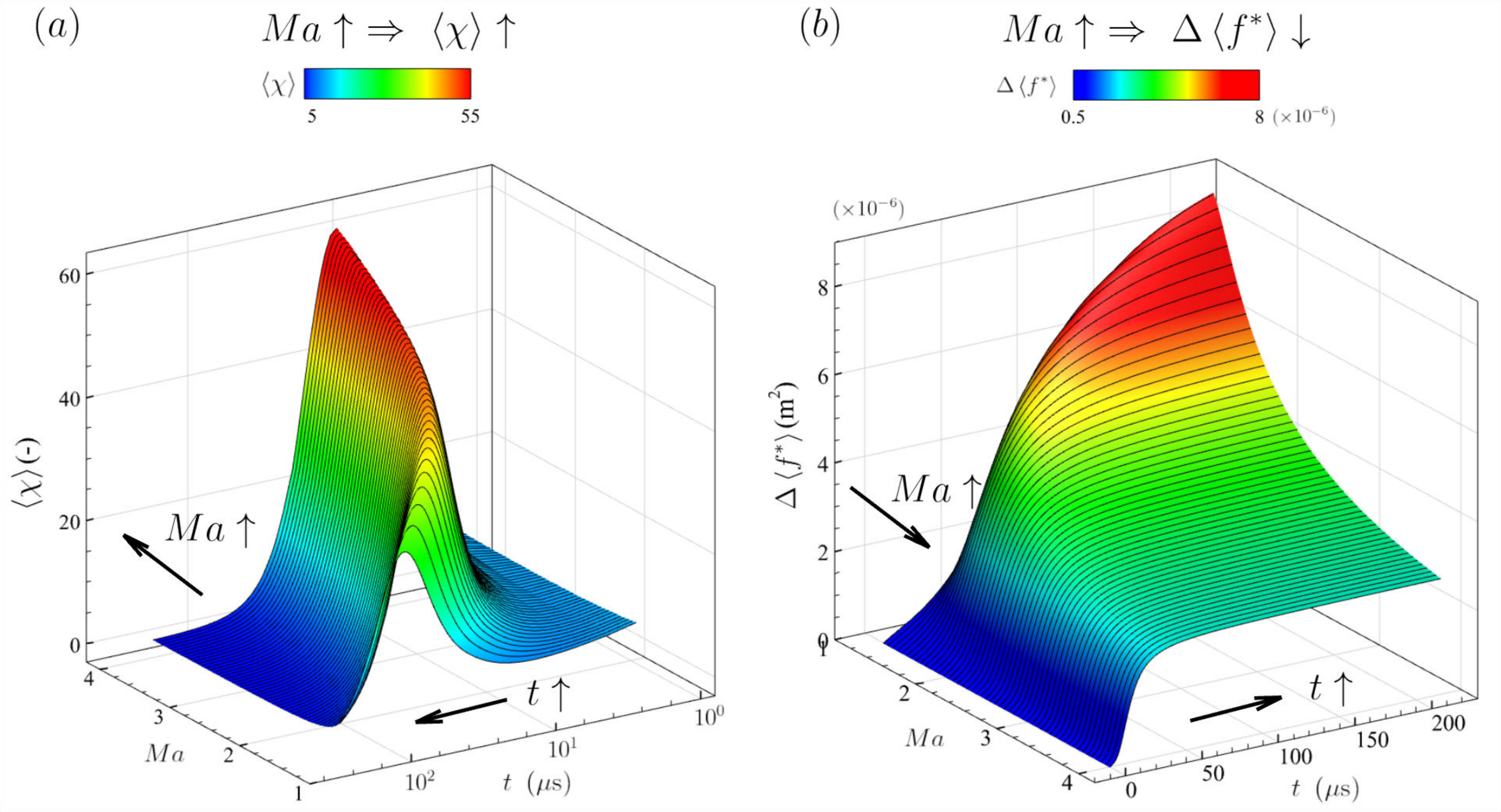}
    \captionsetup{justification=justified, singlelinecheck=false}
    \caption{The predictions by VD mixing model (Eq.~\ref{VD mixing model expression}) for $(a)$ the \textcolor{black}{total SDR} $\left\langle\chi\right\rangle$ and increase of \textcolor{black}{total mixedness} $\Delta\left\langle f^{*}\right\rangle$ within VD SBI for different Mach numbers.}
    \label{Ma number variation for model prediction}
\end{figure}

Using the VD mixing model, the behavior of the theoretical predictions for \textcolor{black}{the total SDR} $\left\langle\chi\right\rangle$ 
and the increase of \textcolor{black}{total mixedness} $\Delta\left\langle f^{*}\right\rangle$ with respect to Mach number can be analyzed. For the undetermined parameters $\left(\varGamma_{t},\eta, t_0, t_{sbv},\sigma\right)$, the total circulation $\varGamma_{t}$ is calculated by Yang's circulation model \citep{yang1994model}:
\begin{equation}
    \varGamma_{t} \approx \varGamma_{YKZ} = \frac{4R}{W_{i}}\frac{p_1'-p_1}{\rho_1'}\lvert At^{+}\rvert.
    \label{YZK model}
\end{equation}
The compression rate $\eta$ is expressed as a function of Mach number \citep{yu2020scaling}:
\begin{equation}
    \eta \approx \frac{\rho_1}{\rho_1'} = \frac{\left(\gamma - 1\right)Ma^2 + 2}{\left(\gamma+1\right)Ma^2}.
\end{equation}
The initial time $t_0$ for $s_1^{ps}$, the emergence time $t_{sbv}$, and the post-shock density ratio $\sigma$ are estimated according to table~\ref{parameters for VD mixing model} as follows:
\begin{equation}
    \frac{\sqrt{4\nu t_0}}{\sqrt{\eta}R} \approx 0.5,\quad \frac{t_{sbv}u_1'}{R} \approx 1.5, \quad \sigma \approx 0.09.
\end{equation}
Once these parameters are determined, the VD mixing model can predict the \textcolor{black}{total SDR} $\left\langle\chi\right\rangle$ 
and the increase of \textcolor{black}{total mixedness} $\Delta\left\langle f^{*}\right\rangle$, for VD SBI cases across different Mach numbers. These predictions are illustrated in Fig.~\ref{Ma number variation for model prediction}.

As shown in Fig.~\ref{Ma number variation for model prediction}, the variation of \textcolor{black}{total SDR} and \textcolor{black}{total mixedness} with Mach number is evident. As the Mach number increases, the \textcolor{black}{total SDR} $\left\langle\chi\right\rangle$ increases due to the growth of total circulation $\varGamma_{t}$ as predicted by Eq.~\ref{YZK model}. However, even with the larger \textcolor{black}{total SDR} $\left\langle\chi\right\rangle$, the mixing process in SBI cases with higher Mach numbers is completed in a shorter time, leading to a decrease in \textcolor{black}{total mixedness} $\Delta\left\langle f^{*}\right\rangle$ with increasing Mach number. These tendencies can be summarized as follows:
\begin{equation}
    \left\{
    \begin{aligned}
        & Ma \uparrow\ \Rightarrow\ \left\langle\chi\right\rangle \uparrow,\\
        &  Ma \uparrow\ \Rightarrow\ \Delta\left\langle f^{*}\right\rangle \downarrow,
    \end{aligned}
    \right.
\end{equation}
\textcolor{black}{which is consistent with the theoretical predictions shown in Fig.~\ref{comparision of VD model and mixedness, SDR in a Mach number range}}. The inherent scaling behavior of these two mixing indicators will be further examined in the subsequent discussion.

\subsection{Scaling analysis of the mixing rate on Pe number}

Based on the VD mixing model applicable to VD SBI across a Mach number range, this section examines the scaling behavior of the time-averaged mixing rate. As described in Section \ref{PS mixing theory}, the mixing process in VD SBI can be divided into two distinct stages: the growth stage and the decay stage, separated by the characteristic time $t_{charac} = \left(\frac{3}{32}\frac{s_0^2}{s_1^2\mathcal{D}}\right)^{\frac{1}{3}}$. Without loss of generality, the time interval $0 < t \leq t_{charac}$ is defined as the growth stage, while $t_{charac} < t \leq 2t_{charac}$ is defined as the decay stage. The time-averaged mixing rate during the mixing process is defined as:
\begin{equation}
    \overline{\left\langle\chi\right\rangle} = \frac{\Delta \left\langle f^* \right\rangle|_{0}^{2t_{charac}}}{2\mathcal{D}t_{charac}}. 
\end{equation}

During the growth stage $(0 < t \leq t_{charac})$, to simplify the structure of the VD mixing model (Eq. \ref{VD mixing model expression}), the principal strain rate induced by the secondary baroclinic effect $s_{1}^{b,\nu}$ is neglected. Under this assumption, Eq. \ref{VD mixing model expression} reduces to:
\begin{equation}
    \left\langle \chi\right\rangle \approx \int_{0}^{\sqrt{2\eta}R} {g\left(\tau^{\mathcal{D}}\right)}^2\sqrt{\frac{6}{\upi\mathcal{D}}} s_1 t^{\frac{1}{2}} {\rm d}r,
\end{equation}
By estimating the parameters in this equation:
\begin{equation}
    \int_{0}^{\sqrt{2\eta}R}{\rm d}r \sim \sqrt{\eta} R, \quad s_1 \sim \frac{\varGamma_{t}}{\eta R^2}
\end{equation}
the expression for $\left\langle \chi \right\rangle$ can be approximated as:
\begin{equation}
    \left\langle \chi\right\rangle \sim g\left(\tau^{\mathcal{D}}\right)^2 \sqrt{\frac{6}{\upi\mathcal{D}}} \frac{\varGamma_t}{\sqrt{\eta}R} t^{\frac{1}{2}}.
\end{equation}
Arisen from the similar post-shock density ratio listed in Table \ref{VD SBI parameters}, the modified function $g(\tau^{\mathcal{D}})$ is assumed to be nearly identical for the VD SBI across the Mach number range of $1.22 \ \sim \ 4.0$. This leads to a further simplification of this estimation:
\begin{equation}
    \left\langle \chi\right\rangle \sim \sqrt{\frac{6}{\upi\mathcal{D}}} \frac{\varGamma_t}{\sqrt{\eta}R} t^{\frac{1}{2}}.
\end{equation}
Using this estimation for the \textcolor{black}{total SDR} $\left\langle \chi\right\rangle$, the increase of \textcolor{black}{total mixedness} $\Delta \left\langle f^* \right\rangle$ in the growth stage can be approximated:
\begin{equation}
    \Delta \left\langle f^* \right\rangle \approx \left\langle T_{SDR}^{f} \right\rangle \sim 8\mathcal{D}\int_{0}^{t_{charac}} \sqrt{\frac{6}{\upi\mathcal{D}}} \frac{\varGamma_t}{\sqrt{\eta}R} t^{\frac{1}{2}} {\rm d}t = \frac{2}{\sqrt{\upi}}\eta R^2.
\end{equation}
Similarly, the approximation for the \textcolor{black}{total SDR} $\left\langle \chi \right\rangle$ in the decay stage $(t_{charac} < t \leq 2t_{charac})$ is presented as:
\begin{equation}
    \left\langle \chi\right\rangle \sim \frac{3}{16}\sqrt{\frac{3}{2\upi}}\frac{\eta^{\frac{5}{2}}R^5}{\mathcal{D}^{\frac{3}{2}}\varGamma_t}t^{-\frac{5}{2}},
\end{equation}
and the increase of \textcolor{black}{total mixedness} $\Delta\left\langle f^{*}\right\rangle$ in this stage is given by:
\begin{equation}
    \Delta \left\langle f^* \right\rangle \approx \left\langle T_{SDR}^{f} \right\rangle \sim 8\mathcal{D}\int_{t_{charac}}^{2t_{charac}} \frac{3}{16}\sqrt{\frac{3}{2\upi}}\frac{\eta^{\frac{5}{2}}R^5}{\mathcal{D}^{\frac{3}{2}}\varGamma_t}t^{-\frac{5}{2}} {\rm d}t = \frac{2}{\sqrt{\upi}}\left(2-\frac{1}{2\sqrt{2}}\right)\eta R^2.
\end{equation}
Therefore, the growth of the \textcolor{black}{total mixedness} throughout the whole mixing process is only in proportion to the compression rate:
\begin{equation}
    \Delta \left\langle f^* \right\rangle \approx \left\langle T_{SDR}^{f} \right\rangle = 8\mathcal{D}\int_{0}^{2t_{charac}}\left\langle \chi\right\rangle {\rm d}t = \frac{2}{\sqrt{\upi}}\left(3-\frac{1}{2\sqrt{2}}\right)\eta R^2 \sim \eta R^2,
\end{equation}
This finding is consist with the conclusion derived from alternative approaches in previous studies \citep{li2019gaussian, yu2020scaling}.
Moreover, for the time period associated within the mixing process $2t_{charac} = \left(\frac{3}{16}\frac{s_0^2}{s_1^2\mathcal{D}}\right)^{\frac{1}{3}}$, using the estimation for the width of scalar strips and the principal strain:
\begin{equation}
    s_0 \sim \sqrt{\eta}R,\quad s_1\sim \frac{\varGamma_{t}}{\eta R^2}
\end{equation}
this time period can be approximated as:
\begin{equation}
    2t_{charac} = \left(\frac{3}{16}\frac{s_0^2}{s_1^2\mathcal{D}}\right)^{\frac{1}{3}} \sim (s_0)^{\frac{2}{3}}(s_1)^{-\frac{2}{3}}{\mathcal{D}}^{-\frac{1}{3}} \sim \eta R^2{\varGamma_{t}}^{-\frac{2}{3}}{\mathcal{D}}^{-\frac{1}{3}},
\end{equation}
which is similar to the estimation for the mixing time reported in related studies \citep{jacobs1992shock,tomkins2008experimental,liu2022mixing}, verifying the reliability of this approximation for the time period.

\begin{figure}
    \centering
    \includegraphics[width=0.9\linewidth]{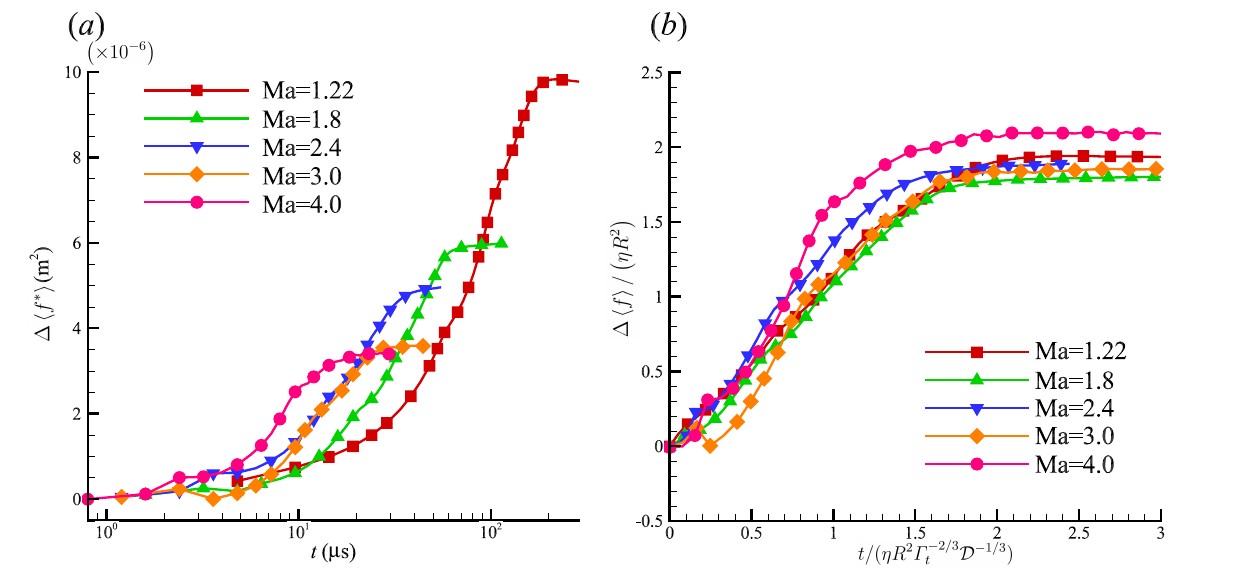}
    \captionsetup{justification=justified, singlelinecheck=false}
    \caption{\textcolor{black}{The evolution of $\left(a\right)$ the increase of total mixedness $\Delta \left\langle f^* \right\rangle$ versus time $t$; $\left(b\right)$ the normalized increase of total mixedness $\Delta \left\langle f^* \right\rangle/(\eta R^2)$ normalized by versus the normalized time $t/(\eta R^2{\varGamma_{t}}^{-\frac{2}{3}}{\mathcal{D}}^{-\frac{1}{3}})$ in the VD SBI cases with different Mach number. }}
    \label{normalization of mixedness}
\end{figure}

\begin{figure}
    \centering
    \includegraphics[width=0.9\linewidth]{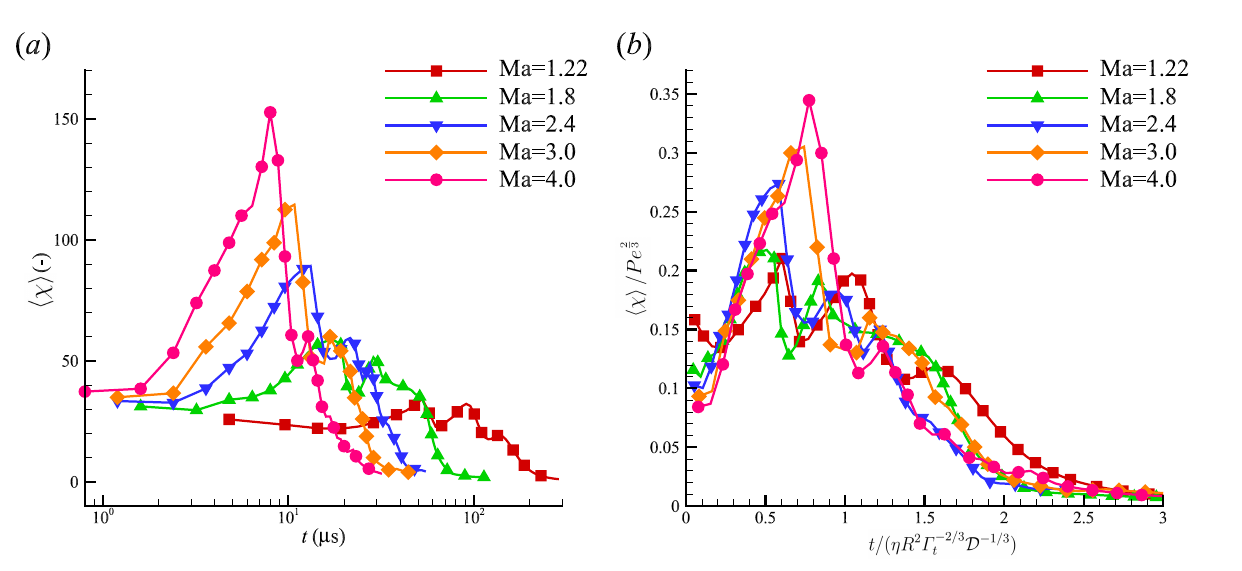}
    \captionsetup{justification=justified, singlelinecheck=false}
    \caption{\textcolor{black}{The evolution of $\left(a\right)$ the total SDR $\left\langle \chi \right\rangle$ versus time $t$; $\left(b\right)$ the normalized total SDR $\Delta \left\langle f^* \right\rangle/Pe^{\frac{2}{3}}$ versus the normalized time $t/(\eta R^2{\varGamma_{t}}^{-\frac{2}{3}}{\mathcal{D}}^{-\frac{1}{3}})$ in the VD SBI cases with different Mach number.}}
    \label{normalization of SDR}
\end{figure}

Thus, the scaling of increase of the mixedness and time period during the mixing process can be summarized as following:
\begin{equation}
    \left\{
    \begin{aligned}
        & \Delta \left\langle f^* \right\rangle \sim \eta R^2,\\
        & 2t_{charac} \sim \eta R^2{\varGamma_{t}}^{-\frac{2}{3}}{\mathcal{D}}^{-\frac{1}{3}}.
    \end{aligned}
    \right.
    \label{estimation for mixedness and mixing time}
\end{equation}
This leads to the scaling behavior of the time-averaged mixing rate: 
\begin{equation}
    \overline{\left\langle \chi\right\rangle} = \frac{\Delta \left\langle f^* \right\rangle}{2\mathcal{D}t_{charac}} \sim \left(\frac{\varGamma_{t}}{\mathcal{D}}\right)^{\frac{2}{3}}, \quad {\rm i.e.} \quad \overline{\left\langle \chi\right\rangle} \sim Pe^{\frac{2}{3}},
    \label{estimation for SDR}
\end{equation}
where $Pe = \frac{\varGamma_t}{\mathcal{D}}$ is the dimensionless P\'{e}clet number \citep{meunier2003vortices}. This dimensionless number quantifies the relative strength of the stretching dynamics compared to the diffusion process.

\begin{figure}
    \centering
    \includegraphics[width=0.6\linewidth]{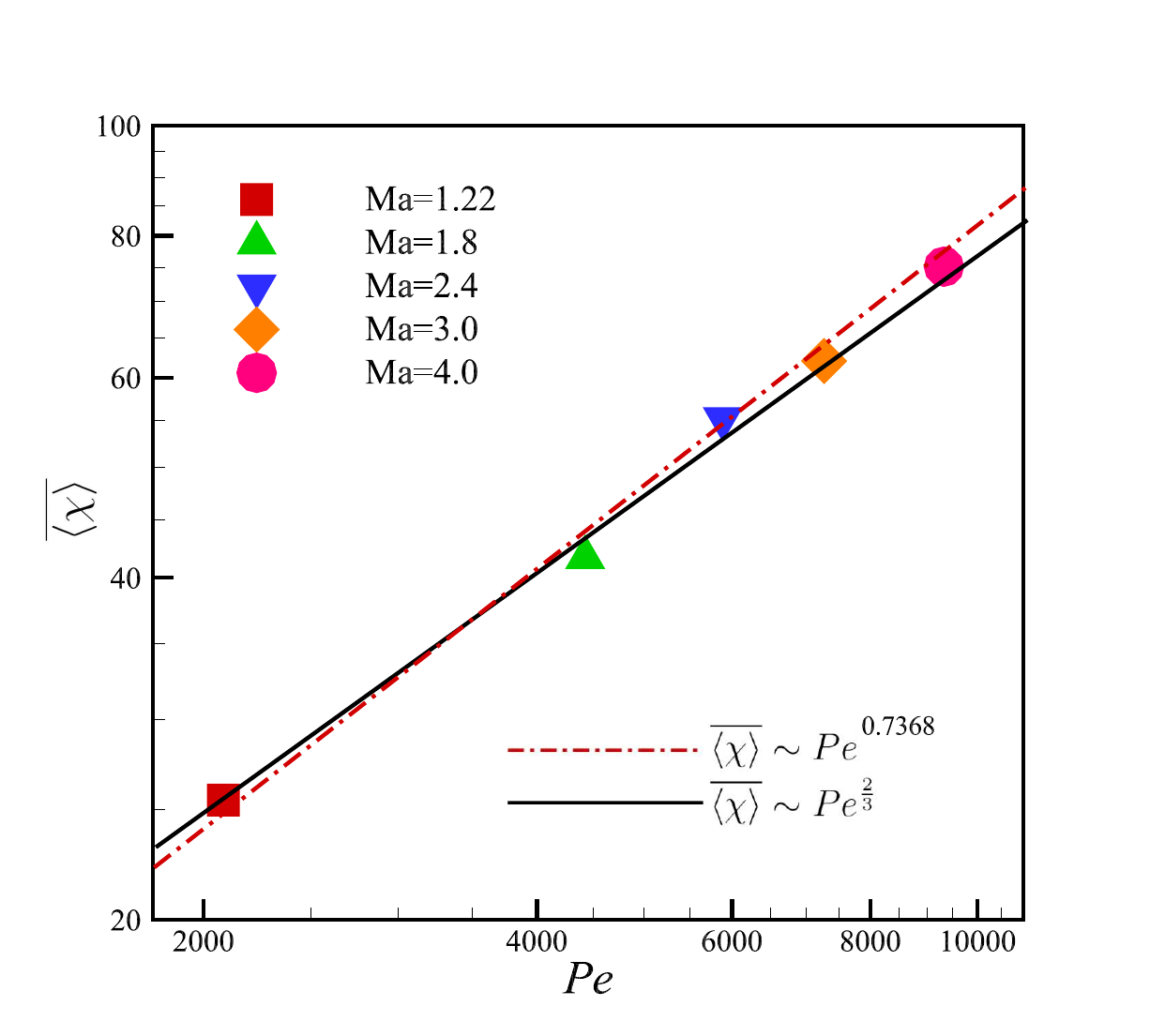}
    \captionsetup{justification=justified, singlelinecheck=false}
    \caption{\textcolor{black}{The scaling of the time-averaged mixing rate, i.e., total SDR $\overline{\left\langle\chi \right\rangle}$ on the dimensionless P\'{e}clet number $Pe$.}}
    \label{SDR scaling validation}
\end{figure}

\begin{table}
\begin{center}
\def~{\hphantom{0}}
  \begin{tabular}{lccc}
    \hline
    $Ma$ & $Pe$ & $\overline{\left\langle\chi \right\rangle}$ & $\overline{\left\langle\chi \right\rangle}/Pe^{\frac{2}{3}}$\\ 
    \hline
    1.22 & 2084.5 & 25.49 & 0.156 \\
    1.8 & 4422.5 & 41.69 & 0.155  \\
    2.4 & 5887.3 & 55.21 & 0.169 \\
    3.0 & \textcolor{black}{7267.6} & \textcolor{black}{62.07} & \textcolor{black}{0.166}  \\
    4.0 & \textcolor{black}{9323.9} & \textcolor{black}{75.15} & \textcolor{black}{0.170}  \\
    \hline
  \end{tabular}
  \captionsetup{justification=justified, singlelinecheck=false}
  \caption{\textcolor{black}{Parameters for validating the scaling law of time-averaged \textcolor{black}{total SDR} $\overline{\left\langle\chi \right\rangle}$, including P\'{e}clet number $Pe$, time-averaged \textcolor{black}{total SDR} $\overline{\left\langle\chi \right\rangle}$, and the normalized time-averaged \textcolor{black}{total SDR} $\overline{\left\langle\chi \right\rangle}/Pe^{\frac{2}{3}}$.}}
  \label{SDR scaling validation table}
  \end{center}
\end{table}

As presented in Fig.~\ref{normalization of mixedness} and Fig.~\ref{normalization of SDR}, the scaling suggested by Eq. \ref{estimation for mixedness and mixing time} and Eq. \ref{estimation for SDR} provides good collapses for the evolution of the normalized increase of the mixedness and the normalized \textcolor{black}{total SDR} over time. The scaling behavior of the time-averaged mixing rate (Eq. \ref{estimation for SDR}) is further validated through Fig.~\ref{SDR scaling validation} and Table \ref{SDR scaling validation table}. \textcolor{black}{In Fig.~\ref{SDR scaling validation}, the least-squares fit curve for cases with varying Mach numbers is $\overline{\left\langle \chi\right\rangle} \sim Pe^{0.7368}$.} This curve closely aligns with the theoretically derived scaling law $\overline{\left\langle \chi\right\rangle} \sim Pe^{\frac{2}{3}}$, with small discrepancies likely arising from oversimplifications made during our derivation. Nevertheless, due to the minimal difference between the least-squares fit curve and the theoretical scaling law, as well as the nearly identical normalized time-averaged \textcolor{black}{total SDR} in Table \ref{SDR scaling validation table}, Eq. \ref{estimation for SDR} provides a reasonable description of the dependence of the time-averaged \textcolor{black}{total SDR} on the P\'{e}clet number. Since the SDR quantifies the mixing rate, and the P\'{e}clet number measures the relative strength of stretching dynamics compared to diffusion, this dependence $\overline{\left\langle \chi\right\rangle} \sim Pe^{\frac{2}{3}}$ highlights the critical role of stretching dynamics in governing the mixing rate within this canonical VD SBI mixing process.

\section{Conclusions and future work}
\label{Conclusions and future work}
\textcolor{black}{This paper begins with the central question in variable-density (VD) mixing research: What is the role of stretching dynamics in VD mixing within the context of canonical RM-type shock-bubble interactions (SBI)? Comparing with our previous investigation on the mixing time to predict the time required to reach a steady mixing state, in this study, the relationship between VD mixing and stretching is systematically investigated by focus on qualitatively characterizing the mixing process, as reflected by evolution of the scalar dissipation rate (SDR), commonly referred to as the mixing rate, and its time integral, denote as mixedness over time. Employing high-resolution simulations, we examined the interaction of a cylindrical helium bubble with a wide range of shock Mach numbers, spanning from 1.22 to 4.}

From the starting point, by analysing the decomposition of the \textcolor{black}{total SDR} $\left\langle \chi\right\rangle$, we identified the stretching term $\left\langle T_{stretch}^{\chi}\right\rangle$ as the primary driver of SDR growth, while the strictly negative diffusion term $\left\langle T_{diff}^{\chi}\right\rangle$ predominantly contributes to the reduction of SDR. This observation reveals that mixing primarily originates from the diffusion process and is substantially enhanced by stretching dynamics, thus preliminarily highlighting the significance of stretching in the mixing process. Furthermore, the stretching rate is defined rigorously originating from the expression of stretching term $\left\langle T_{stretch}^{\chi}\right\rangle$: $R_{stretch} = 2s_{1}\lambda_{alignment}$, which is determined by the principal strain rate $s_{1}$ and scalar gradient alignment $\lambda_{alignment}$. By focusing on the governing equations for the principal strain rate $s_i$ (SDGE-$s_i$) and scalar gradient alignment $\lambda_i$ (SDGE-$\lambda_i$) within the strain rate eigenframe, we comprehensively established a framework for analyzing stretching dynamics through the integration of SDGE-$s_i$ and SDGE-$\lambda_i$. Motivated by the nearly steady single-vortex following the shock's interaction with bubble, the single-vortex stretching dynamics are researched by solving this stretching dynamics framework analytically in a cylindrical coordinate system $(r,\varphi)$ \textcolor{black}{under the assumptions for PS mixing}. The solutions of SDGE-$s_i$ and SDGE-$\lambda_i$ illustrate the algebraic stretching characteristic, which is further confirmed by the Lagrangian particle movement results.

Before investigating the role of stretching dynamics on mixing in VD SBI, we initially explore the relative simple scenario of PS SBI, where density differences are ignored, as the foundation for understanding VD mixing. By setting the origin of the cylindrical coordinate system $(r,\varphi)$ at the vortex center, the azimuthal velocity distribution indicates that mixing in PS SBI is predominantly governed by the large-scale main vortex. Further, by comparing the distribution of the azimuthal averaged principal strain rate $s_1$ and scalar gradient alignment $\lambda_{alignment}$ with single-vortex theoretical solutions, we assert that the stretching dynamics in PS SBI exhibit single-vortex algebraic stretching characteristic. According that the evolution of SDR is also influenced by molecular diffusion, the diffusion process is incorporated by solving the advection-diffusion equation on a series of scalar strips based on the analytical expressions of the single-vortex stretching dynamics. Combining the single-vortex stretching expressions and the modeled diffusion process, we propose a PS mixing model for the \textcolor{black}{total SDR} $\left\langle \chi \right \rangle$, which has good capacity to accurately predict the evolution of $\left\langle \chi \right \rangle$ in PS SBI, therefore the mixing in PS SBI is proved to be governed by the single-vortex algebraic stretching rate. The two stage approximations of this model reveal the existence of a characteristic time $t_{charac}$ and identify two distinct mixing stages divided by this characteristic time: the SDR growth stage and the SDR decay stage.

In the context of VD SBI, we identify that the presence of density variations introduces two significant density-gradient effects that challenge the extension of the PS mixing model to VD mixing. The first effect is a secondary baroclinic effect observed in the governing equations for the principal strain rate $s_1$. This effect amplifies the magnitude of $s_1$, while maintaining the algebraic stretching characteristic observed in single-vortex scenarios. The second effect is the emergence of the density-gradient source term in the VD advection-diffusion equation, which suppresses the diffusion process due to its strictly negative value. We model the secondary baroclinic effect as an additional principal strain  $s_{1}^{b,\nu}$ and the density-source effect as a modified function $g(\tau^{\mathcal{D}})$. Based on these modifications, we propose a VD mixing model for the \textcolor{black}{total SDR} $\left\langle \chi \right\rangle$ on the basis of the PS mixing model. This model is validated to accurately describe the \textcolor{black}{total SDR} $\left\langle \chi \right\rangle$ and its time integral, specifically the increment of \textcolor{black}{total mixedness} $\Delta\left\langle f^{*} \right\rangle$ in VD SBI cases across a wide range of Mach numbers from 1.22 to 4. Consequently, a qualitative relationship between stretching dynamics and VD mixing is established by this model. To further explain the essential role of stretching dynamics in VD mixing, we conduct scaling analysis on the time-averaged mixing rate $\overline{\left\langle \chi \right\rangle}$. We derive that the time-averaged mixing rate $\overline{\left\langle \chi \right\rangle}$ is directly proportional to $Pe^{\frac{2}{3}}$ as: $\overline{\left\langle \chi \right\rangle} \sim Pe^{\frac{2}{3}}$. Here, the dimensionless P\'{e}clet number $Pe = {\varGamma_{t}}/{\mathcal{D}}$ serves as a measure of the relative strength of stretching dynamics compared to the diffusion process, thus highlighting the crucial role of stretching dynamics in mixing through this scaling behavior.

In conclusion, this paper systematically explores the stretching dynamics and its effects on VD mixing within the context of the typical single-vortex scenario in VD SBI. The established relationship between stretching dynamics and the evolution of the mixing rate provides a novel perspective on the fundamental mechanism by which the flow governs the mixing process. The proposed VD mixing model serves as a rapid estimation tool for predicting the mixing rate and the extent of mixing in VD SBI. \textcolor{black}{Furthermore, the findings of this study will pave the way for future research on VD mixing in practical engineering applications, such as oblique shock/jet interactions. Future work will focus on the extension of the current VD mixing model to VD SBI with sharp contact and multi-mode RMI scenarios.}

\backsection[Supplementary materials]{\label{SupMat}The supplementary materials are included in the attachment files.}

\backsection[Funding]{Thanks to the Postdoctoral Fellowship Program (Grade C) of China Postdoctoral Science Foundation (Grant Number GZC20231566), Sichuan Science and Technology Program (2025ZNSFSC0834), Natural Science Foundation of Shanghai (Youth Projects 25ZR1402273) and Natural Science Foundation of China (nos. 91941301) for funding and supporting this research.}

\backsection[Declaration of interests]{The authors report no conflict of interest.}

\appendix
\section{Initial conditions for VD SBI across a wide range of Mach number}
\label{wide range Ma number VD SBI settings}

This section outlines the case settings for the VD SBI across a range of Mach numbers, spanning from 1.22 to 4. The initial conditions for the VD SBI cases are depicted in Fig.~\ref{Initial conditions}, which are identical to those specified for the $Ma = 2.4$ case in Section \ref{Numerical method}. The primary distinction among the cases with varying shock strengths lies in the post-shock conditions $\left(u_1',\rho_1',P_1', T_1'\right)$. According to Rankine-Hugoniot equation, the post-shock air conditions for the VD SBI cases are summarized in Table~\ref{VD SBI parameters}. The simulation results for these SBI cases are also analyzed by the total circulation $\varGamma_{t}$ and compression rate $\eta$.
\begin{table}
\begin{center}
\def~{\hphantom{0}}
  \begin{tabular}{lccccccccc}
    \hline
    $Ma$ & $p_{1}'\,\rm{(Pa)}$ & $T_{1}'\,\rm{(K)}$ & $u_{1}'\,\rm{(m\,s^{-1})}$ & $W_{t}\,\rm{(m\ s^{-1})}$ & $At^{+}\,\left(-\right)$ & $\mathcal{D}\,\rm{(m^{2}\,s^{-1})}$ & $\varGamma_{t}\,\rm{(m^{2}\,s^{-1})}$ & $Re\,\left(-\right)$ & $Pe\,\left(-\right)$\\ 
    \hline
    1.22 & 159036.9 & 334.06 & 114.74 & 1121.89 & -0.789 & $355 \times 10^{-6}$ & 0.74 & 5083.3 & 2084.5 \\
    1.8 & 366015.0 & 448.28 & 356.58 & 1440.00 & -0.831 & $355 \times 10^{-6}$ & 1.57 & 20975.1 & 4422.5 \\
    2.4 & 667391.3 & 596.87 & 568.31 & 1678.08 & -0.845 & $355 \times 10^{-6}$ & 2.10 & 38446.2 & 5887.3 \\
    3.0 & 1046646.5 & 783.41 & 768.11 & 1958.78 & -0.852 & \textcolor{black}{$355 \times 10^{-6}$} & 2.58 & 57473.6 & \textcolor{black}{7267.6} \\
    4.0 & 1873802.9 & 1182.91 & 1074.53 & 2439.15 & -0.852 &  \textcolor{black}{$355 \times 10^{-6}$} & 3.31 & 95307.9 & \textcolor{black}{9323.9} \\
    \hline
  \end{tabular}
  \captionsetup{justification=justified, singlelinecheck=false}
  \caption{Parameters of VD SBI cases within a wide range of $Ma$ number, including the post-shock air pressure $p_{1}'$, post-shock air temperature $T_{1}'$, post-shock air velocity $u_{1}'$, the transmitted shock-wave speed $W_{t}$, which are calculated from the Rankine-Hugoniot equation. The post-shock Atwood number, which presents the difference between the helium bubble density following the shock action $\rho_{2}'$ and the post-shock ambient air density $\rho_{1}'$, is defined as $At^{+} = \left(\rho_{2}'-\rho_{1}'\right)/\left(\rho_{2}'+\rho_{1}'\right)$. $\mathcal{D}$ is the constant Fickian diffusivity. The total circulation $\varGamma_{t}$ and the P\'{e}clet $Pe = \varGamma_{t}/\mathcal{D}$ are also demonstrated in this table to reflect the strength of stretching for different cases. While the Reynolds number is defined as $Re = \varGamma_{t}/\nu$, here $\nu = \mu/\overline{\rho'}$ is the kinematic viscosity, $\overline{\rho'} = \left(\rho_{1}' + \rho_{2}'\right)/2$ is the average density of the post-shock bubble and ambient air. }
  \label{VD SBI parameters}
  \end{center}
\end{table}

\begin{figure}
  \centering
    \includegraphics[clip=true,width=0.95\textwidth]{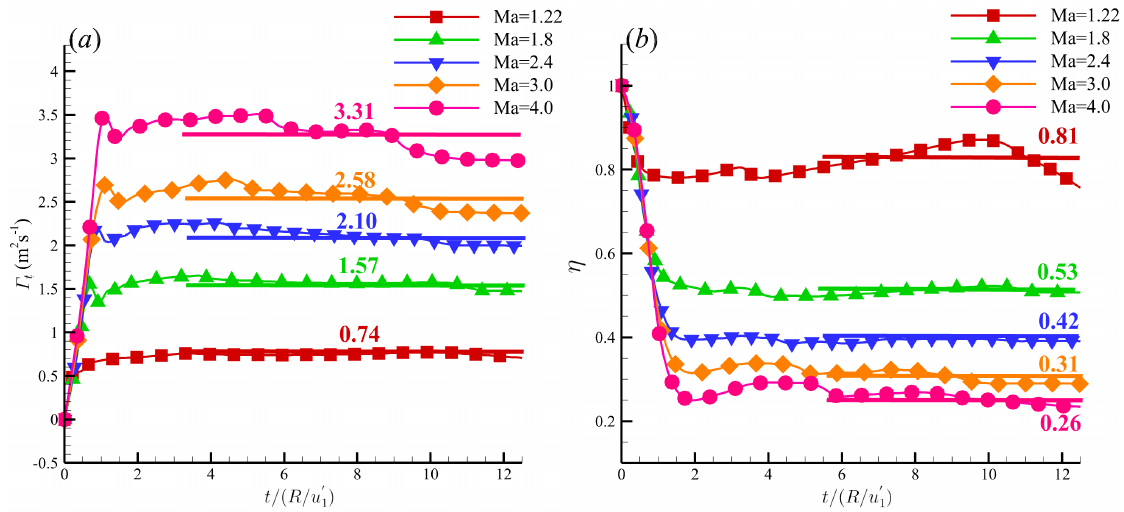}\\
    \captionsetup{justification=justified, singlelinecheck=false}
    \caption{Time evolution of $(a)$ the total circulation $\varGamma_{t}$ and $(b)$ the compression rate $\eta$ for different shock Mach number cases.}
    \label{total circulation and compression rate}
\end{figure}

Figure \ref{total circulation and compression rate} presents the time evolution of these two parameters across different shock Mach numbers. It is observed that, following the shock interaction with the bubble, both the total circulation $\varGamma_{t}$ and the compression rate $\eta$ approximate constant values, indicating the formation of a nearly steady single-vortex. With increasing shock Mach number, the total circulation  $\varGamma_{t}$ increases, while the compression rate $\eta$ decreases, consistent with previous research findings \citep{niederhaus2008computational}.

\section{Grid resolution study}
\label{grid resolution study}

\begin{figure}
    \centering
    \includegraphics[width=0.6\linewidth]{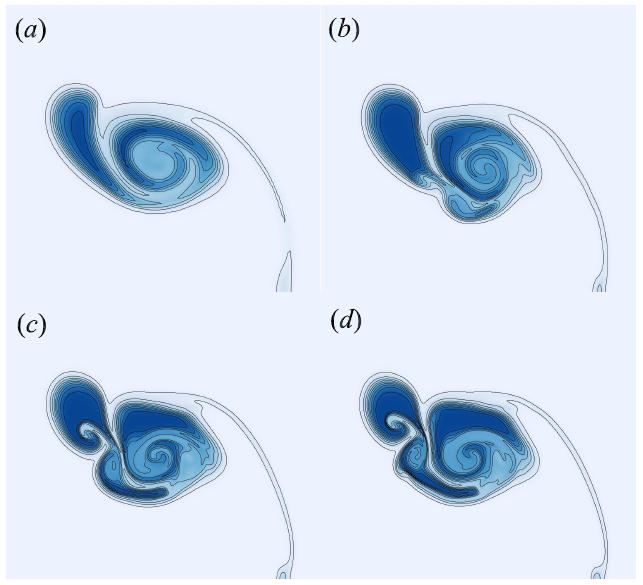}
    \captionsetup{justification=justified, singlelinecheck=false}
    \caption{\textcolor{black}{Grid resolution study on the mass fraction contour at $t=21.6\ {\rm \mu s}$. $(a)$ Mesh-1: $\Delta = 5.0 \times 10^{-5}\ {\rm m}$, $(b)$ Mesh-2: $\Delta = 2.5 \times 10^{-5}\ {\rm m}$, $(c)$ Mesh-3: $\Delta = 1.25 \times 10^{-5}\ {\rm m}$, $(d)$ Mesh-4: $\Delta = 0.625 \times 10^{-5}\ {\rm m}$}}
    \label{Mesh resolution on contour}
\end{figure}

\textcolor{black}{The sensitivity of the mixing indicator SDR to mesh resolution necessitates careful selection of grid size to ensure the resolution of numerical results. In this study, four mesh resolutions with consistent refinement ratio are evaluated: $\Delta = 5.0 \times 10^{-5}\ {\rm m}$, $\Delta = 2.5 \times 10^{-5}\ {\rm m}$, $\Delta = 1.25 \times 10^{-5}\ {\rm m}$ and $\Delta = 0.625 \times 10^{-5}\ {\rm m}$. The interaction of a $Ma = 2.4$ shock with a helium bubble is selected as the representative case. Qualitative comparisons among these mesh resolutions are presented in Fig.~\ref{Mesh resolution on contour}. By comparing the mass fraction contour of four different resolutions, it is evident that that small-scale flow structures are significantly smeared by numerical viscosity and diffusivity in coarser meshes, specifically for grid sizes $\Delta = 5.0 \times 10^{-5}\ {\rm m}$ and $\Delta = 2.5 \times 10^{-5}\ {\rm m}$. In contrast, these structures are well-captured in finer meshes, with grid sizes $\Delta = 1.25 \times 10^{-5}\ {\rm m}$ and $\Delta = 0.625 \times 10^{-5}\ {\rm m}$. Additionally, it can also be found that a high degree of consistency in flow structures between the two finest resolutions, Mesh-3 and Mesh-4, indicating sufficient convergence.}

\begin{figure}
    \centering
    \includegraphics[width=0.8\linewidth]{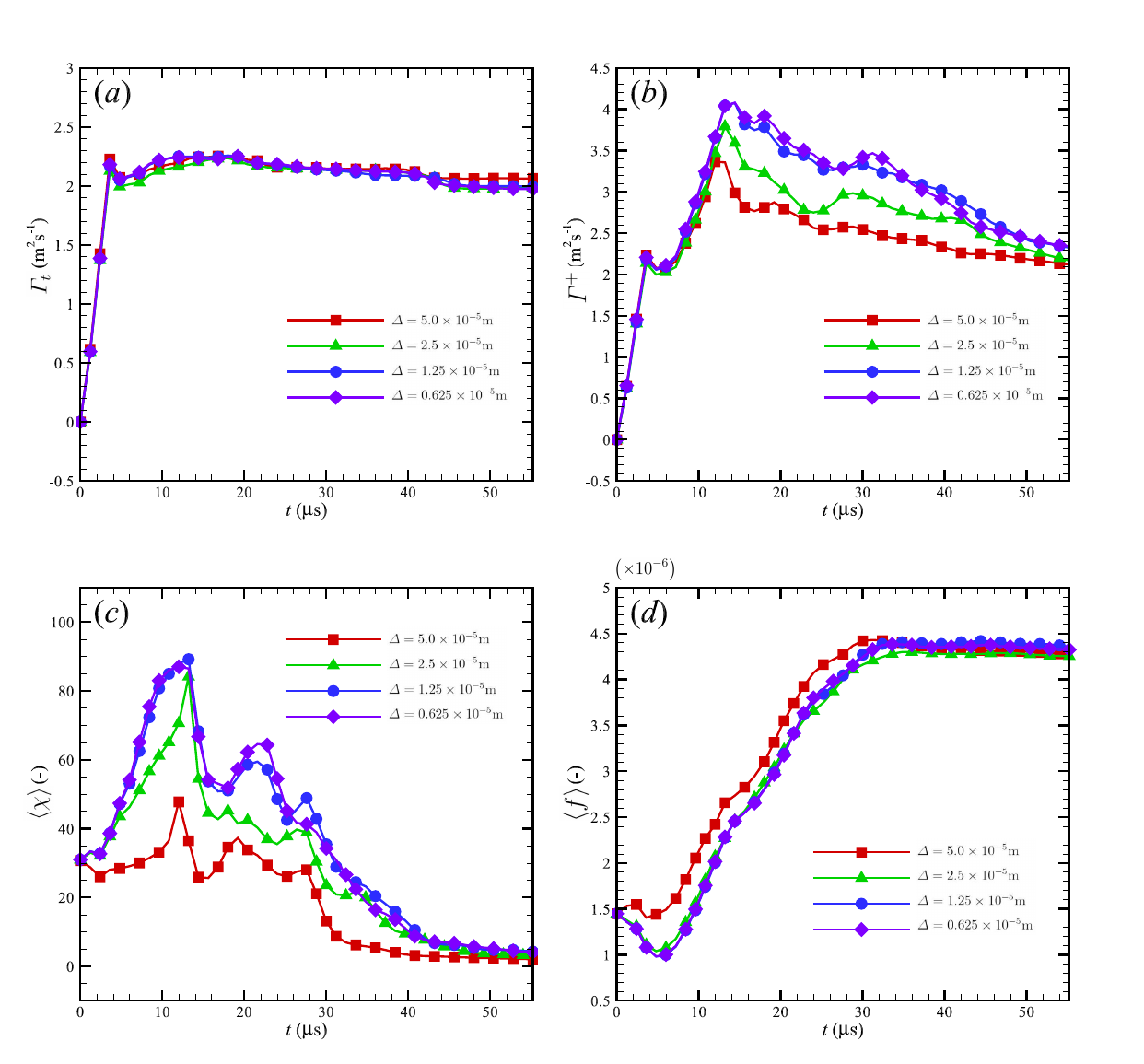}
    \captionsetup{justification=justified, singlelinecheck=false}
    \caption{\textcolor{black}{Comparison of qualitative parameters from different mesh resolutions. $(a)$ Total circulation $\varGamma_{t}$, $(b)$ positive circulation $\varGamma^{+}$, $(c)$ \textcolor{black}{total SDR} $\left\langle \chi \right\rangle$, $(d)$ \textcolor{black}{total mixedness} $\left\langle f \right\rangle$.}}
    \label{Mesh resolution on parameters}
\end{figure}

\textcolor{black}{Figure \ref{Mesh resolution on parameters} presents a comparison of qualitative parameters across different mesh resolutions. Regarding flow characteristics, Fig.~\ref{Mesh resolution on parameters} $(a)$ and Fig.~\ref{Mesh resolution on parameters} $(b)$ demonstrate that both total circulation $\varGamma_{t}$ and positive circulation $\varGamma^{+}$ exhibit similar trends for Mesh-3 and Mesh-4. The mixing characteristics observed in different meshes are evaluated by comparing the \textcolor{black}{total SDR} $\left\langle \chi \right\rangle$ and \textcolor{black}{total mixedness} $\left\langle f \right\rangle$ in Fig.~\ref{Mesh resolution on parameters} $(c)$ and Fig.~\ref{Mesh resolution on parameters} $(d)$, respectively. The close alignment in the evolution of these mixing indicators between Mesh-3 and Mesh-4 confirms the grid independence of the mixing process. Taking into account computational cost and simulation accuracy, we select the mesh resolution of $\Delta = 1.25 \times 10^{-5}\ {\rm m}$ for the present study. This resolution appropriately captures the dynamics of flow evolution and the mixing process with quantitative accuracy.}

\section{Validation of the SDGE-$s_i$ (Eq. \ref{principal strain rate equation}) and SDGE-$\lambda_{i}$ (Eq. \ref{alignment equation})}
\label{Validation of the stretching dynamics framework}

\begin{figure}
    \centering
    \includegraphics[width=1.0\linewidth]{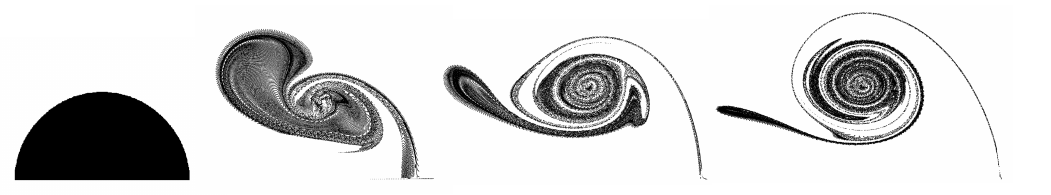}
    \captionsetup{justification=justified, singlelinecheck=false}
    \caption{The results of the Lagrangian particle movement at the moment $t=0$, $t=47\ {\rm \mu s}$, $t=94\ {\rm \mu s}$, and $t=141\ {\rm \mu s}$ for the case with a shock strength of $Ma = 1.22$.}
    \label{partical movement for VD SBI}
\end{figure}

\begin{figure}
  \centering
    \includegraphics[clip=true,width=1.0\linewidth]{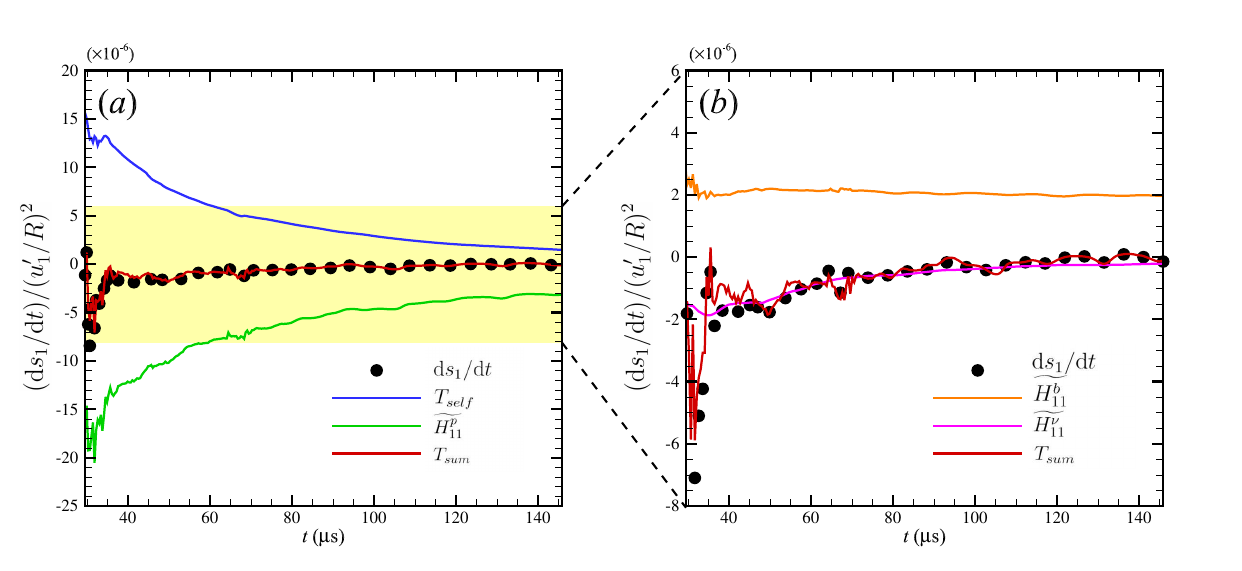}\\
    \captionsetup{justification=justified, singlelinecheck=false}
    \caption{\textcolor{black}{Comparison of the material derivative of the principal strain $s_1$, obtained from Lagrangian particle movement results, with the terms on the right-hand side of Eq.~\ref{SDGE1 in 2D SBI}. Subfigure $(a)$ shows the terms with large magnitudes, while subfigure $(b)$ presents the terms with small magnitudes.}}
    \label{validation of SDGE1}
\end{figure}

\begin{figure}
  \centering
    \includegraphics[clip=true,width=1.0\linewidth]{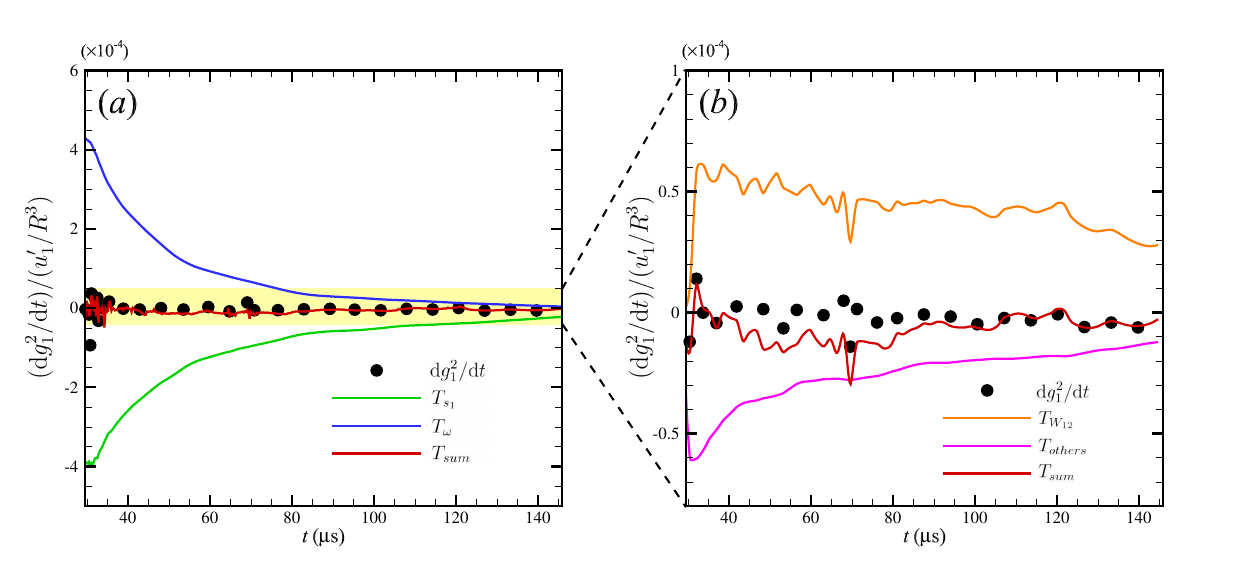}\\
    \captionsetup{justification=justified, singlelinecheck=false}
    \caption{\textcolor{black}{Comparison of the material derivative of the projection of the scalar gradient $g_1^{2}$, which are obtained from Lagrangian particle movement results, with the terms in the right side of Eq.~\ref{SDGE2 in 2D SBI}. Subfigure $(a)$ presents the terms with large magnitude, while subfigure $(b)$ presents the terms with small magnitude.}}
    \label{validation of SDGE2}
\end{figure}

This section validates the stretching dynamics framework, which consists of the SDGE-$s_i$ (Eq. \ref{principal strain rate equation}) and SDGE-$\lambda_{i}$ (Eq. \ref{alignment equation}), by using Lagrangian methods. A cylindrical bubble interacting with a $Ma = 1.22$ shock wave is selected as the representative case. At the initial time, approximately $n = 80600$ passive particles are seeded within the bubble, and their trajectories are computed by solving the advection ordinary differential equation (Eq. \ref{Partical integration}) using a fourth-order Runge–Kutta integrator:
\begin{equation}
    \frac{{\rm d}\boldsymbol{x}}{{\rm d}t} = \boldsymbol{u}\left(\boldsymbol{x},t\right),\boldsymbol{x} = \left(x,y\right), \boldsymbol{u} = \left(u,v\right),
    \label{Partical integration}
\end{equation}
where the evolution of the passive particles is illustrated in Fig. \ref{partical movement for VD SBI}.

For the SDGE-$s_i$ (Eq. \ref{principal strain rate equation}), we first validate the governing equation for the principal strain rate $s_1$ in two-dimensional VD SBI:
\begin{equation}
    \frac{{\rm d}s_1}{{\rm d}t} = -s_1^2 + \frac{1}{4}\omega^2 + \widetilde{H_{11}^{p}} + \widetilde{H_{11}^{b}} + \widetilde{H_{11}^{\nu}},
    \label{SDGE1 in 2D SBI}
\end{equation}
where $T_{self} = -s_1^2 + \frac{1}{4}\omega^2$ represents the self-interaction term. On the left-hand side, the material derivative is computed along the trajectories of the passive particles as:
\begin{equation}
    \frac{{\rm d}s_1}{{\rm d}t} = \frac{1}{n}\sum_{i=1}^{n}\frac{s_1(\boldsymbol{x}_i(t + {\rm d} t),t + {\rm d} t) - s_1(\boldsymbol{x}_i(t),t)}{{\rm d}t},
\end{equation}
while the terms on the right-hand side are evaluated using cubic spline interpolation \citep{chian2014detection,hang2020objective}. \textcolor{black}{Figure \ref{validation of SDGE1} presents the temporal evolution of the material derivative $\frac{{\rm d}s_1}{{\rm d}t}$ and the individual terms on the right-hand side of Eq.~\ref{SDGE1 in 2D SBI}. The small discrepancy between the black symbols, representing  $\frac{{\rm d}s_1}{{\rm d}t}$, and the red line, corresponding to the summation of the right-hand side terms ($T_{sum}$), confirms the validity of Eq. \ref{SDGE1 in 2D SBI}.}

Next, the governing equation for the rotation rate of eigenvectors $\widetilde{W_{12}} = \frac{{\rm d} \boldsymbol{e_2}}{{\rm d}t} \cdot \boldsymbol{e_1}$ is considered. Since $\widetilde{W_{12}}$ is not directly measurable, the governing equation for $\widetilde{W_{12}}$ in two-dimensional VD SBI,
\begin{equation}
    \left(s_2 - s_1\right)\widetilde{W_{12}} = \widetilde{H_{12}^{p}} + \widetilde{H_{12}^{b}} + \widetilde{H_{12}^{\nu}},
    \label{W12 in 2D SBI}
\end{equation}
cannot be directly validated. However, as $\widetilde{W_{12}}$ appears in SDGE-$\lambda_i$ (Eq. \ref{alignment equation}), the validation of Eq. \ref{W12 in 2D SBI} is conducted indirectly by verifying SDGE-$\lambda_i$. Specifically, according to the derivation of SDGE-$\lambda_i$ in the supplementary materials, the verification of SDGE-$\lambda_i$ is equivalent to validating the governing equation for the projection of the scalar gradient $\widetilde{g_1}^2 = \left(\nabla Y \cdot \boldsymbol{e_1}\right)^2$, which is given by:
\begin{equation}
    \frac{{\rm d} \widetilde{g_1}^2}{{\rm d}t} = -2s_1 \widetilde{g_1}^2 - 2\widetilde{W_{12}}\widetilde{g_1} \widetilde{g_2} - \omega \widetilde{g_1} \widetilde{g_2} + 2\widetilde{g_1}\widetilde{T_1^{{diff}}} = T_{s_1} + T_{\omega} + T_{W_{12}} + T_{{others}},
    \label{SDGE2 in 2D SBI}
\end{equation}
where $T_{others}$ accounts for diffusion effect on scalar gradient alignment, and $\widetilde{W_{12}}$ is computed as:
\begin{equation}
    \widetilde{W_{12}} = \frac{\widetilde{H_{12}^{p}} + \widetilde{H_{12}^{b}} + \widetilde{H_{12}^{\nu}}}{s_2 -s_1}.
\end{equation}
\textcolor{black}{Figure \ref{validation of SDGE2} shows the close agreement between the black symbols, representing $\frac{{\rm d}g_1^2}{{\rm d}t}$, and the red line, representing the summation of terms on the right-hand side ($T_{sum}$). This agreement confirms the validity of both Eq. \ref{W12 in 2D SBI} and Eq. \ref{SDGE2 in 2D SBI}. Moreover, it is also evident that the diffusion effect on scalar gradient alignment reflected by $T_{others}$ can be nearly neglected.}

At this stage, both  SDGE-$s_i$ (Eq. \ref{principal strain rate equation}) and SDGE-$\lambda_i$ (Eq. \ref{alignment equation}) in two-dimensional VD SBI have been successfully validated.

\bibliographystyle{jfm}
\bibliography{mybibfile}


\end{document}